\documentclass[review,10pt]{elsarticle}

\usepackage{lineno}
\modulolinenumbers[5]

\usepackage{hyperref} 
\hypersetup{ 
  colorlinks   = true, 
  urlcolor     = blue, 
  linkcolor    = blue, 
  citecolor    = blue, 
} 
\usepackage[fleqn]{amsmath}
\usepackage{amsfonts}
\usepackage{amssymb}
\usepackage{amsthm}

\usepackage{pifont}
\usepackage{natbib}
\usepackage{geometry}
\usepackage{graphicx}
\usepackage{txfonts}

\usepackage[utf8]{inputenc}
\usepackage{caption}
\usepackage{subcaption}
\graphicspath{ {images/} }
\usepackage[usenames, dvipsnames]{color}
\usepackage{color,soul}

\journal{Journal of Computational Physics}

\biboptions{sort&compress}

\begin{document}

\begin{frontmatter}

\title{Liquid-jet instability at high pressures with real-fluid interface thermodynamics}

\author{Jordi Poblador-Ibanez\fnref{myfootnote1}\corref{mycorrespondingauthor}}
\ead{poblador@uci.edu}
\author{William A. Sirignano\fnref{myfootnote3}}
\address{University of California, Irvine, CA 92697-3975, United States}
\fntext[myfootnote1]{Graduate Student Researcher, Department of Mechanical and Aerospace Engineering.}
\fntext[myfootnote3]{Professor, Department of Mechanical and Aerospace Engineering.}


\cortext[mycorrespondingauthor]{Corresponding author}


\begin{abstract}
The injection of liquid fuel at supercritical pressures is a relevant topic in combustion, but usually overlooked. In the past, the wrong assumption whereby the liquid experiences a fast transition to a supercritical state was made, thus neglecting any role of two-phase interface dynamics in the early stages of the atomization process. However, recent studies have shown that local thermodynamic phase equilibrium and mixing between the involved species allow the coexistence of both phases in this pressure range. In this work, a Volume-of-Fluid method adapted to variable-density real fluids is used to solve the low-Mach-number governing equations coupled with a thermodynamic model based on the Soave-Redlich-Kwong equation of state. The mixing process, interface thermodynamics and early deformation of a cool liquid jet composed of \textit{n}-decane surrounded by a hotter gas composed of oxygen at 150 bar are analyzed. Although heat conducts from the hotter gas into the liquid, net condensation can provide the proper local energy balance at high pressures. Then, vaporization and condensation may happen simultaneously at different interface locations. As pressure increases, liquid and gas mixtures become more alike in the vicinity of the interface. Thus, a combination of low surface tension force and gas-like liquid viscosities causes an early growth of surface instabilities. Early results indicate some similarity with high-Weber-number incompressible flows. The role of vortex dynamics on the interface deformation is analyzed by using the \(\lambda_\rho\) dynamical vortex identification method.
\end{abstract}


\end{frontmatter}


\setlength\abovedisplayshortskip{0pt}
\setlength\belowdisplayshortskip{-5pt}
\setlength\abovedisplayskip{-5pt}

\section{Introduction}

High-pressure combustion chambers are used in many propulsion applications. For instance, gas turbines operate in the range of 25 to 40 bar while rocket engines typically operate between 70 and 200 bar. The interest in reaching such high pressures is not only related to flow expansion through a nozzle for propulsive reasons, but also related to an optimization of the combustion efficiency and energy conversion per unit mass of fuel. We can expect that more engines will operate at increasing pressures as time passes. \par 

For liquid fuels, the atomization process whereby the liquid deforms and breaks up into small droplets is crucial. These droplets vaporize and mix with the surrounding oxidizer as the combustion chemical reaction occurs. Understanding this mixing process allows engineers to design combustion chambers, the distribution of injectors and their size and shape. Well-known fuels used in the aforementioned applications, such as Jet-A or RP-1, are hydrocarbon mixtures. Their critical pressures are in the 20-bar range; thus, combustion chambers already operate in near-critical or supercritical pressure conditions for the liquid fuel. \par 

A wide range of experimental and numerical atomization studies exist for subcritical conditions. At these low pressures, the thermodynamic behavior of the interface is simpler and both phases can clearly be identified. However, experiments at high pressure show the occurrence of a thermodynamic transition where the liquid-gas interface is rapidly affected by turbulence and is immersed in a variable-density layer~\cite{mayer1996propellant,h1998atomization,mayer2000injection,oschwald2006injection,chehroudi2012recent,falgout2016gas,desouza2017sub,ayyappan2020study}. Therefore, the identification of a two-phase behavior becomes problematic. \par 

Past works assumed the liquid phase undergoes a fast transition to a gas-like supercritical state~\cite{spalding1959theory,rosner1967liquid}; yet evidence of a two-phase behavior at supercritical pressures exists under the assumption that the interface must be in local thermodynamic equilibrium (LTE)~\cite{hsieh1991droplet,delplanque1993numerical,yang1994vaporization,sirignano1997selected,poblador2018transient}. As pressure increases beyond the liquid critical pressure, the dissolution of lighter gas species into the liquid phase is enhanced and diffusion layers grow quickly on both sides of the interface~\cite{poblador2018transient,davis2019development,poblador2021selfsimilar}. The liquid and gas mixtures look more alike in the vicinity of the interface and present strong variations of fluid properties across mixing regions. Mixture critical properties also change and, in general, the liquid critical pressure near the interface will be higher than the ambient pressure~\cite{poblador2018transient}. Thus, surface tension and an energy of vaporization will persist. \par 

The LTE assumption must be revised under certain circumstances. Dahms and Oefelein~\cite{dahms2013transition,dahms2015liquid,dahms2015non} and Dahms~\cite{dahms2016understanding} found that at supercritical pressures the interface layer might be in non-equilibrium when approaching the mixture critical temperature. In this scenario, the interface layer thickness grows and a diffuse region of a few nanometers thickness exists, rather than a true discontinuity. Other circumstances where LTE might also fail relate to the interface thermal resistivity. If large enough, the interface temperature may not be equal on both sides of the interface and a non-equilibrium region develops~\cite{stierle2020selection}. Nevertheless, the interface could still be considered a discontinuity under LTE for practical purposes. For example, compressive shocks are treated as discontinuities while the shock non-equilibrium layer thickness is at least an order of magnitude greater than the transition region for phase non-equilibrium. \par 

The configuration to be addressed herein with a discontinuity in density presents a tougher computational challenge and has a wider range of engineering applications than the pseudo-two-phase situation where continuity with large density gradients can be found. The domain in a combustor near the fuel injectors is more likely to fit the scenario treated in this paper. \par 

At these supercritical pressures, surface tension forces and liquid viscosity near the interface are reduced but remain with finite values~\cite{yang2000modeling,poblador2021vof}. Therefore, a fast distortion of the interface happens, causing the breakup of very small droplets and a rapid radial development of the two-phase mixture. This behavior is already observed in incompressible liquid jets subject to conditions resembling supercritical pressure environments~\cite{jarrahbashi2014vorticity,jarrahbashi2016early,zandian2017planar,zandian2018understanding,zandian2019length,zandian2019vorticity}.  Overall, traditional visual experimental techniques (e.g., shadowgraphy) might fail in capturing clear two-phase phenomena due to scattering and refraction caused by a cloud of very small droplets submerged in a variable-density layer. Some progress is being made to overcome these difficulties and develop new experimental techniques fit for supercritical environments~\cite{minniti2018ultrashort,minniti2019femtosecond,traxinger2019experimental,klima2020quantification}. \par 

The choice of the proper numerical approach to solve supercritical two-phase flows is a challenge. The interface must be tracked while considering mass and heat exchange between liquid and gas, as well as the liquid local volume change due to fluid compressibility. The governing equations must be solved considering a sharp interface whose solution is governed by local jump conditions and LTE. Moreover, a real-gas thermodynamic model is implemented to address the properties of the high-pressure, non-ideal fluid. Altogether, these requirements for supercritical two-phase flows are numerically expensive; thus, a computationally-efficient method is desired to minimize the cost increase. \par

In this work, the two-phase, low-Mach-number governing equations are solved, coupled with a thermodynamic model based on a volume-corrected Soave-Redlich-Kwong equation of state~\cite{lin2006volumetric}. A Volume-of-Fluid method (VOF) adapted to variable-density liquids with phase change is used to track and maintain a sharp interface~\cite{poblador2021vof}. The algorithm is an extension of the computationally-efficient and mass-conserving incompressible VOF method developed in Baraldi et al.~\cite{baraldi2014mass}, Dodd and Ferrante~\cite{dodd2014fast} and Dodd et al.~\cite{dodd2021vof}. Nevertheless, mass conservation is not satisfied to machine precision in variable-density flows due to the finite resolution of the density field and inaccuracies arising in under-resolved regions. \par 

This numerical foundation is used to analyze the early deformation of supercritical liquid planar jets. Emphasis is made in examining the mixing in both phases and the interface thermodynamics to see how the variations in fluid properties affect the early stages of the atomization process in real liquids. Moreover, the role of vorticity dynamics is studied. \par 

Past works by Jarrahbashi et al.~\cite{jarrahbashi2014vorticity,jarrahbashi2016early} and Zandian et al.~\cite{zandian2017planar,zandian2018understanding,zandian2019length,zandian2019vorticity} have used vorticity dynamics to analyze in detail the generation of liquid structures and the breakup process of various incompressible liquid configurations (e.g., round jet or planar jet). It was found that the cascade of liquid structures (e.g., formation and stretching of lobes) can be explained by analyzing the interaction between hairpin vortices and Kelvin-Helmholtz vortices. Various atomization domains were identified under characteristic deformation patterns depending on the flow configuration. \par 

A similar but preliminary study is performed in the present work without going into the detailed interaction between vortical structures. Vortices are identified by using the dynamical vortex identification method \(\lambda_\rho\)~\cite{yao2018toward}, which is a compressible extension of the incompressible identification method \(\lambda_2\) by Jeong and Hussain~\cite{jeong1995identification}.

The structure of this paper is as follows. First, the governing equations simplified for low-Mach-number flows are presented. The matching relations (i.e., jump conditions and LTE) that link the solutions in both phases are also introduced. Then, a brief summary of the thermodynamic model (i.e., equation of state and other correlations to determine transport properties) is presented. An overview of the numerical methodology is given, with extensive details available in Poblador-Ibanez and Sirignano~\cite{poblador2021vof} and other mentioned works. Features of two-dimensional, symmetric, temporal planar jets are shown and analyzed. Lastly, some three-dimensional results are shown to corroborate the viability of the method in realistic configurations. \par 

\section{Governing equations}

The problem configurations analyzed in this work belong to a low-Mach-number environment. Compressibility effects are linked to species and energy mixing under the high-pressure environment. Therefore, pressure variations are only responsible for fluid motion and have a negligible impact on the fluid properties. Moreover, only binary configurations are considered here. Initially, the liquid phase is composed of \textit{n}-decane (i.e., \(Y_2=Y_F=1\)) while the gas phase is composed of oxygen (i.e., \(Y_1=Y_O=1\)). The mass fractions of both species are related as \(\sum_{i=1}^{N=2} Y_i=Y_O+Y_F = 1\). \par

The low-Mach-number governing equations for a binary configuration are the continuity equation, Eq. (\ref{eqn:cont}), the momentum equation, Eq. (\ref{eqn:mom}), the species continuity equation, Eq. (\ref{eqn:spcont}), and the energy equation, Eq. (\ref{eqn:energy}).

\begin{equation}
\label{eqn:cont}
\frac{\partial \rho}{\partial t} + \nabla \cdot (\rho\vec{u})=0
\end{equation}
\begin{equation}
\label{eqn:mom}
\frac{\partial}{\partial t}(\rho \vec{u})+\nabla \cdot (\rho \vec{u}\vec{u}) = -\nabla p + \nabla \cdot \bar{\bar{\tau}}
\end{equation}
\begin{equation}
\label{eqn:spcont}
\frac{\partial}{\partial t}(\rho Y_O) + \nabla\cdot(\rho Y_O \vec{u}) = \nabla \cdot (\rho D_m \nabla Y_O)
\end{equation}
\begin{equation}
\label{eqn:energy}
\frac{\partial}{\partial t}(\rho h) + \nabla\cdot(\rho h \vec{u}) = \nabla \cdot \bigg(\frac{\lambda}{c_p}\nabla h \bigg) + \sum_{i=1}^{N=2} \nabla \cdot \Bigg(\bigg[\rho D_m - \frac{\lambda}{c_p}\bigg]h_i \nabla Y_i\Bigg)
\end{equation}

\(\rho\) and \(\vec{u}\) are the fluid density and velocity, respectively. In Eq. (\ref{eqn:mom}), \(p\) is the dynamic pressure and the viscous stress dyad, \(\bar{\bar{\tau}}\), is evaluated as \(\bar{\bar{\tau}}=\mu [\nabla\vec{u}+\nabla\vec{u}^\text{T}-\frac{2}{3}(\nabla\cdot\vec{u})\bar{\bar{I}}]\), where \(\mu\) represents the dynamic viscosity of the fluid and \(\bar{\bar{I}}\) represents the identity dyad. For simplicity, a Newtonian fluid under the Stokes' hypothesis is considered. However, due to the non-ideal fluid behavior at very high pressures, models to estimate the bulk viscosity or second coefficient of viscosity~\cite{jaeger2018bulkvisc} should be considered in the future for a more general treatment of the term. \par

Because of the binary nature of the analyzed mixtures, only one species continuity equation has to be solved. As shown in Eq. (\ref{eqn:spcont}), the equation addresses the transport of the oxygen mass fraction, \(Y_O\). Diffusion is modeled with a mass-based, high-pressure, non-ideal Fickian diffusion coefficient, \(D_m\). Similar to the Stokes' hypothesis, this modeling can be revised in the future to study the effects of using more complex models to estimate mass diffusion (i.e., generalized Maxwell-Stefan formulation for multi-component mixtures) or including thermo-diffusion effects (i.e., Soret effect). \par

The energy equation is written as an enthalpy transport equation where compressible effects have been neglected due to the low-Mach-number environment analyzed here (i.e., pressure gradient terms and viscous dissipation). The temporal variations of pressure are also neglected. The temperature gradient is replaced by \(\lambda \nabla T = (\lambda/c_p)\nabla h - \sum_{i=1}^{N=2}(\lambda/c_p)h_i\nabla Y_i\). For a non-ideal fluid, the mixture enthalpy, \(h\), is not exactly equal to the weighted sum of each species' enthalpy at the same temperature and pressure. For the convection and conduction terms in Eq.~(\ref{eqn:energy}), the proper formulation for mixture enthalpy at high pressures is used. The term for energy transport via mass diffusion demands the partial derivative of mixture enthalpy with respect to mass fraction, which here is referred to as partial enthalpy or \(h_i\). Note that this definition is not exactly the standard definition of partial enthalpy as the species' enthalpy at the same temperature and pressure as the mixture. Only for the ideal case, both approaches would be equivalent. Fickian diffusion is still considered in the term for energy transport via mass diffusion. \(\lambda\) and \(c_p\) are the thermal conductivity and the specific heat at constant pressure, respectively. \par 

Matching relations are needed to connect the solutions of the governing equations in both phases across the liquid-gas interface. \(\vec{n}\) and \(\vec{t}\) are the interface normal unit vector pointing towards the gas phase and the interface tangential unit vector, respectively. Liquid and gas properties at the interface are identified with the subscripts \(l\) and \(g\), respectively. \par 

The relations for the normal and tangential components of the velocity across the interface are

\begin{equation}
\label{eqn:veljump1}
(\vec{u}_g-\vec{u}_l) \cdot \vec{n} = \bigg(\frac{1}{\rho_g}-\frac{1}{\rho_l}\bigg)\dot{m}'
\end{equation}

\noindent
and

\begin{equation}
\label{eqn:veljump2}
\vec{u}_g \cdot \vec{t} = \vec{u}_l \cdot \vec{t}
\end{equation}

\noindent
where the tangential component is continuous but a velocity jump normal to the interface exists in the presence of phase change. Here, \(\dot{m}'\) represents the mass flux per unit area across the interface due to phase change and it is positive when vaporization occurs and negative with condensation. \par 

Similarly, a momentum jump normal to the interface is caused by surface tension, mass exchange and different normal viscous stresses on both sides of the interface. A pressure jump appears and is given by

\begin{equation}
\label{eqn:momjump1}
p_l - p_g = \sigma \kappa + (\bar{\bar{\tau}}_l \cdot \vec{n}) \cdot \vec{n} - (\bar{\bar{\tau}}_g \cdot \vec{n} ) \cdot \vec{n} +\bigg(\frac{1}{\rho_g}-\frac{1}{\rho_l}\bigg)(\dot{m}')^2
\end{equation}

In Eq. (\ref{eqn:momjump1}), \(\sigma\) represents the surface tension coefficient and \(\kappa=\nabla\cdot\vec{n}\) is the interface curvature, defined positive with a convex liquid shape. Since fluid properties vary along the interface, there exists a gradient of the surface tension coefficient as well. Therefore, a momentum jump tangential to the interface also exists, which is given by

\begin{equation}
\label{eqn:momjump2}
(\bar{\bar{\tau}}_g \cdot \vec{n})\cdot \vec{t}-(\bar{\bar{\tau}}_l \cdot \vec{n}) \cdot \vec{t} = \nabla_s \sigma \cdot \vec{t}
\end{equation}

\noindent
where \(\nabla_s=\nabla-\vec{n}(\vec{n}\cdot\nabla)\) represents the surface gradient. \par 

The surface tension force acts differently in Eqs. (\ref{eqn:momjump1}) and (\ref{eqn:momjump2}). The term \(\sigma\kappa\) in Eq. (\ref{eqn:momjump1}) tends to minimize the surface area. For two-dimensional structures it also smooths the interface. Smoothing can also occur in three dimensions, but surface tension might also cause ligament thinning and neck formation that leads to liquid breakup. On the other hand, the gradient of the surface tension coefficient in Eq. (\ref{eqn:momjump2}), \(d\sigma/ds\) where \(s\) is the distance along the interface, drives the flow towards regions of higher surface tension coefficient along the interface. In three dimensions, two tangential directions \(s_1\) and \(s_2\) are considered. \par

Jump conditions for the species continuity and energy equations are given by Eqs. (\ref{eqn:spcontmatch}) and  (\ref{eqn:energymatch}), respectively. Eq. (\ref{eqn:energymatch}) has been further simplified for a binary configuration.

\begin{equation}
\label{eqn:spcontmatch}
\dot{m}'(Y_{O,g}-Y_{O,l}) = (\rho D_m \nabla Y_O)_g \cdot \vec{n} - (\rho D_m \nabla Y_O)_l \cdot \vec{n}
\end{equation}
\begin{equation}
\label{eqn:energymatch}
\begin{split}
\dot{m}'(h_g-h_l) = \bigg(\frac{\lambda}{c_p}\nabla h\bigg)_g \cdot \vec{n} - \bigg(\frac{\lambda}{c_p}\nabla h\bigg)_l \cdot \vec{n} &+ \Bigg[\bigg(\rho D_m - \frac{\lambda}{c_p}\bigg)(h_O-h_F) \nabla Y_O\Bigg]_g \cdot \vec{n} \\ &- \Bigg[\bigg(\rho D_m - \frac{\lambda}{c_p}\bigg)(h_O-h_F) \nabla Y_O\Bigg]_l \cdot \vec{n}
\end{split}
\end{equation}

Lastly, phase equilibrium (i.e., LTE) provides a necessary thermodynamic closure to solve the interface state that satisfies the interface matching relations. Here, the species' fugacity, \(f_i\), is used to express an equality in chemical potential for each species on both sides of the interface~\cite{soave1972equilibrium,poling2001properties}. In general, fugacity is a function of temperature, pressure and composition. However, the pressure jump across the interface (see Eq. (\ref{eqn:momjump1})) has little effect on the LTE solution. Based on the low-Mach-number assumption, the interface pressure is assumed to be constant (for the purpose of thermodynamic analysis) and equal to the ambient or thermodynamic pressure. Using the fugacity coefficient, \(\Phi_i=f_i/pX_i\), the condition for phase equilibrium is expressed as \(X_{li}\Phi_{li}=X_{gi}\Phi_{gi}\), where \(X_{li}\) is the mole fraction of species \(i\) in the liquid phase and \(X_{gi}\) is the mole fraction of species \(i\) in the gas phase. Eqs. (\ref{eqn:spcontmatch}) and (\ref{eqn:energymatch}) are written in terms of the equilibrium mass fraction of oxygen in each phase (i.e., \(Y_{O,g}\) and \(Y_{O,l}\)). \par

The interface thickness is neglected and temperature is assumed to be continuous across it. This hypothesis simplifies the solution of phase equilibrium and the mixture composition on each side of the interface can be easily obtained. As reported in Dahms and Oefelein~\cite{dahms2013transition,dahms2015liquid} and Dahms~\cite{dahms2016understanding}, the interface non-equilibrium thickness is of the order of nanometers. Compared to the fast growth of mixing regions of the order of micrometers~\cite{poblador2018transient,davis2019development,poblador2021selfsimilar}, the treatment of the interface as a discontinuity is justified. Nevertheless, this approach should be revised in configurations where the mixture approaches the mixture critical temperature~\cite{dahms2015non} or for large interfacial thermal resistivity~\cite{stierle2020selection}. In these cases, the non-equilibrium region widens and fluid properties may experience a diffuse transition from one phase to the other. \par 

\section{Thermodynamic model}

A volume-corrected Soave-Redlich-Kwong (SRK) equation of state~\cite{lin2006volumetric} is used as a foundation to develop a thermodynamic model able to represent non-ideal fluid states for both phases. The volumetric correction is necessary since the original SRK equation of state~\cite{soave1972equilibrium} can present large density errors for dense fluids (i.e., liquids) compared to experimental measurements~\cite{yang2000modeling,prausnitz2004thermodynamics}. Using the compressibility factor, \(Z\), the volume-corrected SRK equation of state is expressed as

\begin{equation}
\label{eqn:SRKEoS}
Z^3+(3B_{*}-1)Z^2+\big[B_{*}(3B_{*}-2)+A-B-B^2\big]Z+B_{*}(B_{*}^{2}-B_{*}+A-B-B^2)-AB=0
\end{equation}

\noindent
with

\begin{equation}
Z = \frac{p}{\rho RT} \quad ; \quad A = \frac{a(T)p}{R_{u}^{2}T^2} \quad ; \quad B = \frac{bp}{R_uT} \quad ; \quad B_{*} = \frac{c(T)p}{R_uT}
\end{equation}

The parameters of this equation of state are a temperature-dependent cohesive energy parameter, \(a(T)\), a molecular volumetric parameter, \(b\), and a temperature-dependent volume correction, \(c(T)\), which aims to recover the experimental molar volume at the critical point~\cite{lin2006volumetric}. \(R\) and \(R_u\) are the specific gas constant and the universal gas constant, respectively. The binary interaction parameter of the mixing rules, \(k_{ij}\), is set to 0. The authors could not find available data for the oxygen-decane mixture used in this work. An analysis with available data for nitrogen-decane mixtures~\cite{soave2010srk} suggests that \(k_{ij}\) can be reasonably neglected assuming oxygen and nitrogen are similar components. The solution of this cubic equation can be obtained with analytical expressions and provides the density of the fluid mixture, \(\rho\). \par 

This equation of state is used together with high-pressure correlations to build a thermodynamic model to determine fluid and transport properties. Specific details on the development of necessary thermodynamic expressions based on the SRK equation of state (e.g., evaluation of mixture enthalpy) are available in Davis et al.~\cite{davis2019development}. Viscosity and thermal conductivity are obtained using the multi-parameter correlations from Chung et al.~\cite{chung1988generalized} and the mass diffusion coefficient for non-ideal fluids is obtained from Leahy-Dios and Firoozabadi~\cite{leahy2007unified}. The Macleod-Sugden correlation is used to estimate the surface tension coefficient as suggested in Poling et al.~\cite{poling2001properties}. For low-Mach-number flows, pressure remains uniform and equal to the ambient pressure everywhere in the thermodynamic model. \par 

\section{Numerical methods}

The computational approach used in this work is detailed and its validation discussed in Poblador-Ibanez and Sirignano~\cite{poblador2021vof}. Each modeling block (e.g., thermodynamic model, interface tracking solver) has been validated independently. Moreover, the consistency of the fully-coupled numerical approach at high pressures is shown by analyzing a one-dimensional transient flow~\cite{poblador2018transient} and by performing various grid-independence studies with two- and three-dimensional configurations. For the sake of completeness, a summary of the methodology is presented in this section. \par 

A VOF computational foundation is used to solve the low-Mach-number governing equations for two-phase flows at supercritical pressures. The incompressible VOF method presented in Baraldi et al.~\cite{baraldi2014mass} and further extended in Dodd and Ferrante~\cite{dodd2014fast} and Dodd et al.~\cite{dodd2021vof} is used as a basis to develop a VOF method for variable-density liquids with phase change~\cite{poblador2021vof}. This specific VOF approach is selected due to its computational efficiency and its mass-conserving properties in incompressible flows.  \par 

The advection of the volume fraction, \(C\), is performed by integrating in space and time Eq. (\ref{eqn:characVoF2}). \(\chi\) is an indicator function with \(\chi=0\) in the gas phase and \(\chi=1\) in the liquid phase, which acts as reference phase. The volume fraction is related to the indicator function by integrating \(\chi\) over the cell volume, \(V_0\), as \(C = \frac{1}{V_0}\iiint_{V_0} \chi dV\). That is, \(C\) represents the fraction of the cell volume occupied by the liquid phase.

\begin{equation}
\label{eqn:characVoF2}
\frac{\partial \chi}{\partial t}  + \nabla \cdot (\chi \vec{u}_l) = \chi \nabla \cdot \vec{u}_l - \frac{\dot{m}}{\rho_l}
\end{equation}

In Eq. (\ref{eqn:characVoF2}), the liquid phase advects with a phase-wise liquid velocity, \(\vec{u}_l\), and it also accounts for the local volume expansion (or compression) and the volume added (or subtracted) due to mass exchange at the interface caused by phase change. \(\rho_l\) is the interface liquid density and \(\dot{m}\) is the mass flux per unit volume due to phase change. It relates to the mass flux per unit area as \(\dot{m} = \dot{m}'\delta_\Gamma\), where \(\delta_\Gamma\) is obtained from the concept of interfacial surface area density~\cite{palmore2019volume} and is responsible to activate the phase change effect only at interface cells (i.e., cells with \(0<C<1\)). \par 

Details on the advection algorithm to solve Eq.~(\ref{eqn:characVoF2}) are available in Poblador-Ibanez and Sirignano~\cite{poblador2021vof}. A sharp interface is maintained by using a Piecewise Linear Interface Construction (PLIC)~\cite{youngs1982time} and by carefully including the interface in the discretization of the governing equations. In three dimensions, the linearity of the PLIC method is maintained by locally reconstructing the interface with a planar surface. The geometrical information of the interface is obtained from the volume fraction distribution. The Mixed-Youngs-Centered (MYC) method~\cite{aulisa2007interface} is used to evaluate the normal unit vector, \(\vec{n}\), and an improved Height Function (HF) method~\cite{lopez2009improved} is used to estimate the interface curvature. \par 

The non-conservative forms of the species continuity and energy equations are discretized using finite differences with a first-order explicit time integration. Convective terms are discretized using a one-sided hybrid first- and second-order upwinding scheme to maintain numerical stability and boundedness of the solution (i.e., \(0\leq Y_O\leq 1\)). Diffusive terms are discretized using second-order central differences. The interface solution is directly embedded in the numerical stencils; thus, a one-sided approach is taken where each phase is solved independently using phase-wise variables. The inclusion of the interface might reduce the spatial accuracy near the interface or lead to scenarios where the numerical stencils are poorly defined. Moreover, the interface solution itself may cause nonphysical solutions near under-resolved regions. Solutions to these problems are proposed in Poblador-Ibanez and Sirignano~\cite{poblador2021vof}. \par 

On the other hand, a one-fluid approach is implemented for the continuity and momentum equations. The conservative form of the momentum equation is discretized using the finite-volume method. Fluid properties are volume-averaged at interface cells using the volume fraction of the liquid (i.e., \(\phi=\phi_g+(\phi_l-\phi_g)C\), where \(\phi\) is any fluid property such as density or viscosity). To satisfy the momentum jump conditions, the Continuum Surface Force (CSF) model~\cite{brackbill1992continuum} is considered for flows with variable surface tension~\cite{kothe1996volume,seric2018direct}. \par 

The one-fluid momentum equation reads

\begin{equation}
\label{eqn:momwithsurf}
\frac{\partial}{\partial t}(\rho \vec{u})+\nabla \cdot (\rho \vec{u}\vec{u}) = -\nabla p + \nabla \cdot \bar{\bar{\tau}} + \frac{\rho}{\langle\rho\rangle}\bigg(\sigma\kappa\nabla C + \nabla_s\sigma|\nabla C|\bigg)
\end{equation}

\noindent
where a density scaling, \(\langle\rho\rangle=\frac{1}{2}(\rho_G+\rho_L)\), is implemented within the CSF model to obtain a body force per unit volume independent of fluid density~\cite{brackbill1992continuum,kothe1996volume}. \(\rho_G\) and \(\rho_L\) are the pure gas and pure liquid densities, respectively. \par 

The pressure-velocity coupling is addressed with the predictor-projection method by Chorin~\cite{chorin1967numerical}. The momentum equation without the pressure gradient term is integrated forward in time with a first-order explicit step as

\begin{equation}
\label{eqn:mom_pred}
\vec{u}^p = \frac{\rho^n\vec{u}^n}{\rho^{n+1}} + \frac{\Delta t}{\rho^{n+1}}\bigg[ -\nabla\cdot \big(\rho\vec{u}\vec{u}\big)^n + \nabla\cdot\bar{\bar{\tau}}^n + \frac{\rho^{n+1}}{\langle\rho\rangle}\bigg(\sigma^{n+1}\kappa^{n+1}\nabla C^{n+1} + \nabla_s\sigma^{n+1}|\nabla C^{n+1}|\bigg)\bigg]
\end{equation}

\noindent
where \(\vec{u}^p\) is the predicted one-fluid velocity. Convective terms are discretized using the SMART algorithm~\cite{gaskell1988curvature} and viscous terms are discretized using central differences. For the discretization of the viscous term, phase-wise velocities are used to avoid artificial pressure spikes near the interface~\cite{dodd2021vof}. The surface tension force is evaluated implicitly. That is, the new location of the interface at \(t^{n+1}\), as well as the new interface properties, are used. Further clarification of the step-by-step algorithm is available in Poblador-Ibanez and Sirignano~\cite{poblador2021vof}. \par 

For low-Mach-number flows, the pressure field is obtained from a pressure Poisson equation (PPE) using the split pressure-gradient technique~\cite{dodd2014fast} as

\begin{equation}
\label{eqn:ppe2}
\nabla^2 p^{n+1} = \nabla \cdot \bigg[ \bigg( 1-\frac{\rho_0}{\rho^{n+1}}\bigg)\nabla \hat{p} \bigg] + \frac{\rho_0}{\Delta t}\bigg(\nabla\cdot\vec{u}^p-\nabla\cdot\vec{u}^{n+1}\bigg)
\end{equation}

\noindent
where \(\rho_0 = \text{min}(\rho) \equiv \rho_G \) and \(\hat{p}=2p^n-p^{n-1}\) is an explicit linear extrapolation in time of the pressure field. The low-Mach-number and diffusive nature of the problems analyzed in this work result in the lowest density always being the pure gas density, \(\rho_G\). The one-fluid continuity constraint, \(\nabla\cdot\vec{u}^{n+1}\), is obtained as presented in Duret et al.~\cite{duret2018pressure} (see Eq. (\ref{eqn:divnewtime})). For the sake of a cleaner notation, the \(n\)+1 superscript has been dropped from the terms on the right hand side of the equation.

\begin{equation}
\label{eqn:divnewtime}
\nabla\cdot\vec{u}^{n+1}=-(1-C)\frac{1}{\rho_g}\frac{D\rho_g}{Dt}-C\frac{1}{\rho_l}\frac{D\rho_l}{Dt} + \dot{m}\bigg(\frac{1}{\rho_g}-\frac{1}{\rho_l}\bigg)
\end{equation}

The advantage of using the split pressure-gradient approach is that Eq. (\ref{eqn:ppe2}) becomes a constant-coefficient PPE for \(p^{n+1}\) which can be solved with a computationally efficient Discrete Fourier Transform method~\cite{dodd2014fast,costa2018fft}. Once the pressure field is obtained, the predicted velocity is corrected to get the one-fluid velocity field as

\begin{equation}
\label{eqn:mom_proj2}
\vec{u}^{n+1} = \vec{u}^p - \Delta t \bigg[ \frac{1}{\rho_0}\nabla p^{n+1} + \bigg(\frac{1}{\rho^{n+1}}-\frac{1}{\rho_0}\bigg)\nabla \hat{p} \bigg]
\end{equation}

Within each phase, fluid compressibilities are evaluated using the thermodynamic model and the solution of the governing equations for species mass fraction and enthalpy. For a binary mixture, they are obtained as

\begin{equation}
\label{eqn:rhochange2}
-\frac{1}{\rho}\frac{D\rho}{Dt}  = \frac{1}{c_p\bar{v}}\frac{\partial \bar{v}}{\partial T}\bigg|_{Y_i}\frac{Dh}{Dt} + \Bigg(\frac{\rho}{W_1}\frac{\partial \bar{v}}{\partial X_1}\bigg|_{T,X_{j\neq i}} -\frac{\rho}{W_2}\frac{\partial \bar{v}}{\partial X_2}\bigg|_{T,X_{j\neq i}} - \frac{h_O-h_F}{c_p\bar{v}}\frac{\partial \bar{v}}{\partial T}\bigg|_{Y_i}\Bigg)\frac{DY_O}{Dt}
\end{equation} 

In Eq. (\ref{eqn:rhochange2}), the thermodynamic derivatives are evaluated at constant pressure and at time \(n\)+1, although it is not shown for a cleaner notation. \(\bar{v}\) is the mixture molar volume and \(W_1\) and \(W_2\) are the molecular weights of species 1 and species 2, respectively. This equation can be easily used in liquid and gas cells, but its implementation in interface cells is not straightforward. The extrapolation techniques from Aslam~\cite{aslam2004partial} are adapted to VOF methods and used to extrapolate the compressibility of each phase across the interface in a narrow region of about \(2\Delta x\) (see Poblador-Ibanez and Sirignano~\cite{poblador2021vof}). A linear extrapolation is preferred, but for stability and consistency reasons a constant extrapolation might be desirable. The phase-wise fluid compressibilities can be linked to phase-wise velocities (i.e., \(\nabla\cdot\vec{u}_l=-\frac{1}{\rho_l}\frac{D\rho_l}{Dt}\)). \(\vec{u}_l\) and \(\vec{u}_g\) are extrapolated across the interface in order to match the extrapolated phase-wise fluid compressibilities~\cite{poblador2021vof,dodd2021vof}. \par 
The numerical approach presented in this section requires information on the interface local solution of the system of equations composed by the jump conditions and LTE. A normal-proble technique is used, whereby a probe is extended perpendicular to the interface into each phase. Mass fraction and enthalpy values are linearly interpolated onto the probe and used to evaluate their normal gradients into the interface with a one-sided, second-order scheme. An iterative solver similar to that presented in Poblador-Ibanez and Sirignano~\cite{poblador2018transient} can be used to solve the system of equations and obtain the local solution of the interface properties. \par  

The performance of this numerical model is discussed in greater detail in Poblador-Ibanez and Sirignano~\cite{poblador2021vof}. Mass conservation is analyzed by comparing the total amount of mass exchanged across the interface with the difference in total liquid mass at any given time with respect to the initial liquid mass. In other words, the change in the amount of liquid mass over the computation should correspond to the cumulative mass exchanged across the interface. Due to the finite resolution of the evolution of interface properties and the density field, mass is not conserved to machine error as in the incompressible case and the error depends on the mesh size. A two-dimensional capillary wave at supercritical pressures shows mass errors of about 1\% after an initial relaxation period with a sufficiently fine mesh. This is a negligible error when compared to the total amount of liquid mass. During the analyzed times (i.e., up to 70 \(\mu\)s), the total mass exchanged across the interface totals about 0.46\% of the total liquid mass. \par 

However, for problems with higher and continuous interface deformation (i.e., liquid injection), mass conservation errors as defined in the previous paragraph increase as smaller liquid structures are generated and interface and density fields are resolved more poorly. For the two-dimensional case presented in this paper, these errors are less than 50\% until \(t=2\) \(\mu\)s, when substantial deformation starts to occur. After that, the error increases as even smaller liquid structures evolve. Nevertheless, the error still represents a negligible amount of the total liquid mass. Here, the total mass exchanged across the interface is about 0.42\% of the total liquid mass after 8 \(\mu\)s. As a clarification, the error is linked to a poorer spatial resolution of the density field and the interface, but it does not affect the ability of the numerical method to predict vaporization or condensation rates with reasonable accuracy in smaller liquid structures. Therefore, the amount of fuel vapor that reaches the gas phase can be predicted accurately. \par 

Another issue of this numerical model relates to the handling of under-resolved regions. The extrapolations of phase-wise compressibilities and their respective phase-wise velocities require a non-conflicting and smooth interface (relative to the mesh size) in order to converge properly. To allow extensive time-marching in the simulations, these extrapolations are not performed around high-curvature regions and very thin features with respect to the mesh size. Instead, the liquid and gas phases are assumed to have the same one-fluid velocity and the fluid compressibility is neglected. The impact of this assumption in the evolution of the liquid deformation is not expected to be critical as discussed in Poblador-Ibanez and Sirignano~\cite{poblador2021vof}. For instance, high-curvature regions already have a geometrically ill-defined interface due to a poor mesh resolution that may dominate the liquid deformation progression. The treatment of these under-resolved regions, as well as the convergence criteria imposed on the extrapolation equations, may have an impact on mass conservation. Future work should address the improvement of the model in under-resolved regions. \par 

\section{Results}

A temporal study of a planar \textit{n}-decane liquid jet immersed into hotter oxygen with an ambient pressure of 150 bar (i.e., supercritical for the liquid hydrocarbon) is analyzed. A symmetric and periodic behavior is assumed for computational-cost reasons, although substantial deformation will deviate from periodicity and an antisymmetric mode may develop~\cite{zandian2018understanding}. Nevertheless, only the early-stage deformations are analyzed. Periodic boundary conditions are used in the streamwise and spanwise directions; symmetry boundary conditions are imposed in the center plane of the jet and outflow boundary conditions are set at the gas domain top boundary as seen, for example, in Figure \ref{fig:2djet_T}. \par

First, a two-dimensional configuration is studied where the jet half-thickness is 10 \(\mu\)m and an initial sinusoidal perturbation is imposed at the interface in the streamwise direction with amplitude 0.5 \(\mu\)m and wavelength 30 \(\mu\)m. The domain size is 30 x 30 \(\mu\)m and is discretized with a uniform mesh of 600 x 600 nodes (i.e., \(\Delta=0.05\) \(\mu\)m). This mesh size provides enough resolution to capture the liquid structures without entering into the non-continuum region or the non-equilibrium region of the interface of the order of a few nanometers. \par 

Even though surface tension forces act differently in the two-dimensional configuration (i.e., the surface area minimization in two dimensions inhibits the thinning of ligaments and breakup of droplets), it serves to present the main mixing features and interface thermodynamics appearing at supercritical pressures. \par 

The three-dimensional configuration has a similar initial interface shape, but another sinusoidal perturbation is superimposed in the spanwise direction with a 20 \(\mu\)m wavelength. Three different cases are shown with different initial amplitude of the spanwise perturbation: 0.1 \(\mu\)m, 0.3 \(\mu\)m and 0.5 \(\mu\)m. Here, the domain size is 30 x 30 x 20 \(\mu\)m with a uniform mesh of 600 x 600 x 400 nodes, which again corresponds to \(\Delta=0.05\) \(\mu\)m. \par 

The selected perturbation wavelengths are in line with those reported as being the fastest-growing perturbations in a similar supercritical mixture configuration but with an axisymmetric liquid jet~\cite{poblador2019axisymmetric}. \par 

Both configurations start with pure liquid \textit{n}-decane at 450 K surrounded by a gas composed of pure oxygen at 550 K. This temperature range ensures that the interface is sufficiently far away from the mixture critical point predicted by the SRK equation of state (i.e., about 580 K at 150 bar). Therefore, the interface can be assumed to be at LTE and the mesh size of \(\Delta=0.05\) \(\mu\)m is still sufficiently coarse as to treat the interface as a discontinuity~\cite{dahms2013transition,dahms2015liquid,dahms2015non,dahms2016understanding}. The sharp initial condition relaxes fast and a diffusion-controlled mixing develops during the early stages. A velocity distribution is imposed in the streamwise direction varying from 0 m/s in the liquid to 30 m/s in the gas within a thin region of a few micrometers (i.e., 4 \(\mu\)m) following a hyperbolic tangent profile. \par 

The low-Mach-number assumption is justified by looking at the Mach number of the gas phase. The development of a compressible pressure equation (i.e., in wave-like form) shows that compressible terms scale with \(M^2\). Using the gas freestream velocity and the thermodynamic model to estimate the speed of sound of the gas phase (approximately 450 m/s), it is found that \(M^2\sim\mathcal{O}(10^{-3})\). Thus, wave phenomena can be reasonably neglected. This limit might be pushed up to \(M^2\sim\mathcal{O}(10^{-2})\) before revising the low-Mach-number formulation presented in this work. This upper limit would correspond to a faster jet velocity around 100 m/s. \par

\subsection{Interface development and mixing in two dimensions}

\begin{figure*}[h!]
\centering
\begin{subfigure}{.45\textwidth}
  \centering
  \includegraphics[width=1.0\linewidth]{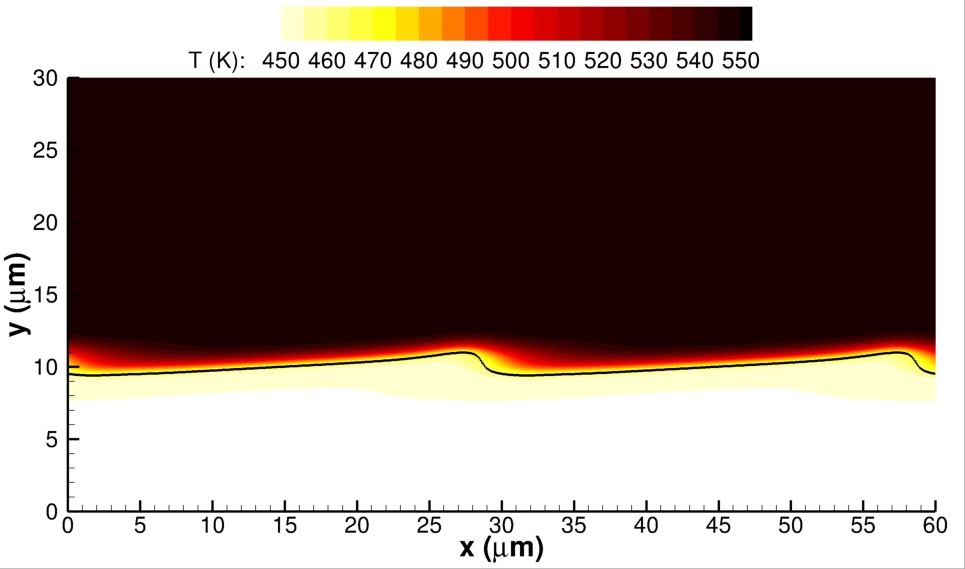}
  \caption{\label{subfig:T_1mus}\(T\) at \(t=1\) \(\mu\)s}
\end{subfigure}%
\begin{subfigure}{.45\textwidth}
  \centering
  \includegraphics[width=1.0\linewidth]{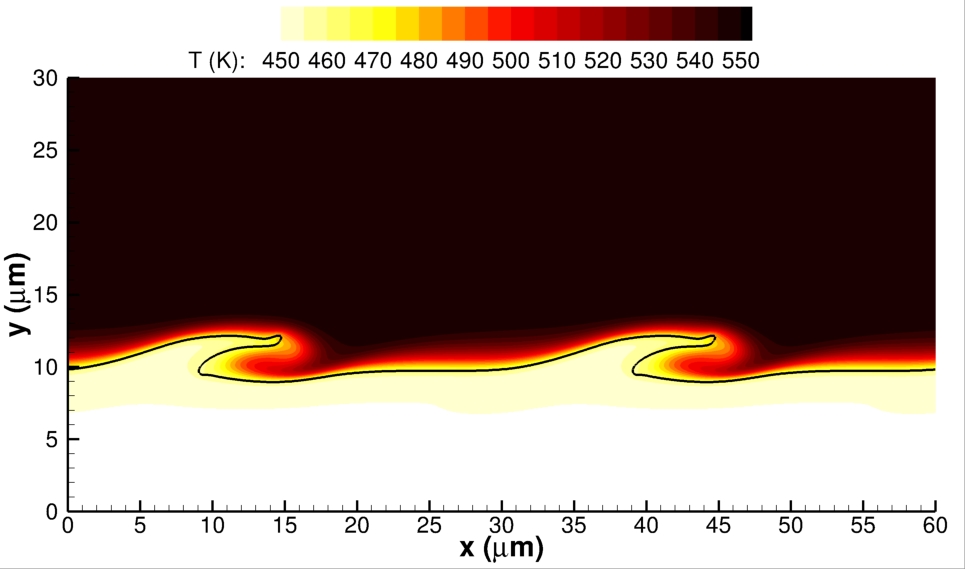}
  \caption{\label{subfig:T_2mus}\(T\) at \(t=2\) \(\mu\)s}
\end{subfigure}%
\\
\begin{subfigure}{.45\textwidth}
  \centering
  \includegraphics[width=1.0\linewidth]{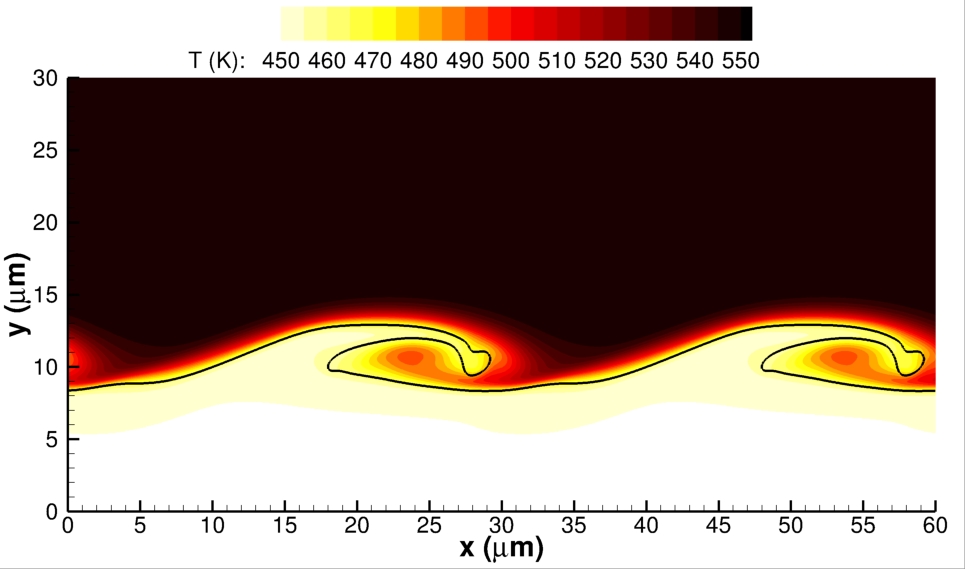}
  \caption{\label{subfig:T_3mus}\(T\) at \(t=3\) \(\mu\)s}
\end{subfigure}%
\begin{subfigure}{.45\textwidth}
  \centering
  \includegraphics[width=1.0\linewidth]{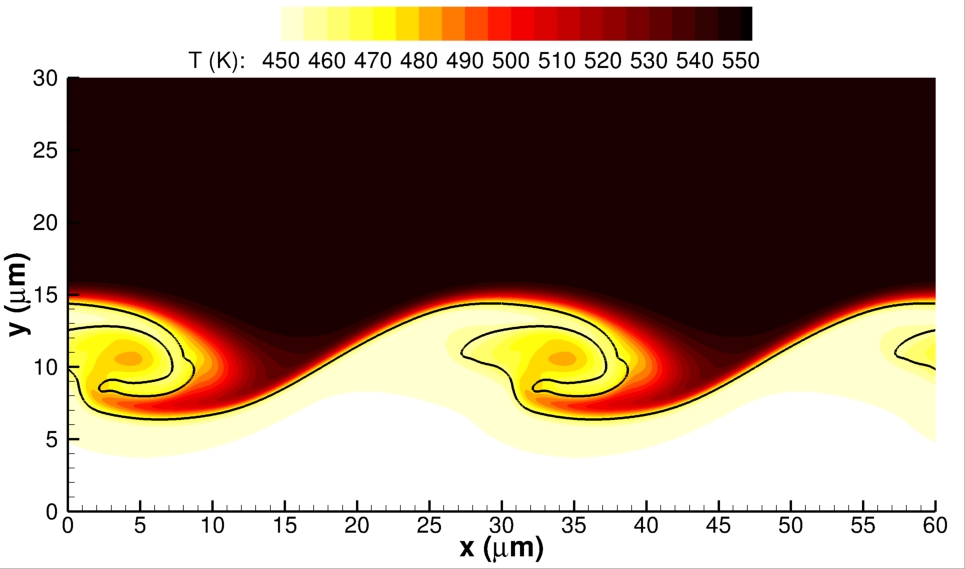}
  \caption{\label{subfig:T_4mus}\(T\) at \(t=4\) \(\mu\)s}
\end{subfigure}%
\\
\begin{subfigure}{.45\textwidth}
  \centering
  \includegraphics[width=1.0\linewidth]{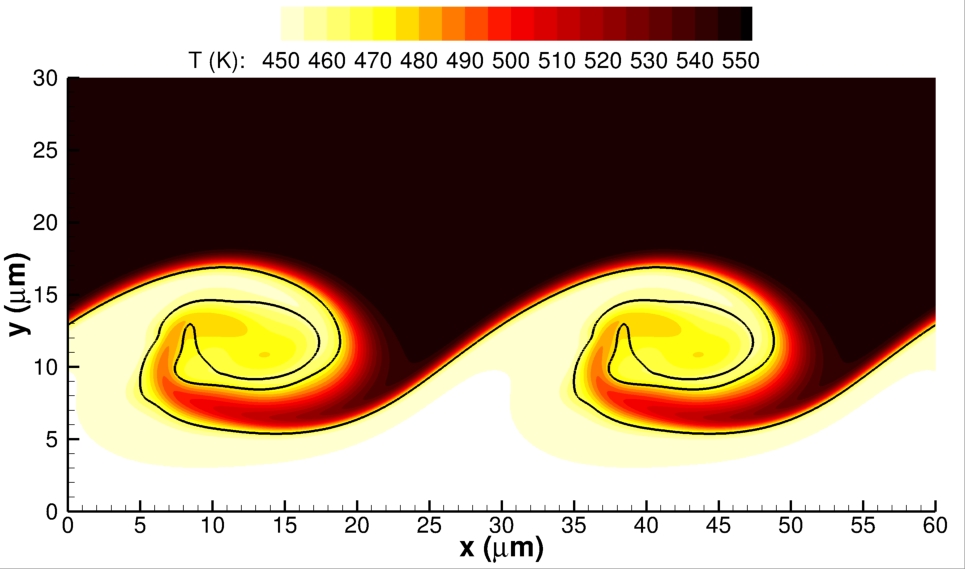}
  \caption{\label{subfig:T_5mus}\(T\) at \(t=5\) \(\mu\)s}
\end{subfigure}%
\begin{subfigure}{.45\textwidth}
  \centering
  \includegraphics[width=1.0\linewidth]{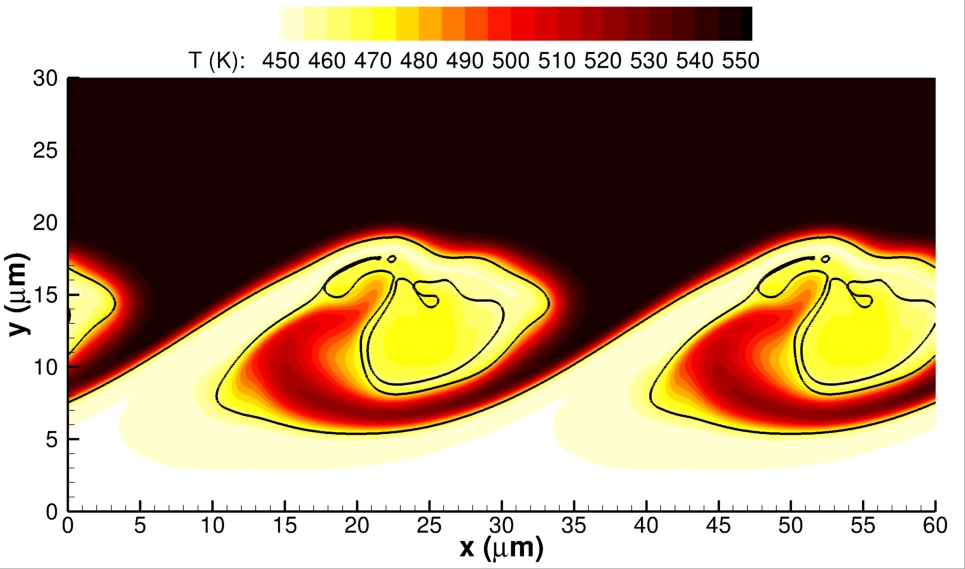}
  \caption{\label{subfig:T_6mus}\(T\) at \(t=6\) \(\mu\)s}
\end{subfigure}%
\\
\begin{subfigure}{.45\textwidth}
  \centering
  \includegraphics[width=1.0\linewidth]{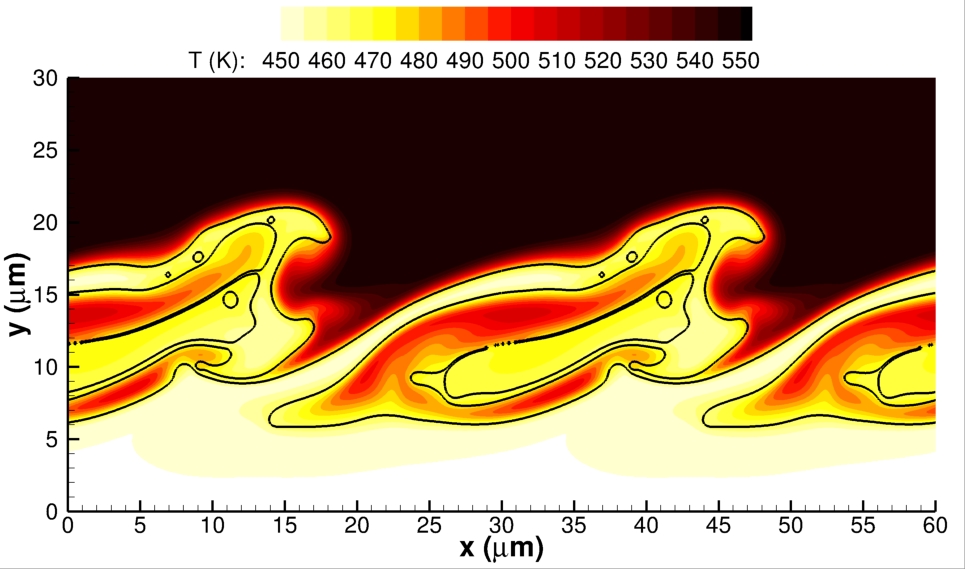}
  \caption{\label{subfig:T_7mus}\(T\) at \(t=7\) \(\mu\)s}
\end{subfigure}%
\begin{subfigure}{.45\textwidth}
  \centering
  \includegraphics[width=1.0\linewidth]{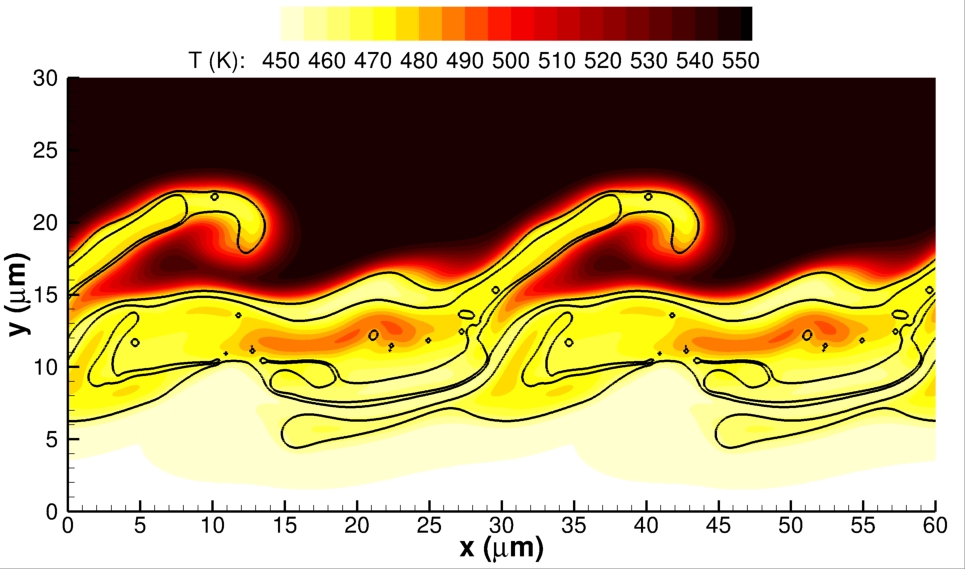}
  \caption{\label{subfig:T_8mus}\(T\) at \(t=8\) \(\mu\)s}
\end{subfigure}%
\caption{\label{fig:2djet_T}Temperature plots for the two-dimensional planar jet at 150 bar. The interface location is highlighted with a solid black curve representing the isocontour with \(C=0.5\).}
\end{figure*}

\begin{figure*}[h!]
\centering
\begin{subfigure}{.45\textwidth}
  \centering
  \includegraphics[width=1.0\linewidth]{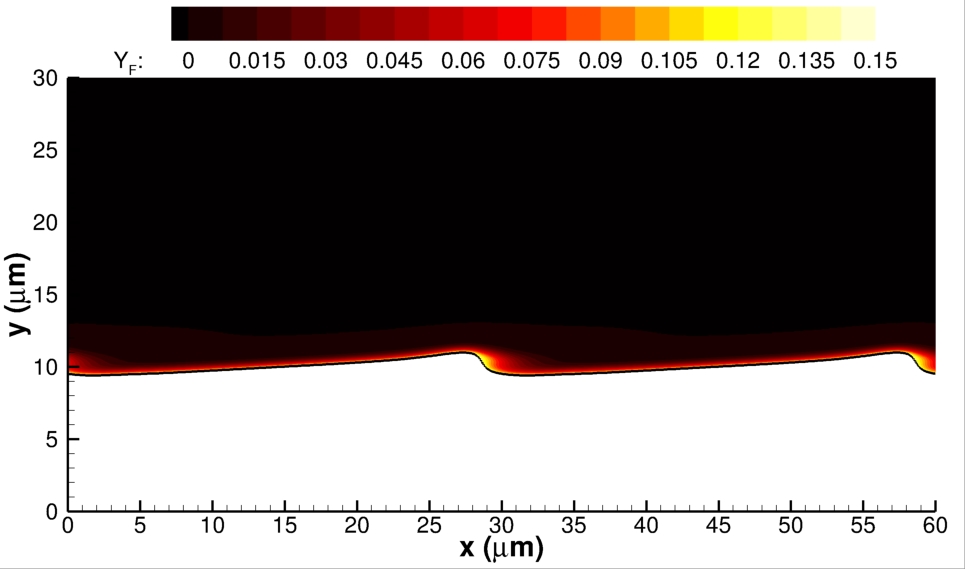}
  \caption{\label{subfig:YFg_1mus}\(Y_F\) at \(t=1\) \(\mu\)s}
\end{subfigure}%
\begin{subfigure}{.45\textwidth}
  \centering
  \includegraphics[width=1.0\linewidth]{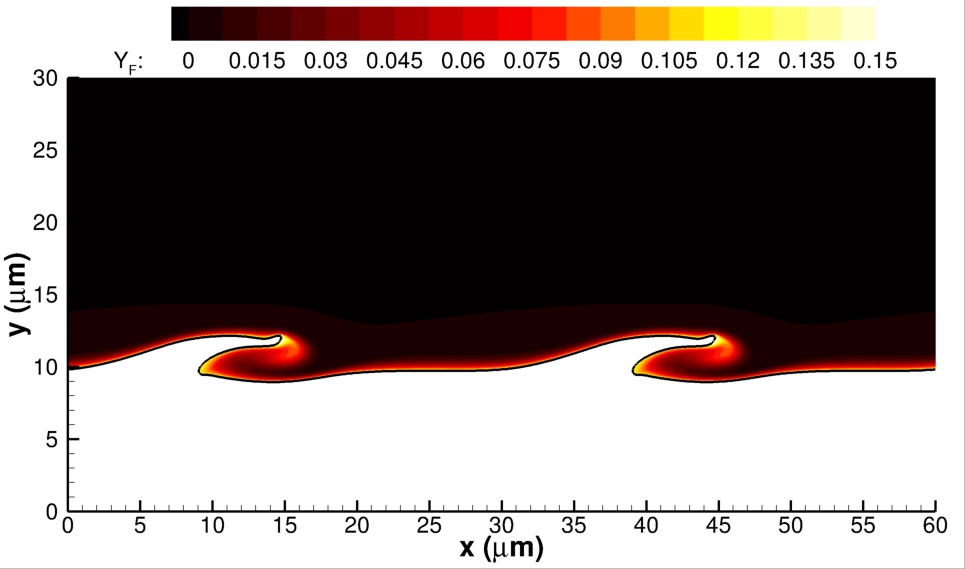}
  \caption{\label{subfig:YFg_2mus}\(Y_F\) at \(t=2\) \(\mu\)s}
\end{subfigure}%
\\
\begin{subfigure}{.45\textwidth}
  \centering
  \includegraphics[width=1.0\linewidth]{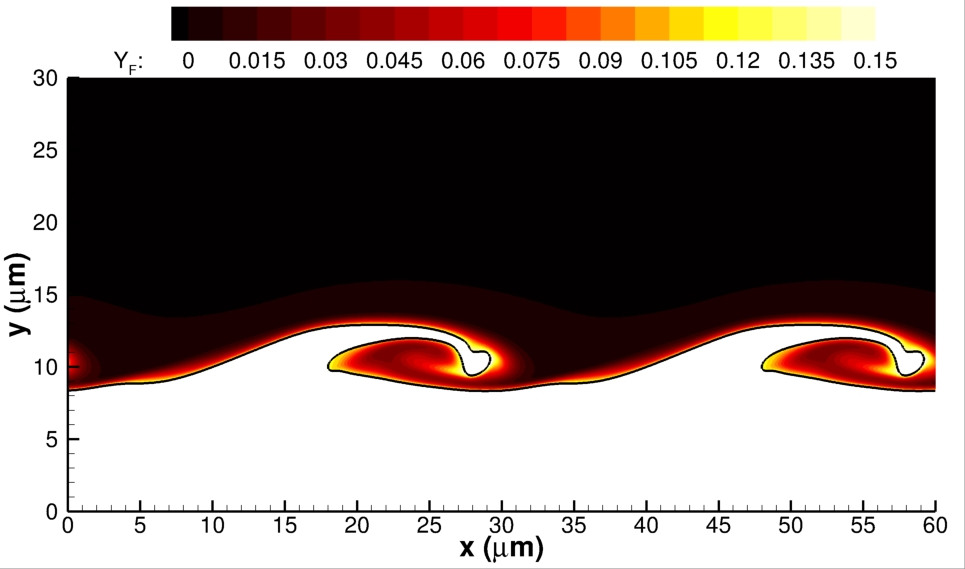}
  \caption{\label{subfig:YFg_3mus}\(Y_F\) at \(t=3\) \(\mu\)s}
\end{subfigure}%
\begin{subfigure}{.45\textwidth}
  \centering
  \includegraphics[width=1.0\linewidth]{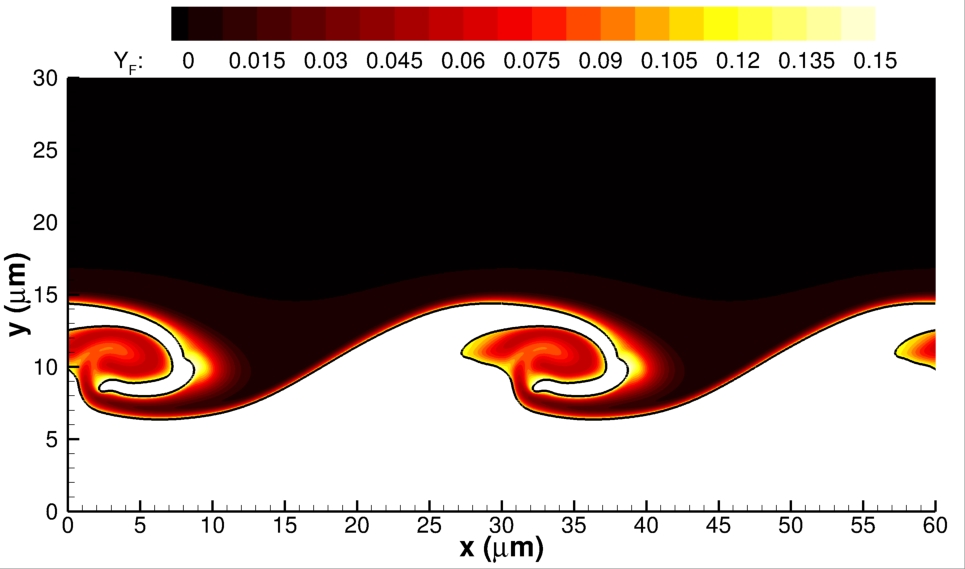}
  \caption{\label{subfig:YFg_4mus}\(Y_F\) at \(t=4\) \(\mu\)s}
\end{subfigure}%
\\
\begin{subfigure}{.45\textwidth}
  \centering
  \includegraphics[width=1.0\linewidth]{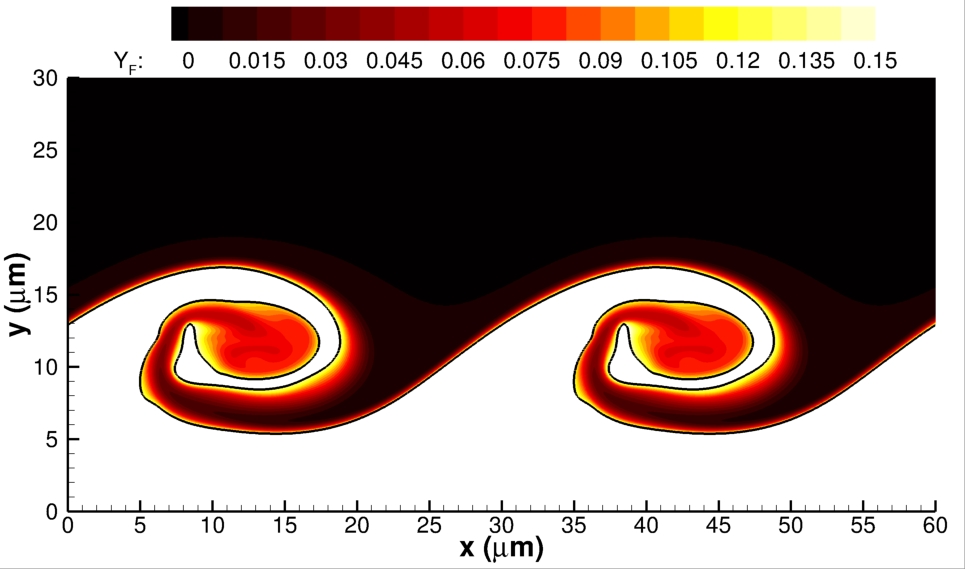}
  \caption{\label{subfig:YFg_5mus}\(Y_F\) at \(t=5\) \(\mu\)s}
\end{subfigure}%
\begin{subfigure}{.45\textwidth}
  \centering
  \includegraphics[width=1.0\linewidth]{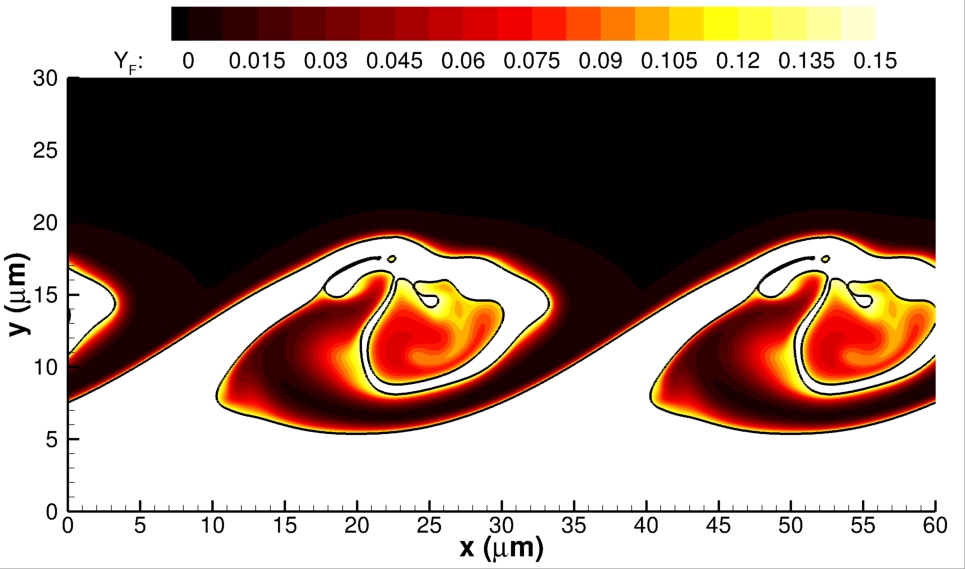}
  \caption{\label{subfig:YFg_6mus}\(Y_F\) at \(t=6\) \(\mu\)s}
\end{subfigure}%
\\
\begin{subfigure}{.45\textwidth}
  \centering
  \includegraphics[width=1.0\linewidth]{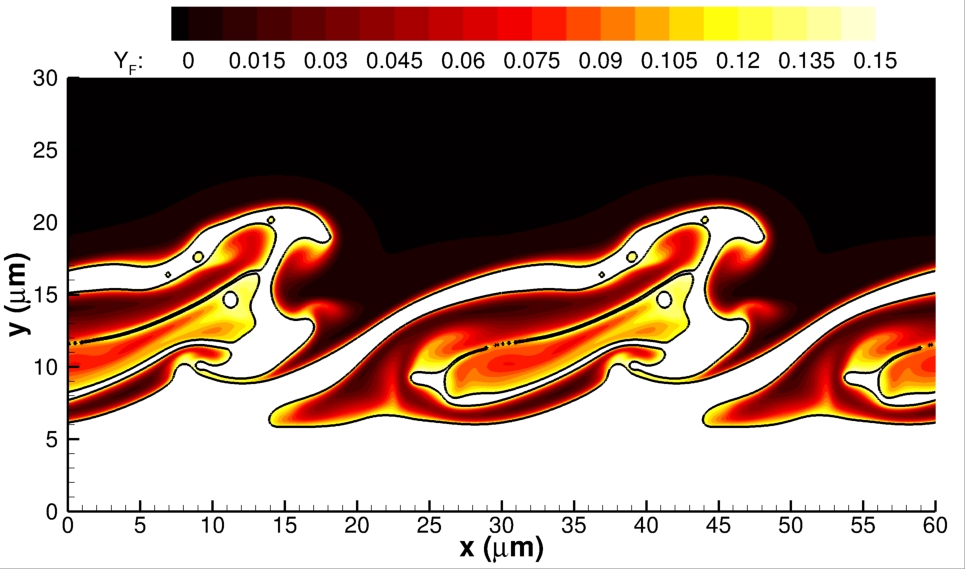}
  \caption{\label{subfig:YFg_7mus}\(Y_F\) at \(t=7\) \(\mu\)s}
\end{subfigure}%
\begin{subfigure}{.45\textwidth}
  \centering
  \includegraphics[width=1.0\linewidth]{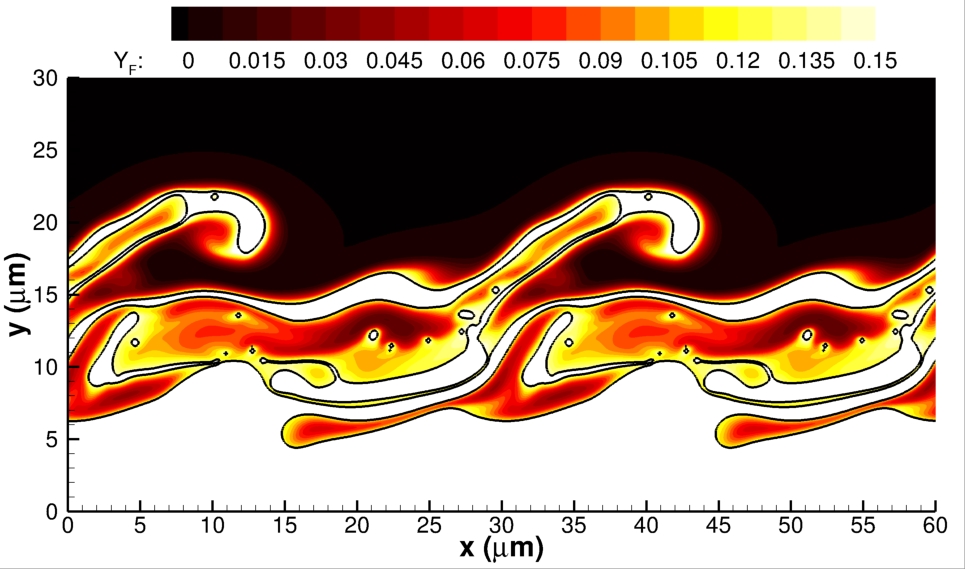}
  \caption{\label{subfig:YFg_8mus}\(Y_F\) at \(t=8\) \(\mu\)s}
\end{subfigure}%
\caption{\label{fig:2djet_YFg}\textit{n}-decane mass fraction plots in the gas phase for the two-dimensional planar jet at 150 bar. The interface location is highlighted with a solid black curve representing the isocontour with \(C=0.5\).}
\end{figure*}

\begin{figure*}[h!]
\centering
\begin{subfigure}{.45\textwidth}
  \centering
  \includegraphics[width=1.0\linewidth]{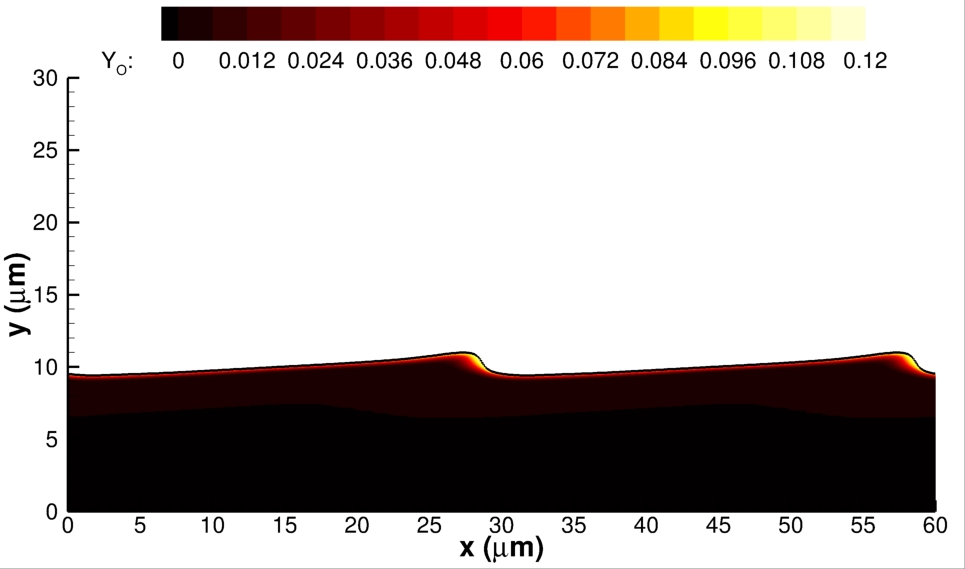}
  \caption{\label{subfig:YOl_1mus}\(Y_O\) at \(t=1\) \(\mu\)s}
\end{subfigure}%
\begin{subfigure}{.45\textwidth}
  \centering
  \includegraphics[width=1.0\linewidth]{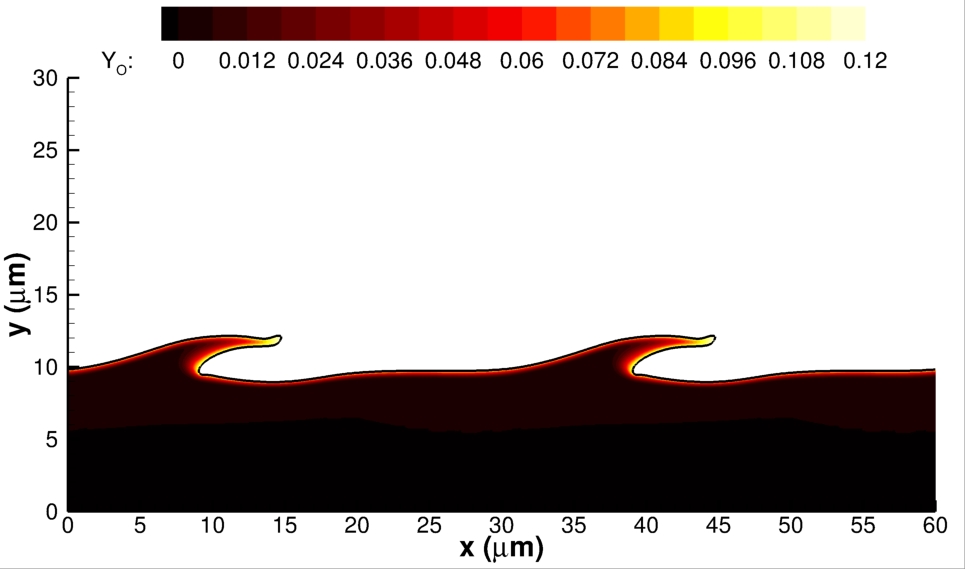}
  \caption{\label{subfig:YOl_2mus}\(Y_O\) at \(t=2\) \(\mu\)s}
\end{subfigure}%
\\
\begin{subfigure}{.45\textwidth}
  \centering
  \includegraphics[width=1.0\linewidth]{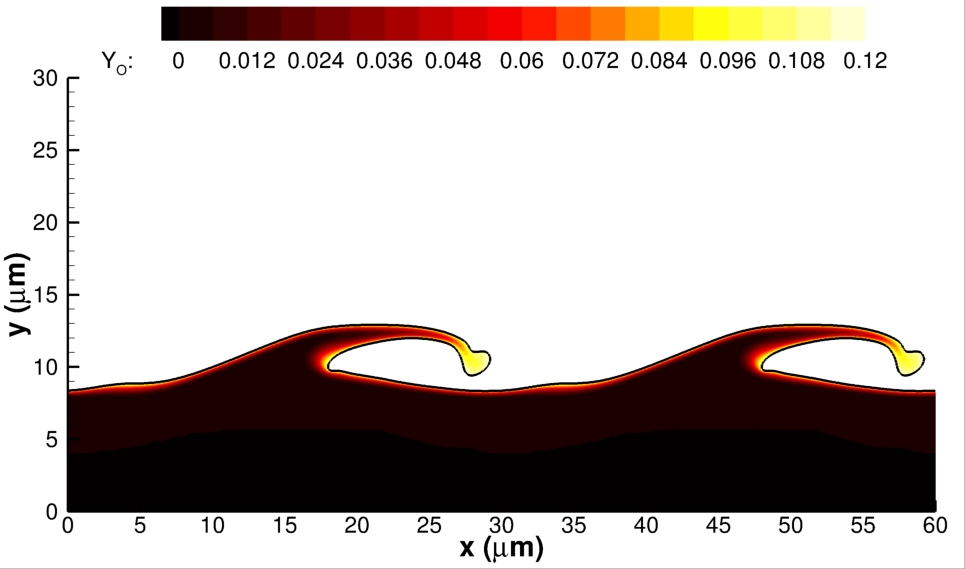}
  \caption{\label{subfig:YOl_3mus}\(Y_O\) at \(t=3\) \(\mu\)s}
\end{subfigure}%
\begin{subfigure}{.45\textwidth}
  \centering
  \includegraphics[width=1.0\linewidth]{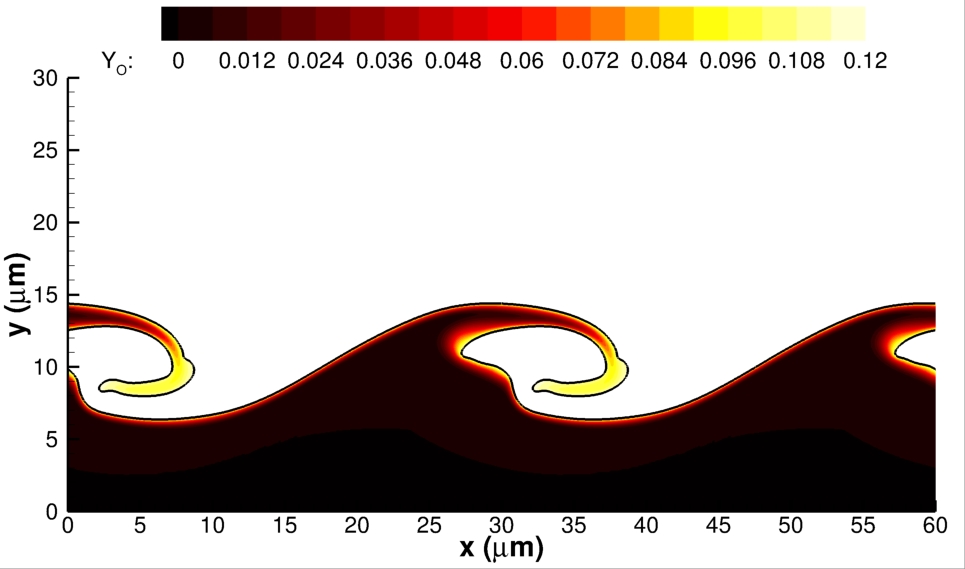}
  \caption{\label{subfig:YOl_4mus}\(Y_O\) at \(t=4\) \(\mu\)s}
\end{subfigure}%
\\
\begin{subfigure}{.45\textwidth}
  \centering
  \includegraphics[width=1.0\linewidth]{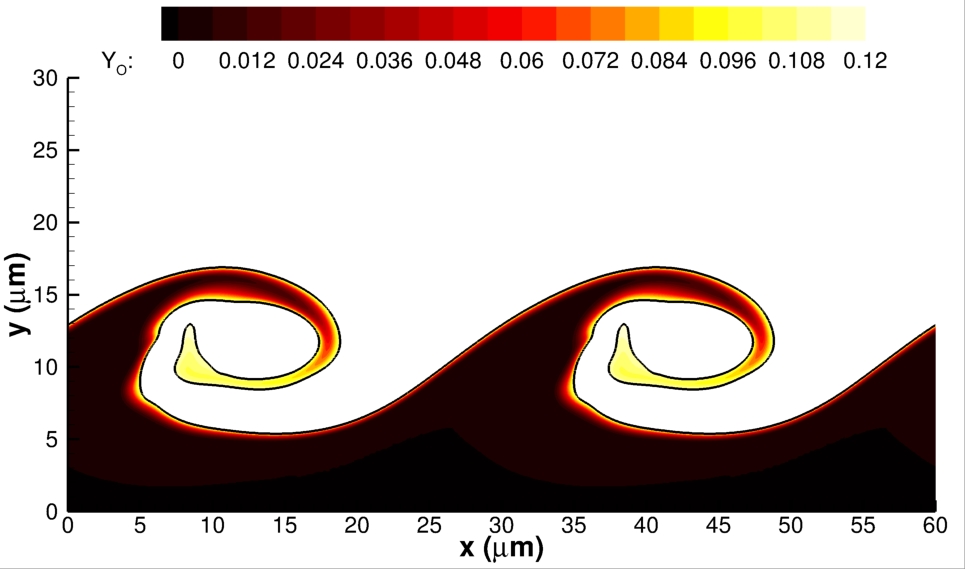}
  \caption{\label{subfig:YOl_5mus}\(Y_O\) at \(t=5\) \(\mu\)s}
\end{subfigure}%
\begin{subfigure}{.45\textwidth}
  \centering
  \includegraphics[width=1.0\linewidth]{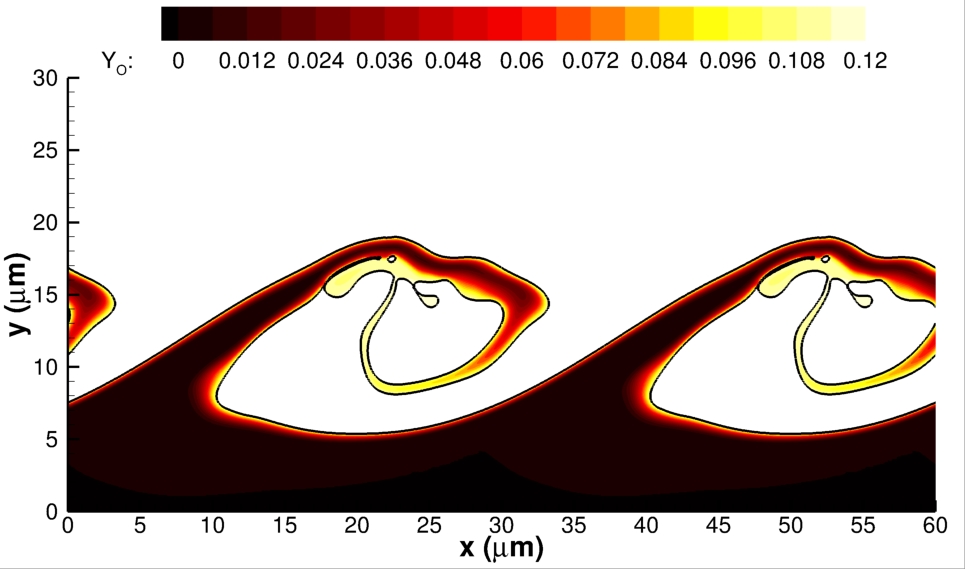}
  \caption{\label{subfig:YOl_6mus}\(Y_O\) at \(t=6\) \(\mu\)s}
\end{subfigure}%
\\
\begin{subfigure}{.45\textwidth}
  \centering
  \includegraphics[width=1.0\linewidth]{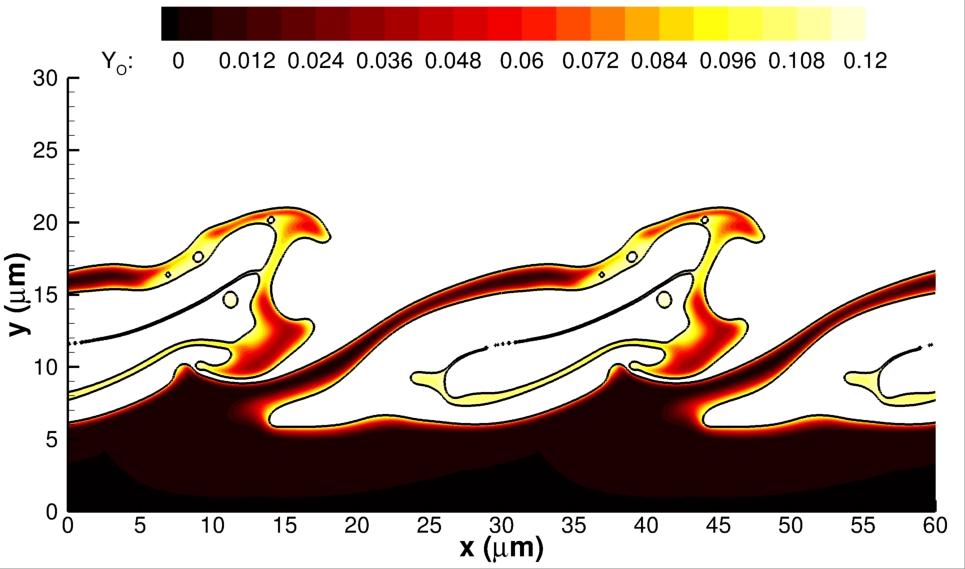}
  \caption{\label{subfig:YOl_7mus}\(Y_O\) at \(t=7\) \(\mu\)s}
\end{subfigure}%
\begin{subfigure}{.45\textwidth}
  \centering
  \includegraphics[width=1.0\linewidth]{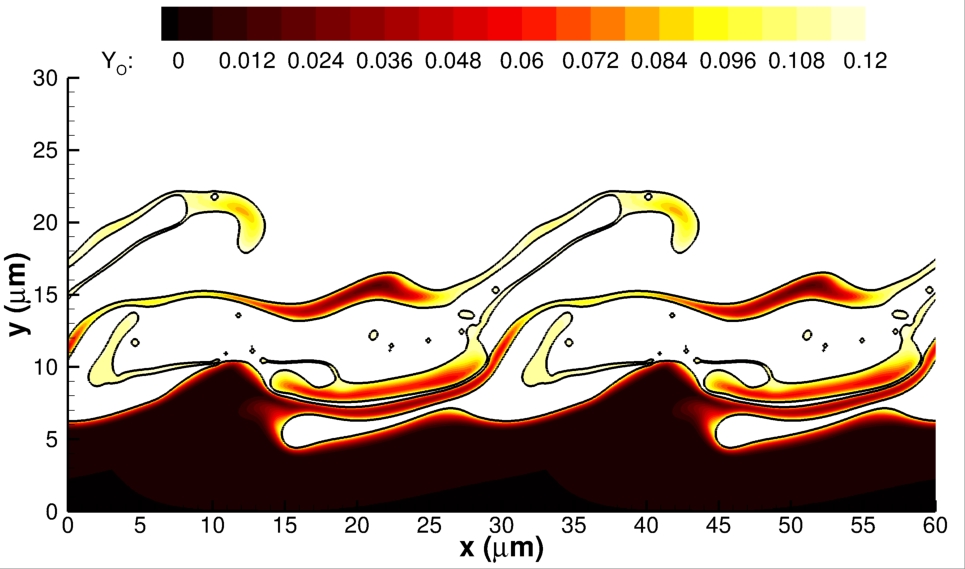}
  \caption{\label{subfig:YOl_8mus}\(Y_O\) at \(t=8\) \(\mu\)s}
\end{subfigure}%
\caption{\label{fig:2djet_YOl}Oxygen mass fraction plots in the liquid phase for the two-dimensional planar jet at 150 bar. The interface location is highlighted with a solid black curve representing the isocontour with \(C=0.5\).}
\end{figure*}

\begin{figure*}[h!]
\centering
\begin{subfigure}{.45\textwidth}
  \centering
  \includegraphics[width=1.0\linewidth]{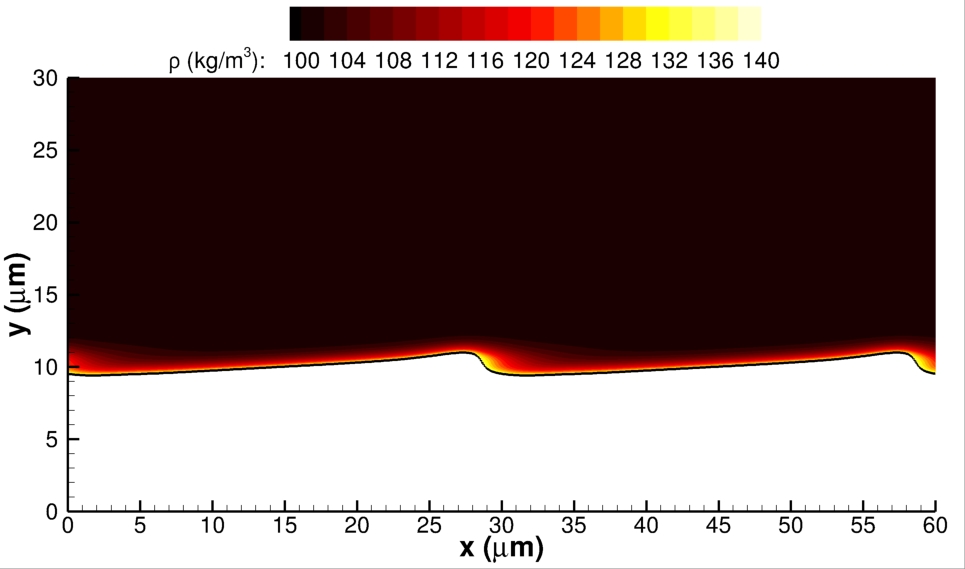}
  \caption{\label{subfig:DENg_1mus}\(\rho\) at \(t=1\) \(\mu\)s}
\end{subfigure}%
\begin{subfigure}{.45\textwidth}
  \centering
  \includegraphics[width=1.0\linewidth]{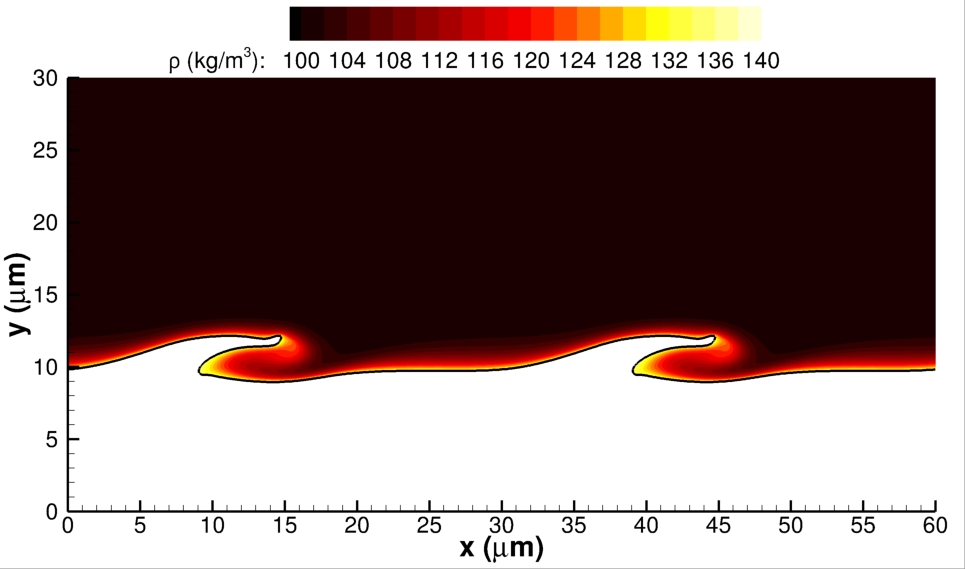}
  \caption{\label{subfig:DENg_2mus}\(\rho\) at \(t=2\) \(\mu\)s}
\end{subfigure}%
\\
\begin{subfigure}{.45\textwidth}
  \centering
  \includegraphics[width=1.0\linewidth]{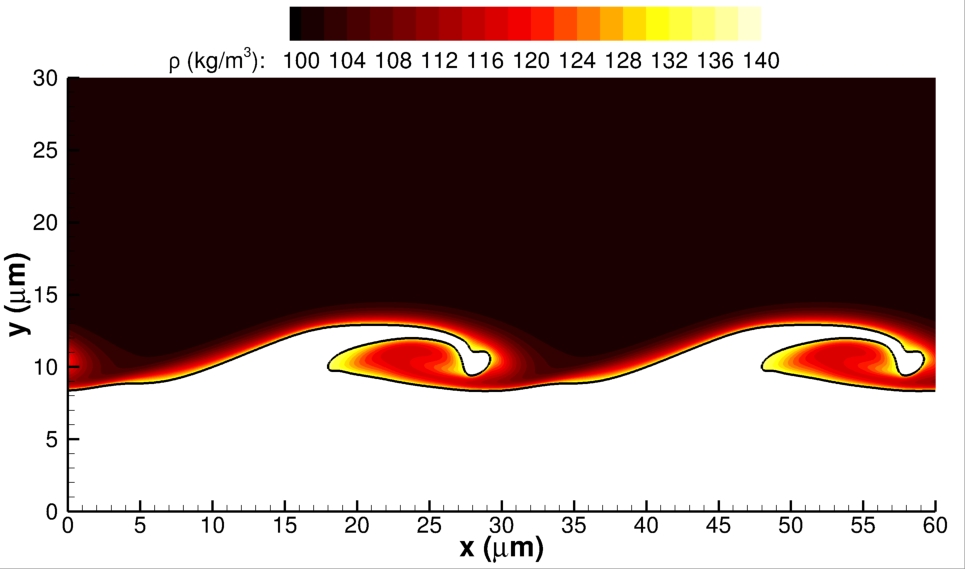}
  \caption{\label{subfig:DENg_3mus}\(\rho\) at \(t=3\) \(\mu\)s}
\end{subfigure}%
\begin{subfigure}{.45\textwidth}
  \centering
  \includegraphics[width=1.0\linewidth]{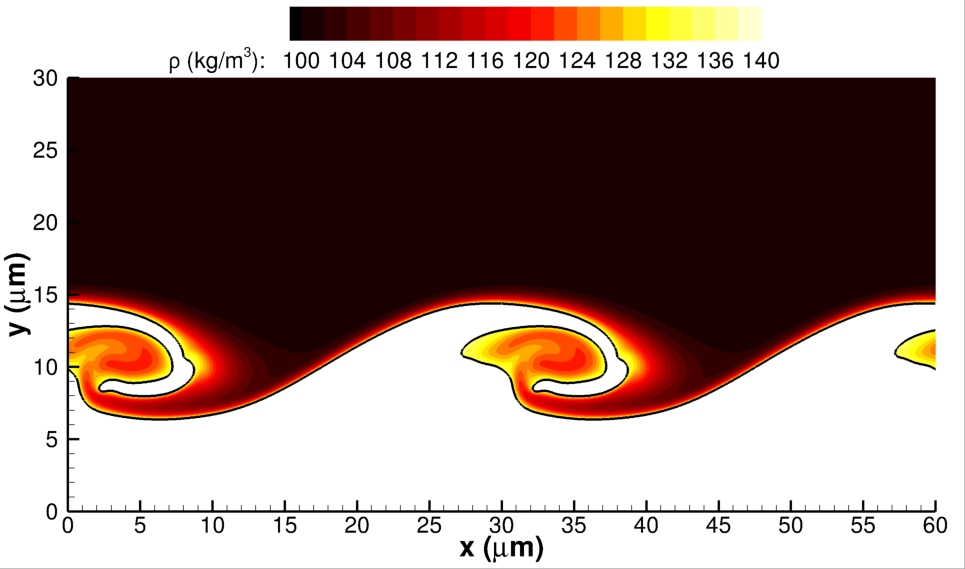}
  \caption{\label{subfig:DENg_4mus}\(\rho\) at \(t=4\) \(\mu\)s}
\end{subfigure}%
\\
\begin{subfigure}{.45\textwidth}
  \centering
  \includegraphics[width=1.0\linewidth]{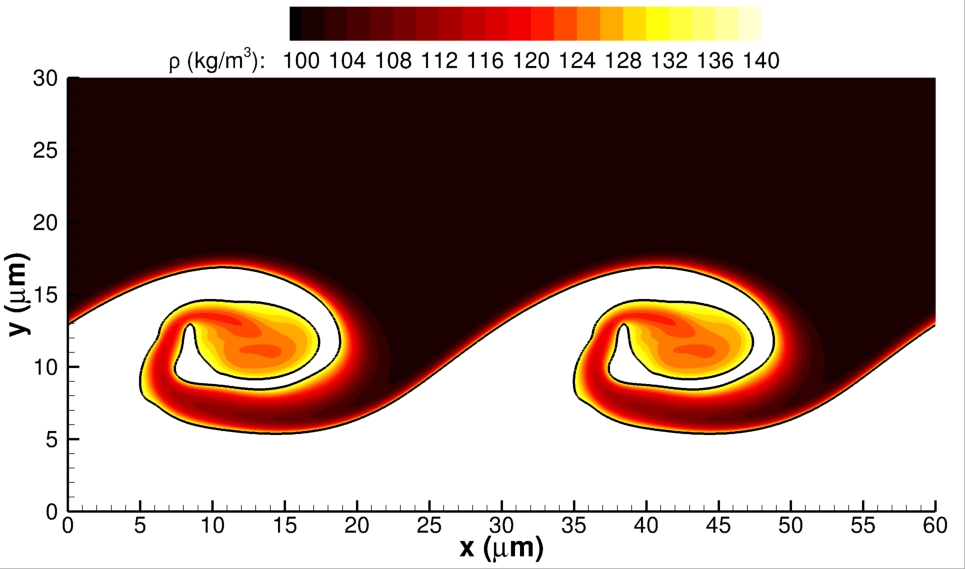}
  \caption{\label{subfig:DENg_5mus}\(\rho\) at \(t=5\) \(\mu\)s}
\end{subfigure}%
\begin{subfigure}{.45\textwidth}
  \centering
  \includegraphics[width=1.0\linewidth]{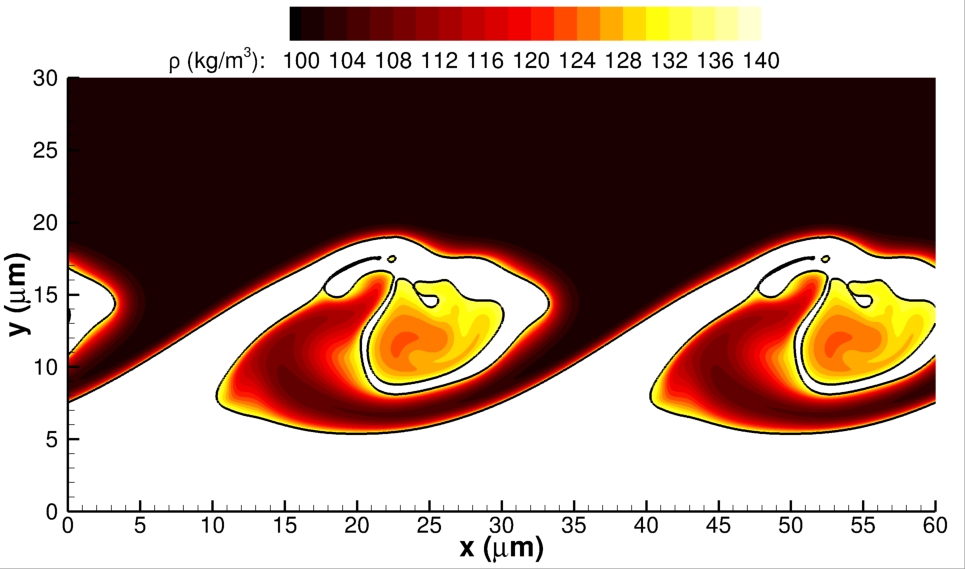}
  \caption{\label{subfig:DENg_6mus}\(\rho\) at \(t=6\) \(\mu\)s}
\end{subfigure}%
\\
\begin{subfigure}{.45\textwidth}
  \centering
  \includegraphics[width=1.0\linewidth]{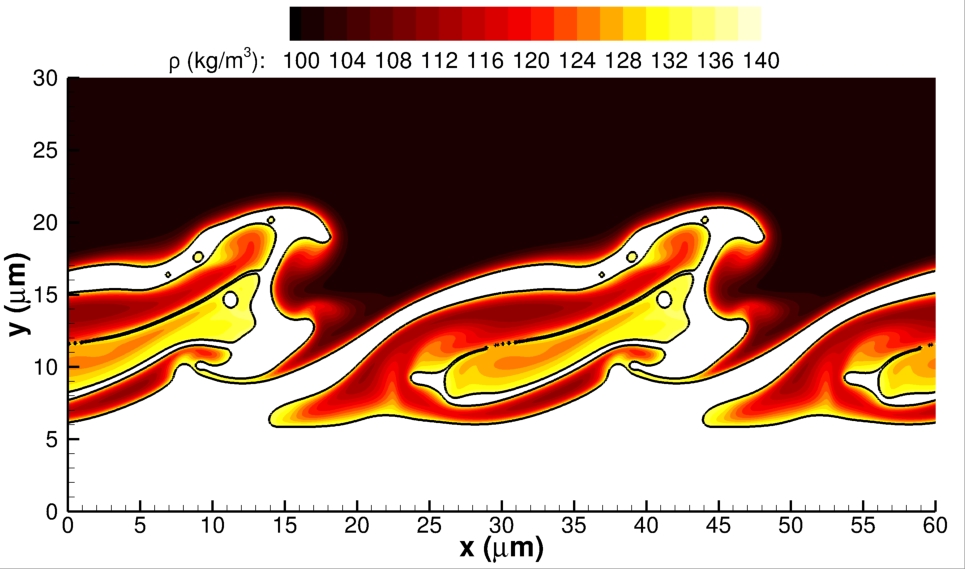}
  \caption{\label{subfig:DENg_7mus}\(\rho\) at \(t=7\) \(\mu\)s}
\end{subfigure}%
\begin{subfigure}{.45\textwidth}
  \centering
  \includegraphics[width=1.0\linewidth]{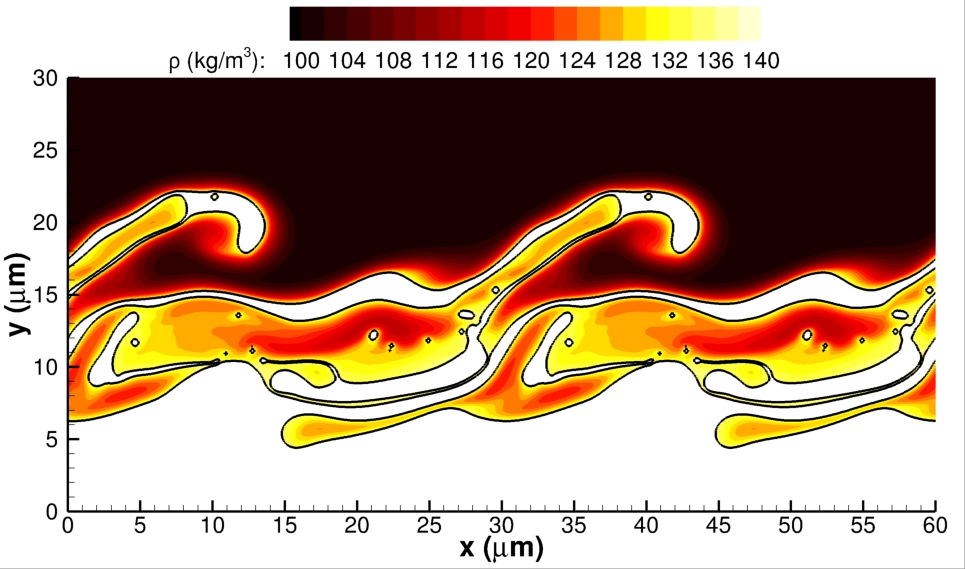}
  \caption{\label{subfig:DENg_8mus}\(\rho\) at \(t=8\) \(\mu\)s}
\end{subfigure}%
\caption{\label{fig:2djet_DENg}Density plots in the gas phase for the two-dimensional planar jet at 150 bar. The interface location is highlighted with a solid black curve representing the isocontour with \(C=0.5\).}
\end{figure*}

\begin{figure*}[h!]
\centering
\begin{subfigure}{.45\textwidth}
  \centering
  \includegraphics[width=1.0\linewidth]{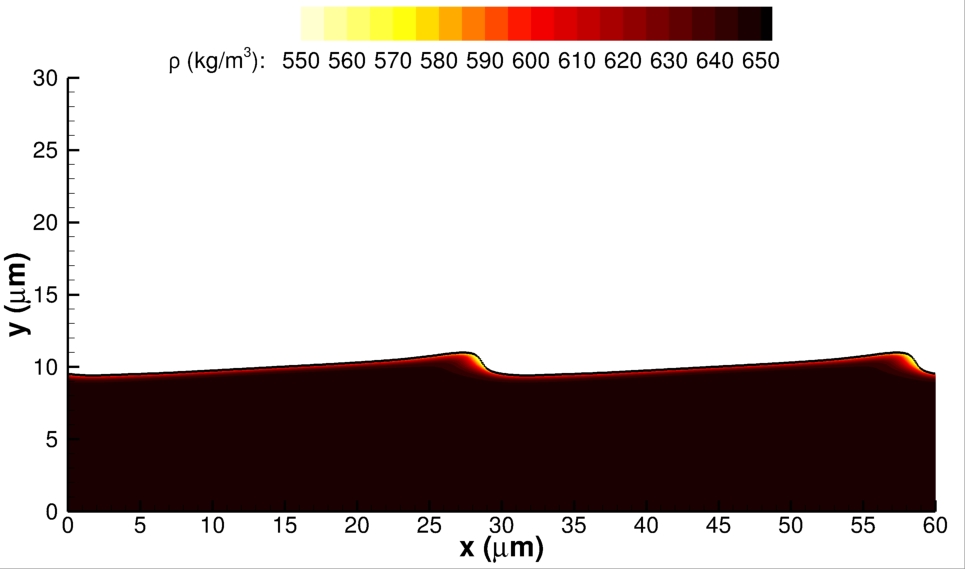}
  \caption{\label{subfig:DENl_1mus}\(\rho\) at \(t=1\) \(\mu\)s}
\end{subfigure}%
\begin{subfigure}{.45\textwidth}
  \centering
  \includegraphics[width=1.0\linewidth]{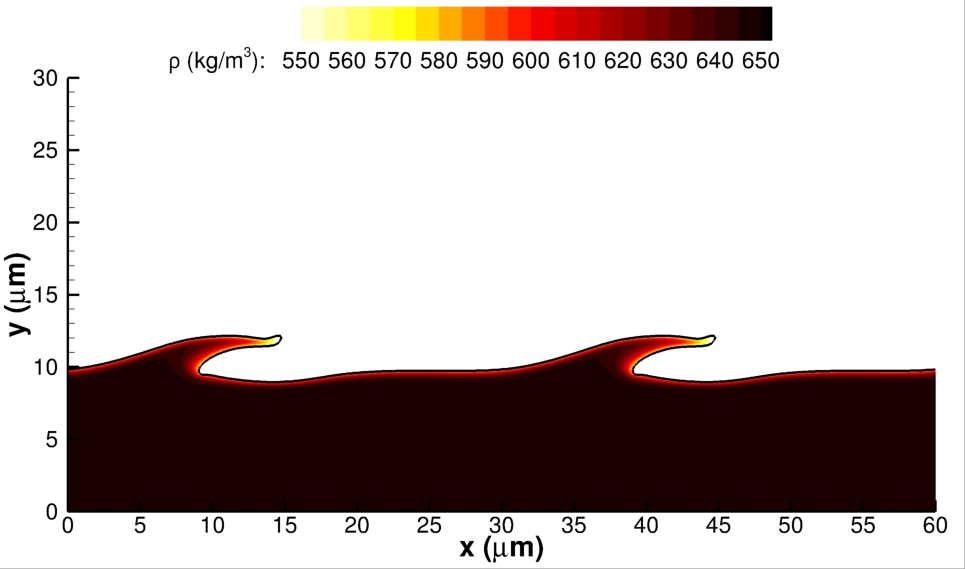}
  \caption{\label{subfig:DENl_2mus}\(\rho\) at \(t=2\) \(\mu\)s}
\end{subfigure}%
\\
\begin{subfigure}{.45\textwidth}
  \centering
  \includegraphics[width=1.0\linewidth]{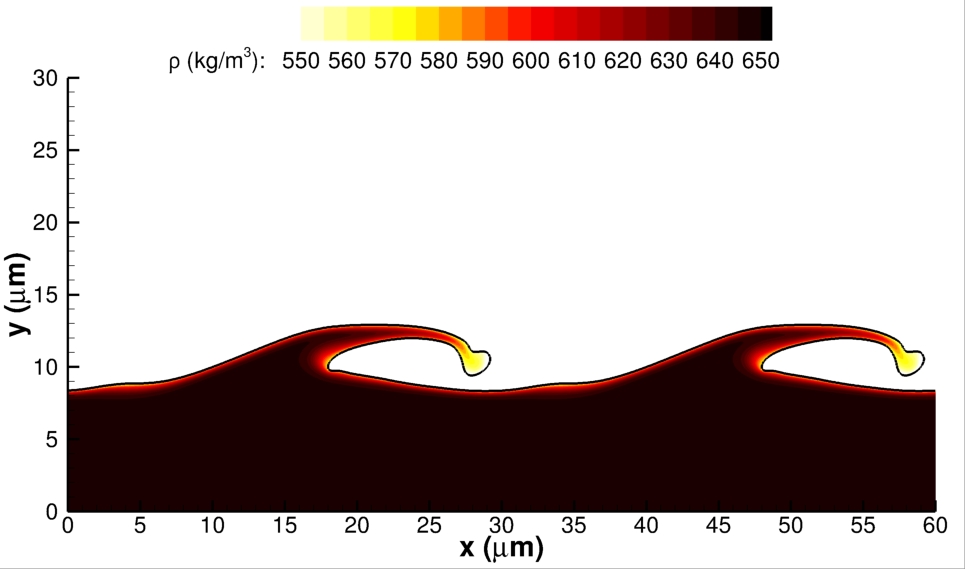}
  \caption{\label{subfig:DENl_3mus}\(\rho\) at \(t=3\) \(\mu\)s}
\end{subfigure}%
\begin{subfigure}{.45\textwidth}
  \centering
  \includegraphics[width=1.0\linewidth]{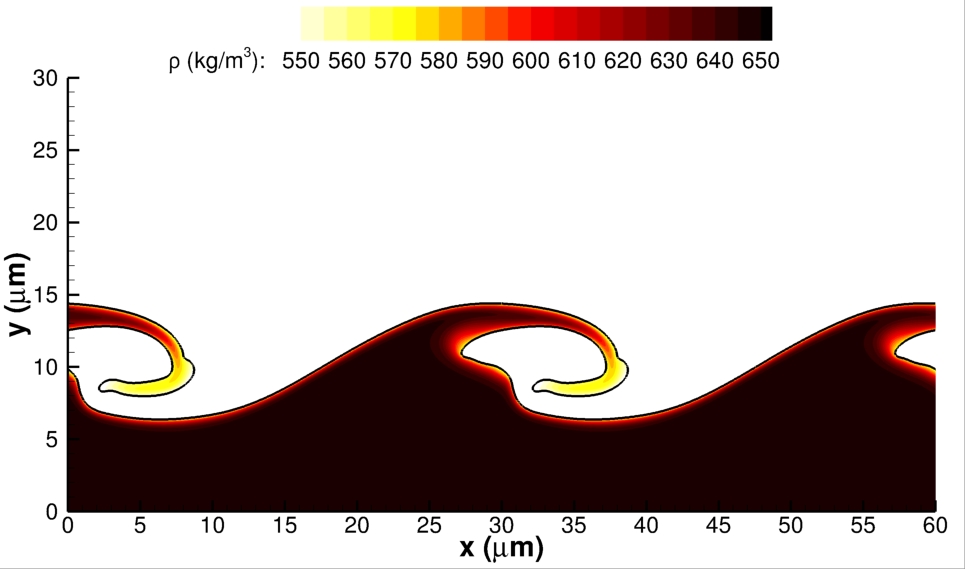}
  \caption{\label{subfig:DENl_4mus}\(\rho\) at \(t=4\) \(\mu\)s}
\end{subfigure}%
\\
\begin{subfigure}{.45\textwidth}
  \centering
  \includegraphics[width=1.0\linewidth]{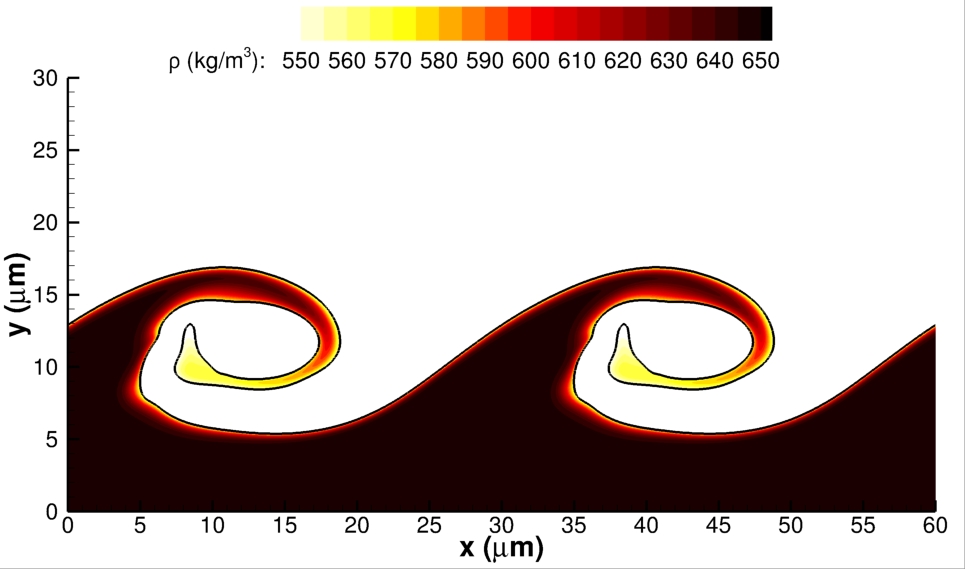}
  \caption{\label{subfig:DENl_5mus}\(\rho\) at \(t=5\) \(\mu\)s}
\end{subfigure}%
\begin{subfigure}{.45\textwidth}
  \centering
  \includegraphics[width=1.0\linewidth]{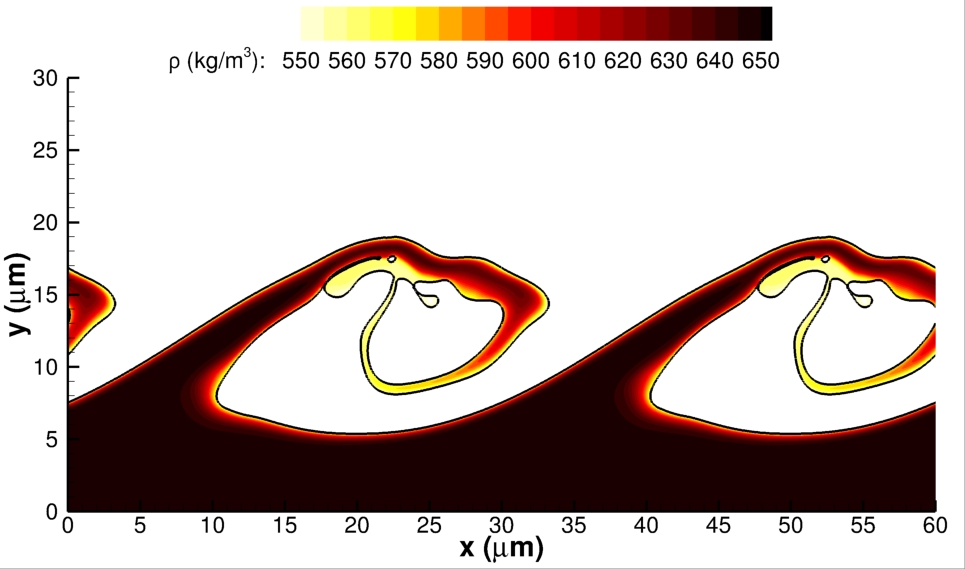}
  \caption{\label{subfig:DENl_6mus}\(\rho\) at \(t=6\) \(\mu\)s}
\end{subfigure}%
\\
\begin{subfigure}{.45\textwidth}
  \centering
  \includegraphics[width=1.0\linewidth]{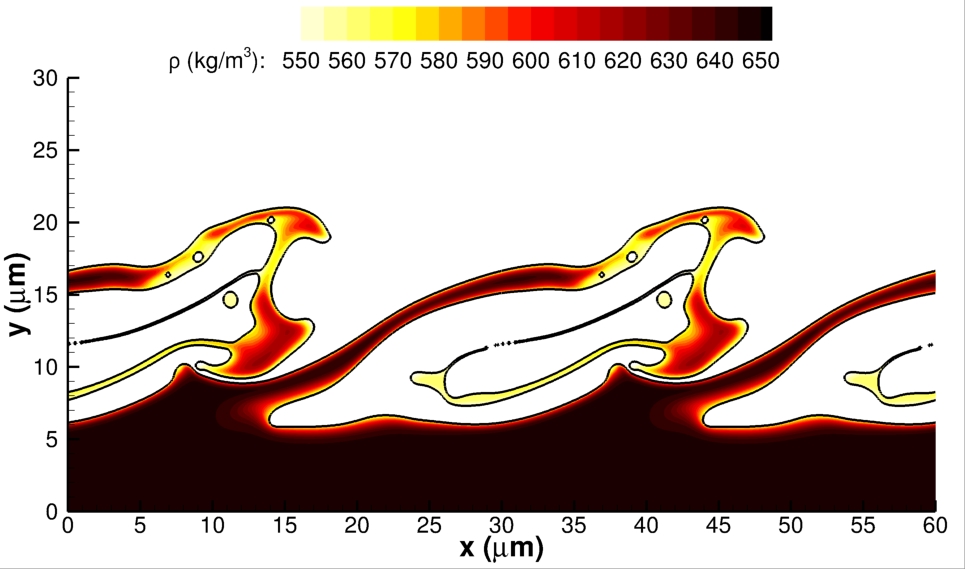}
  \caption{\label{subfig:DENl_7mus}\(\rho\) at \(t=7\) \(\mu\)s}
\end{subfigure}%
\begin{subfigure}{.45\textwidth}
  \centering
  \includegraphics[width=1.0\linewidth]{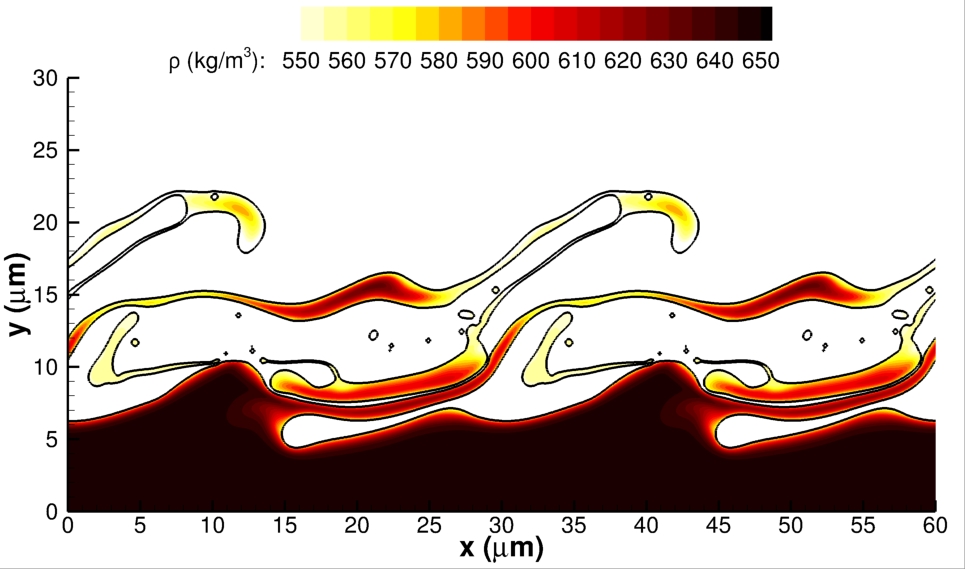}
  \caption{\label{subfig:DENl_8mus}\(\rho\) at \(t=8\) \(\mu\)s}
\end{subfigure}%
\caption{\label{fig:2djet_DENl}Density plots in the liquid phase for the two-dimensional planar jet at 150 bar. The interface location is highlighted with a solid black curve representing the isocontour with \(C=0.5\).}
\end{figure*}

\begin{figure*}[h!]
\centering
\begin{subfigure}{.43\textwidth}
  \centering
  \includegraphics[width=1.0\linewidth]{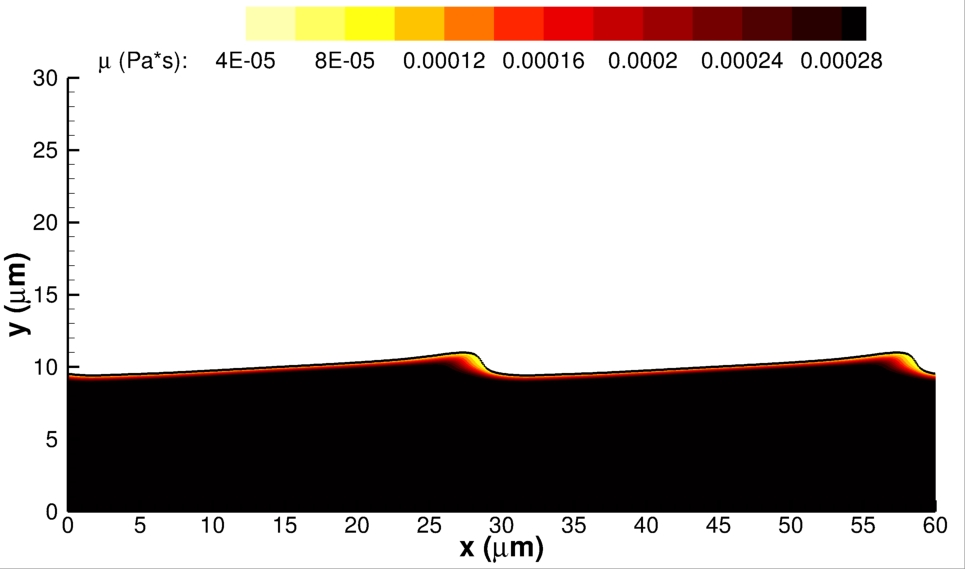}
  \caption{\label{subfig:VIS_1mus}\(\mu\) at \(t=1\) \(\mu\)s}
\end{subfigure}%
\begin{subfigure}{.43\textwidth}
  \centering
  \includegraphics[width=1.0\linewidth]{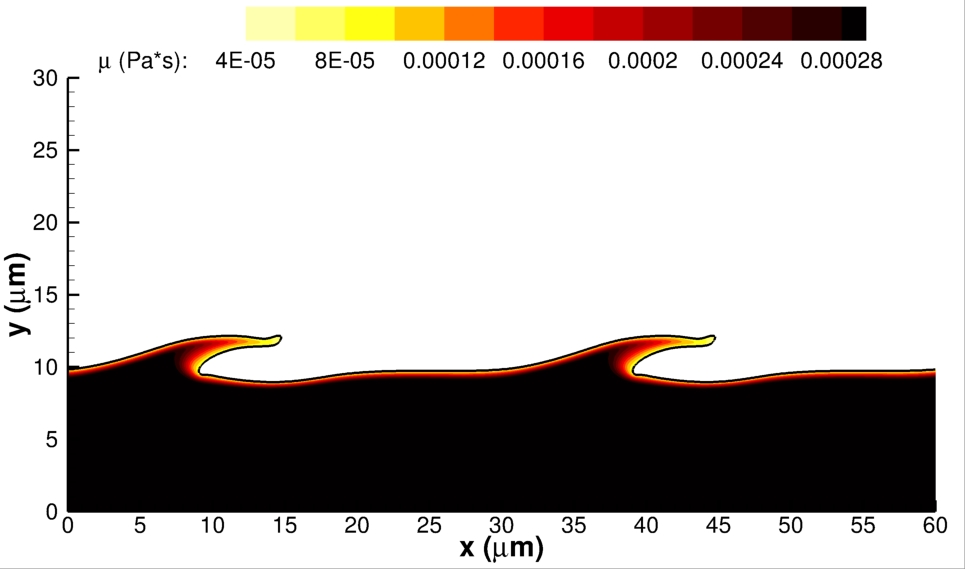}
  \caption{\label{subfig:VIS_2mus}\(\mu\) at \(t=2\) \(\mu\)s}
\end{subfigure}%
\\
\begin{subfigure}{.43\textwidth}
  \centering
  \includegraphics[width=1.0\linewidth]{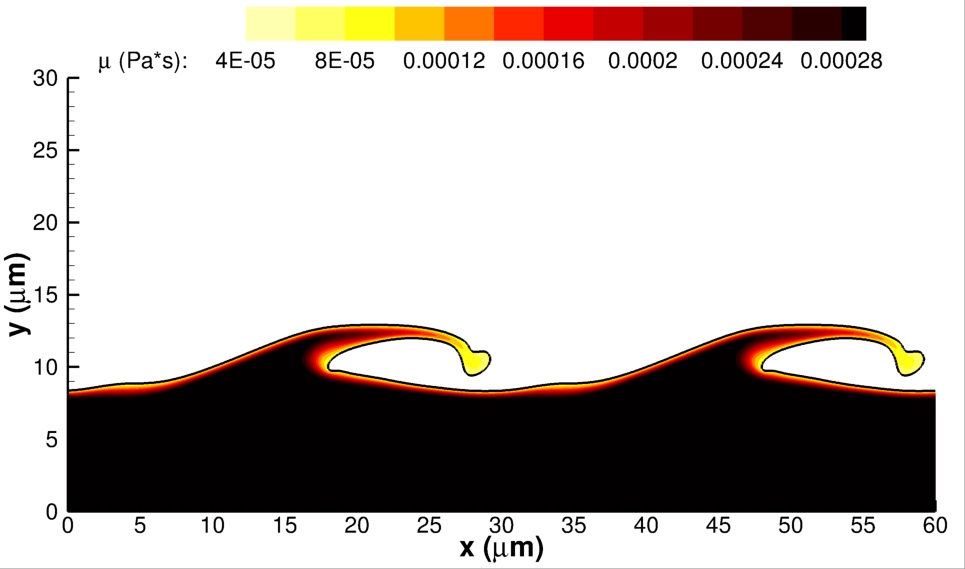}
  \caption{\label{subfig:VIS_3mus}\(\mu\) at \(t=3\) \(\mu\)s}
\end{subfigure}%
\begin{subfigure}{.43\textwidth}
  \centering
  \includegraphics[width=1.0\linewidth]{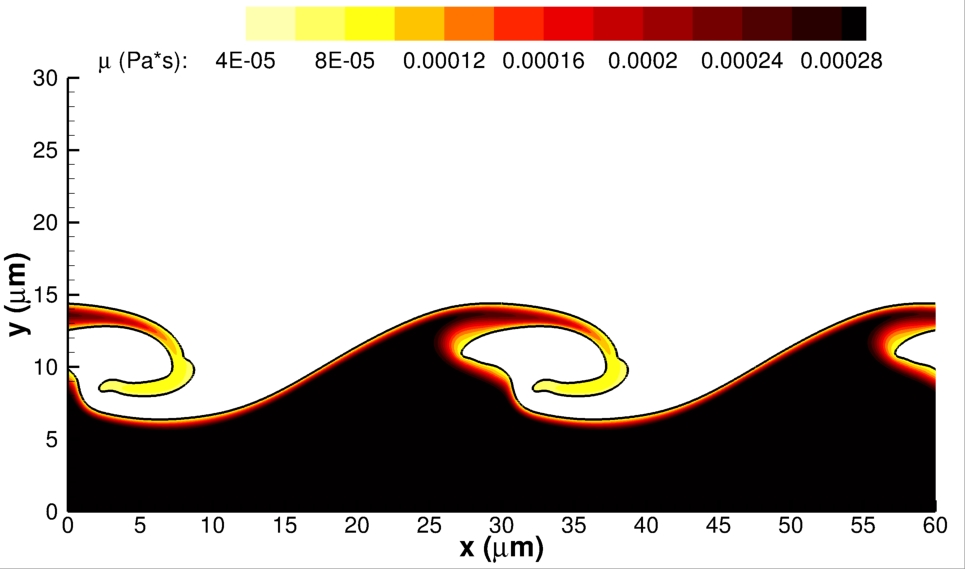}
  \caption{\label{subfig:VIS_4mus}\(\mu\) at \(t=4\) \(\mu\)s}
\end{subfigure}%
\\
\begin{subfigure}{.43\textwidth}
  \centering
  \includegraphics[width=1.0\linewidth]{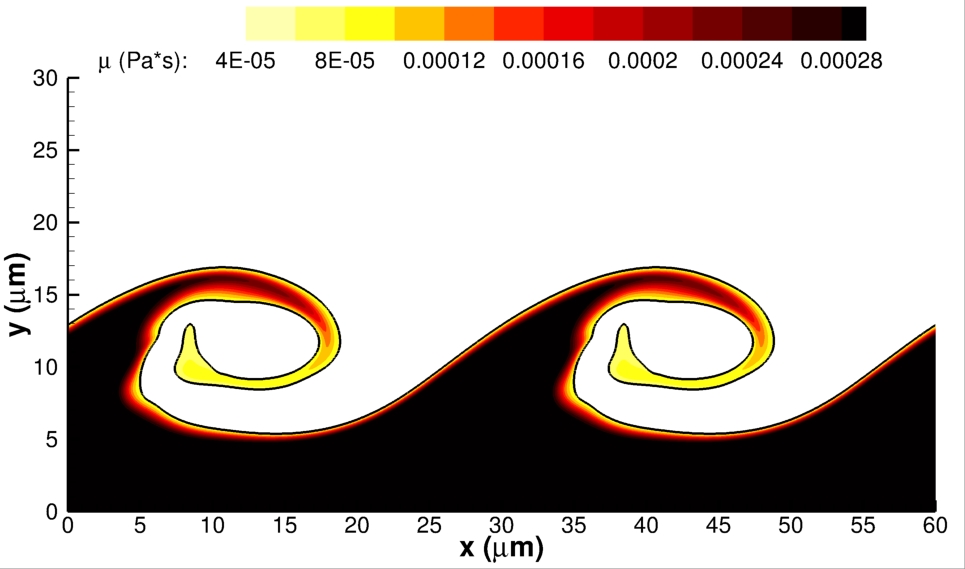}
  \caption{\label{subfig:VIS_5mus}\(\mu\) at \(t=5\) \(\mu\)s}
\end{subfigure}%
\begin{subfigure}{.43\textwidth}
  \centering
  \includegraphics[width=1.0\linewidth]{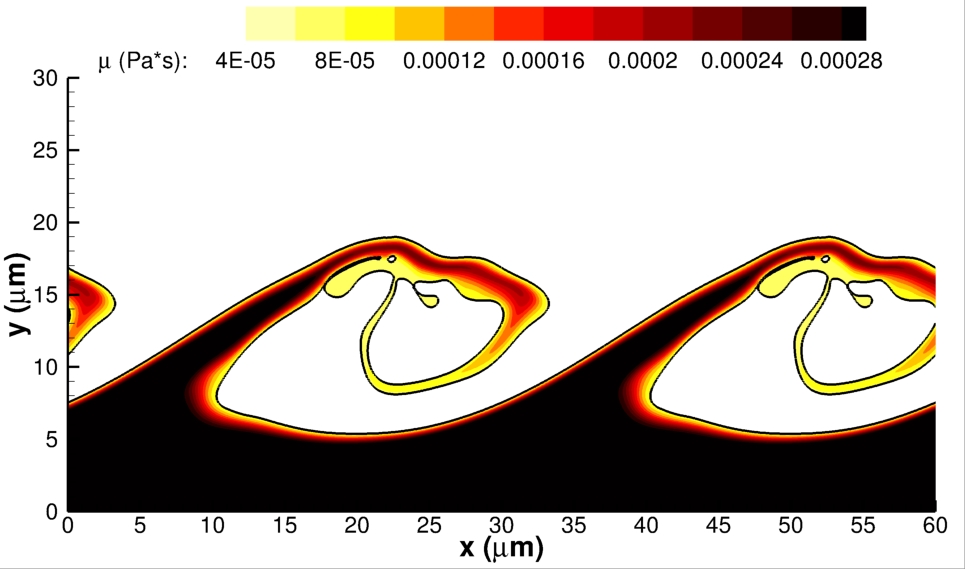}
  \caption{\label{subfig:VIS_6mus}\(\mu\) at \(t=6\) \(\mu\)s}
\end{subfigure}%
\\
\begin{subfigure}{.43\textwidth}
  \centering
  \includegraphics[width=1.0\linewidth]{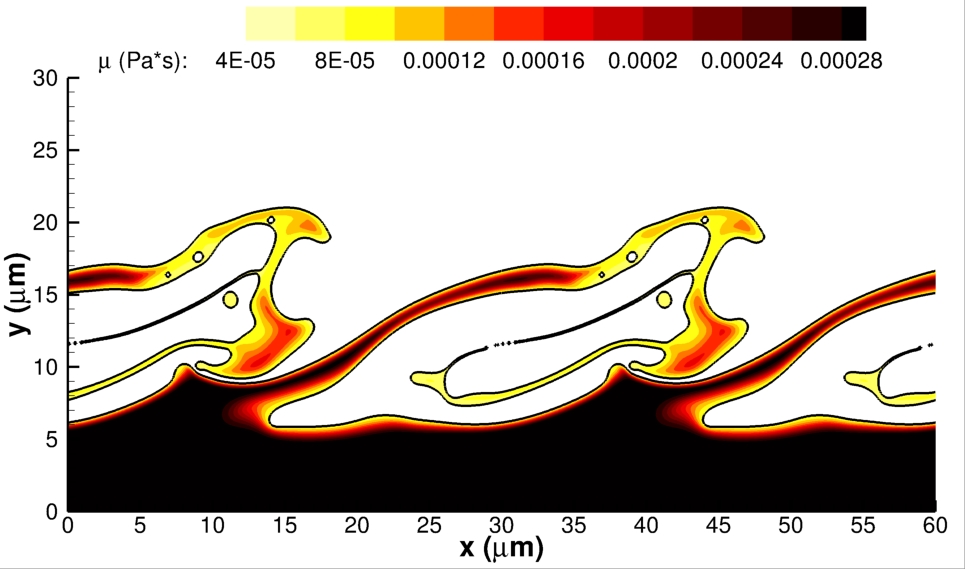}
  \caption{\label{subfig:VIS_7mus}\(\mu\) at \(t=7\) \(\mu\)s}
\end{subfigure}%
\begin{subfigure}{.43\textwidth}
  \centering
  \includegraphics[width=1.0\linewidth]{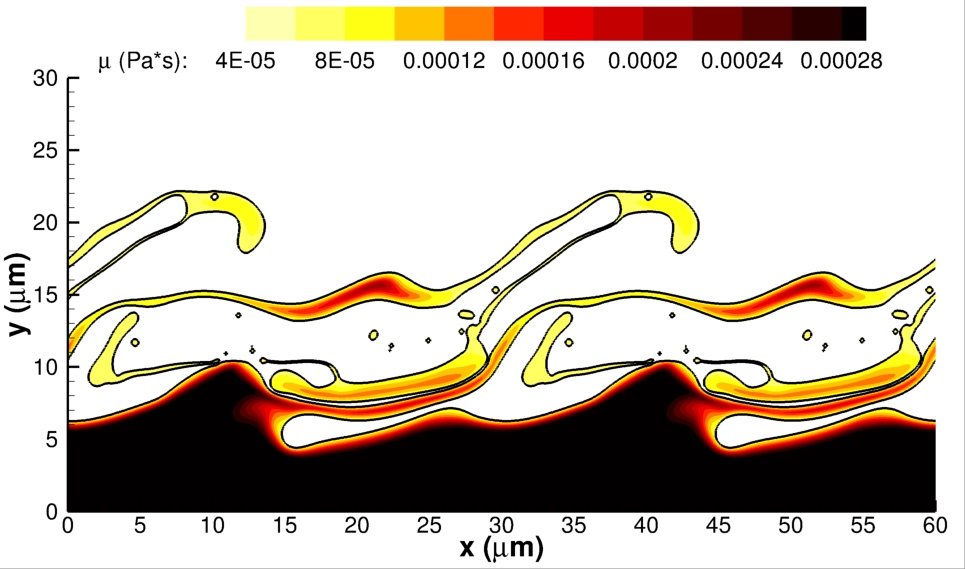}
  \caption{\label{subfig:VIS_8mus}\(\mu\) at \(t=8\) \(\mu\)s}
\end{subfigure}%
\caption{\label{fig:2djet_VIS}Viscosity plots in the liquid phase for the two-dimensional planar jet at 150 bar. At this high pressure, the viscosity of the gas mixture remains within the range of 2.8-3.4x10\(^{-5}\) Pa*s. The interface location is highlighted with a solid black curve representing the isocontour with \(C=0.5\).}
\end{figure*}

Figures \ref{fig:2djet_T} to \ref{fig:2djet_VIS} present the temporal evolution of the two-dimensional jet. The plots show the evolution of the temperature field, mass fraction of both species, density and viscosity. To clearly show the liquid perturbation at all times, the domain has been extended using the periodic behavior in the streamwise direction. \par 

\begin{figure*}[h!]
\centering
\begin{subfigure}{.43\textwidth}
  \centering
  \includegraphics[width=1.0\linewidth]{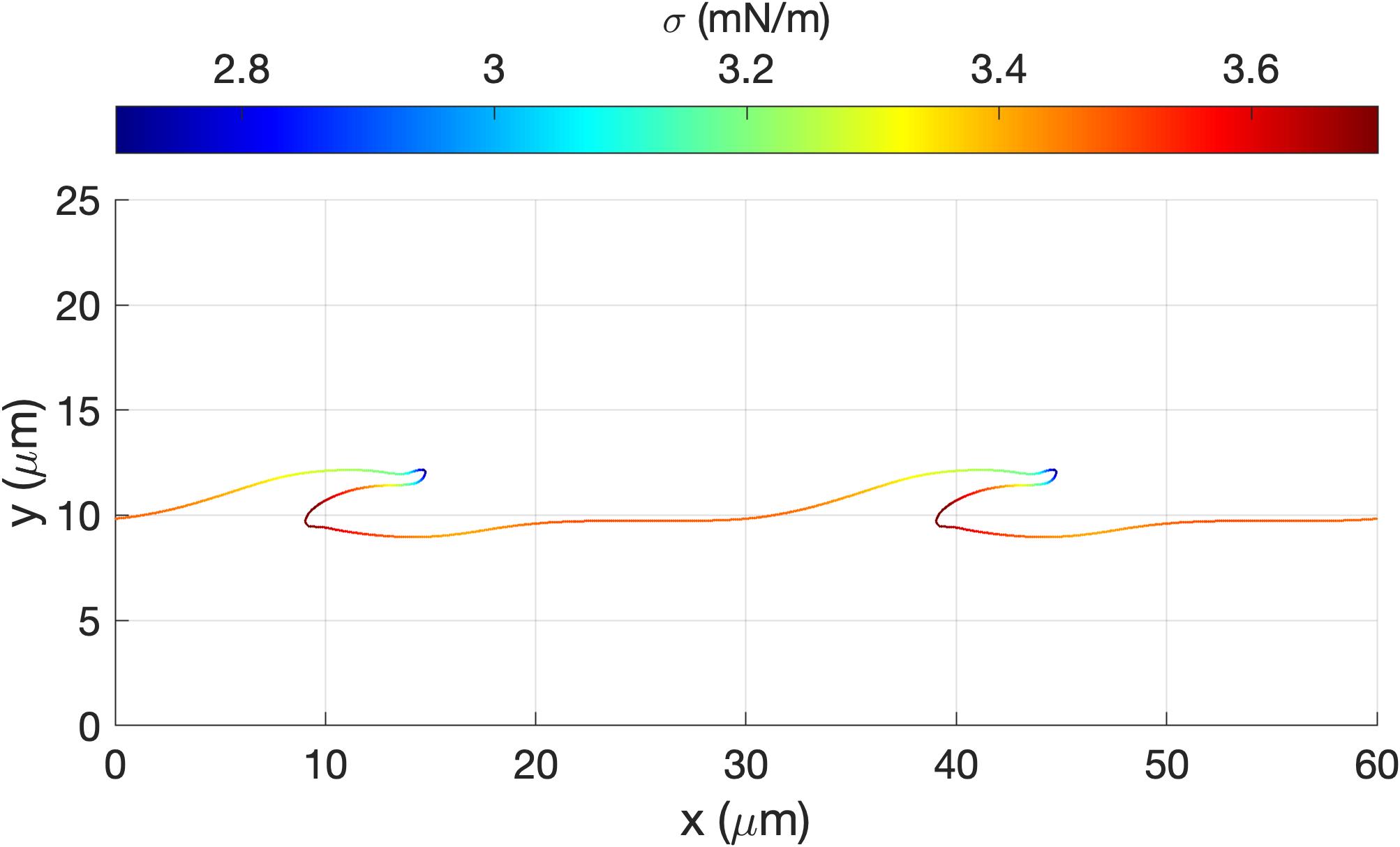}
  \caption{\label{subfig:sigma_2mus_inter}\(\sigma\) at \(t=2\) \(\mu\)s}
\end{subfigure}%
\begin{subfigure}{.43\textwidth}
  \centering
  \includegraphics[width=1.0\linewidth]{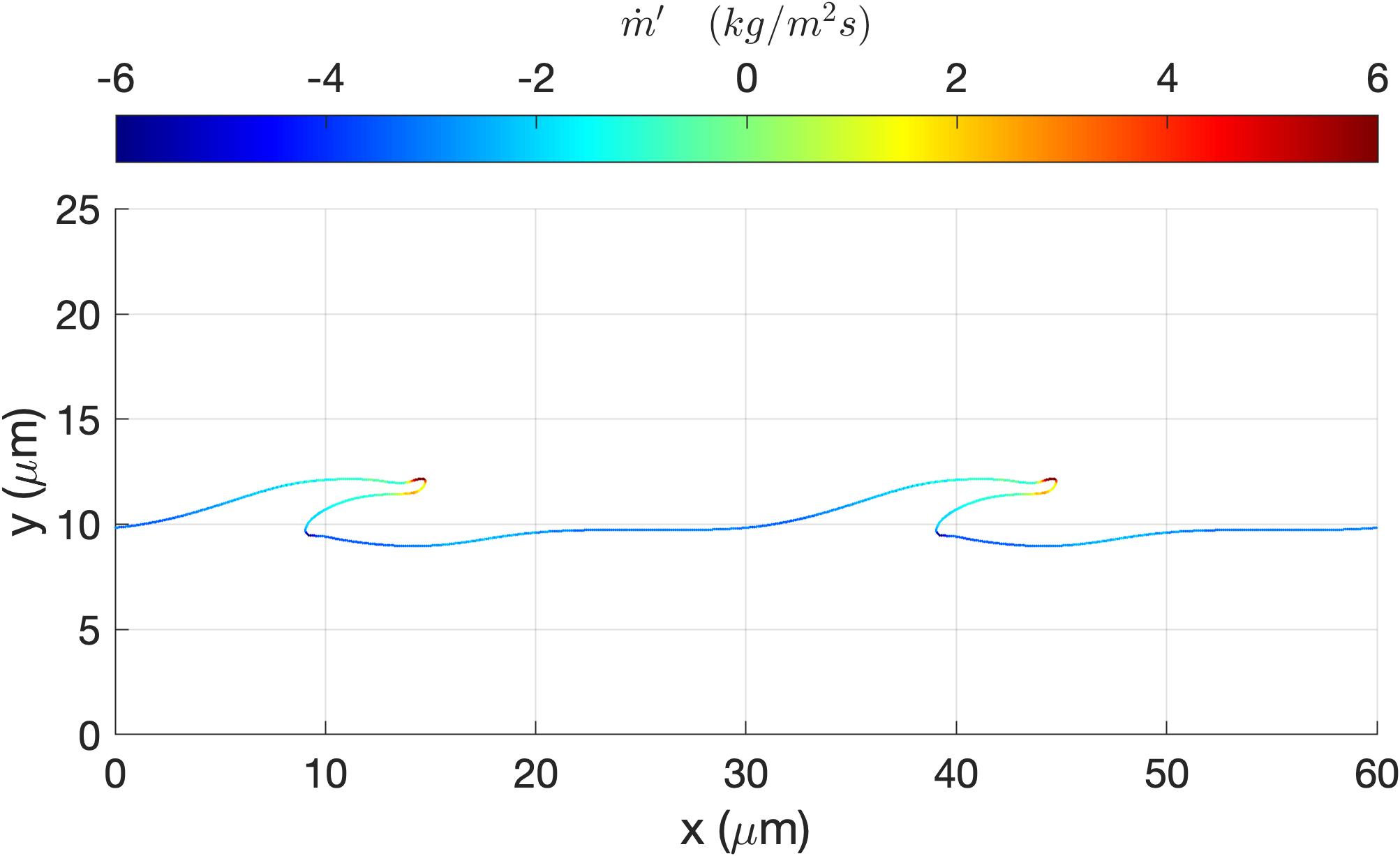}
  \caption{\label{subfig:mflux_2mus_inter}\(\dot{m}'\) at \(t=2\) \(\mu\)s}
\end{subfigure}%
\\
\begin{subfigure}{.43\textwidth}
  \centering
  \includegraphics[width=1.0\linewidth]{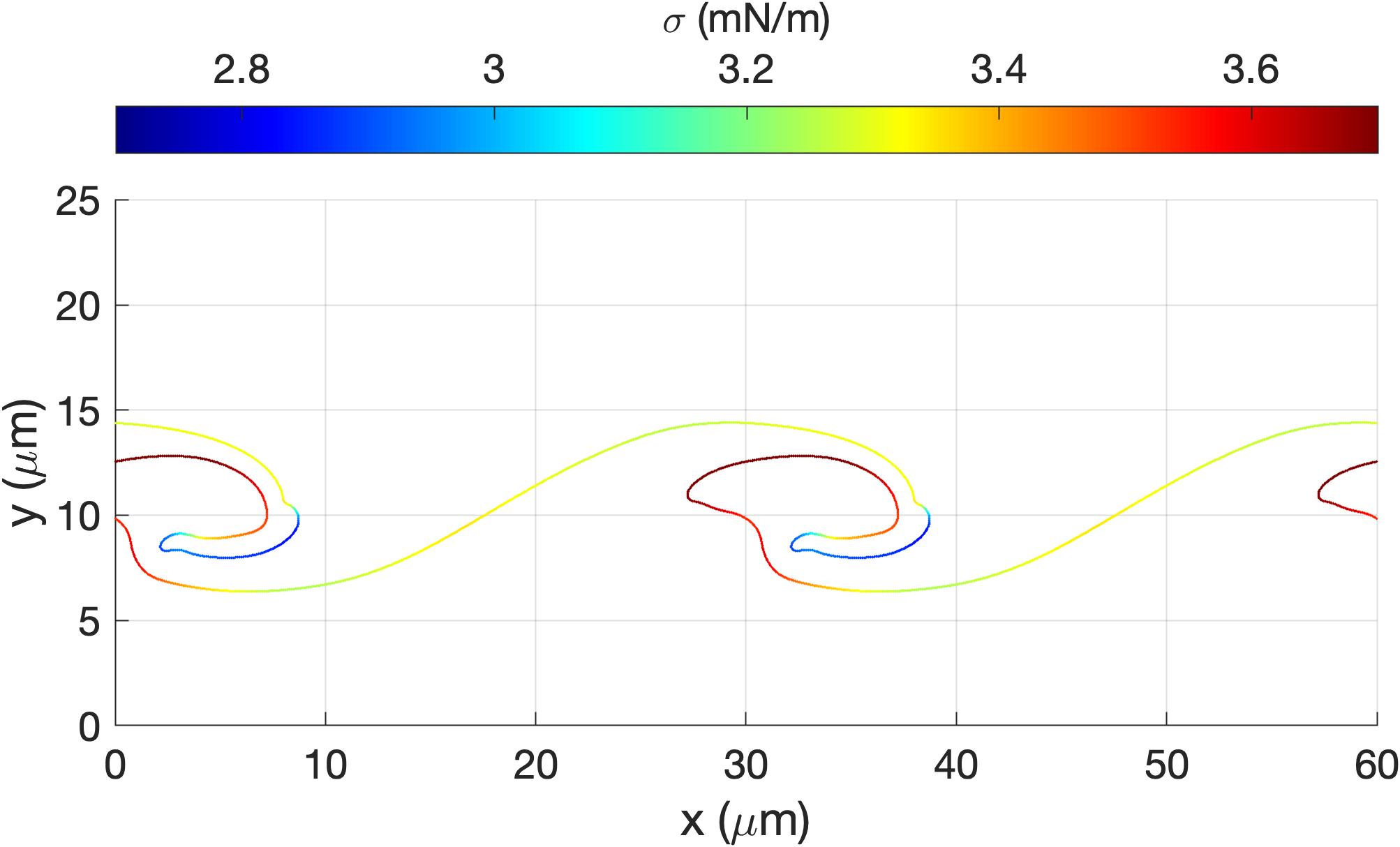}
  \caption{\label{subfig:sigma_4mus_inter}\(\sigma\) at \(t=4\) \(\mu\)s}
\end{subfigure}%
\begin{subfigure}{.43\textwidth}
  \centering
  \includegraphics[width=1.0\linewidth]{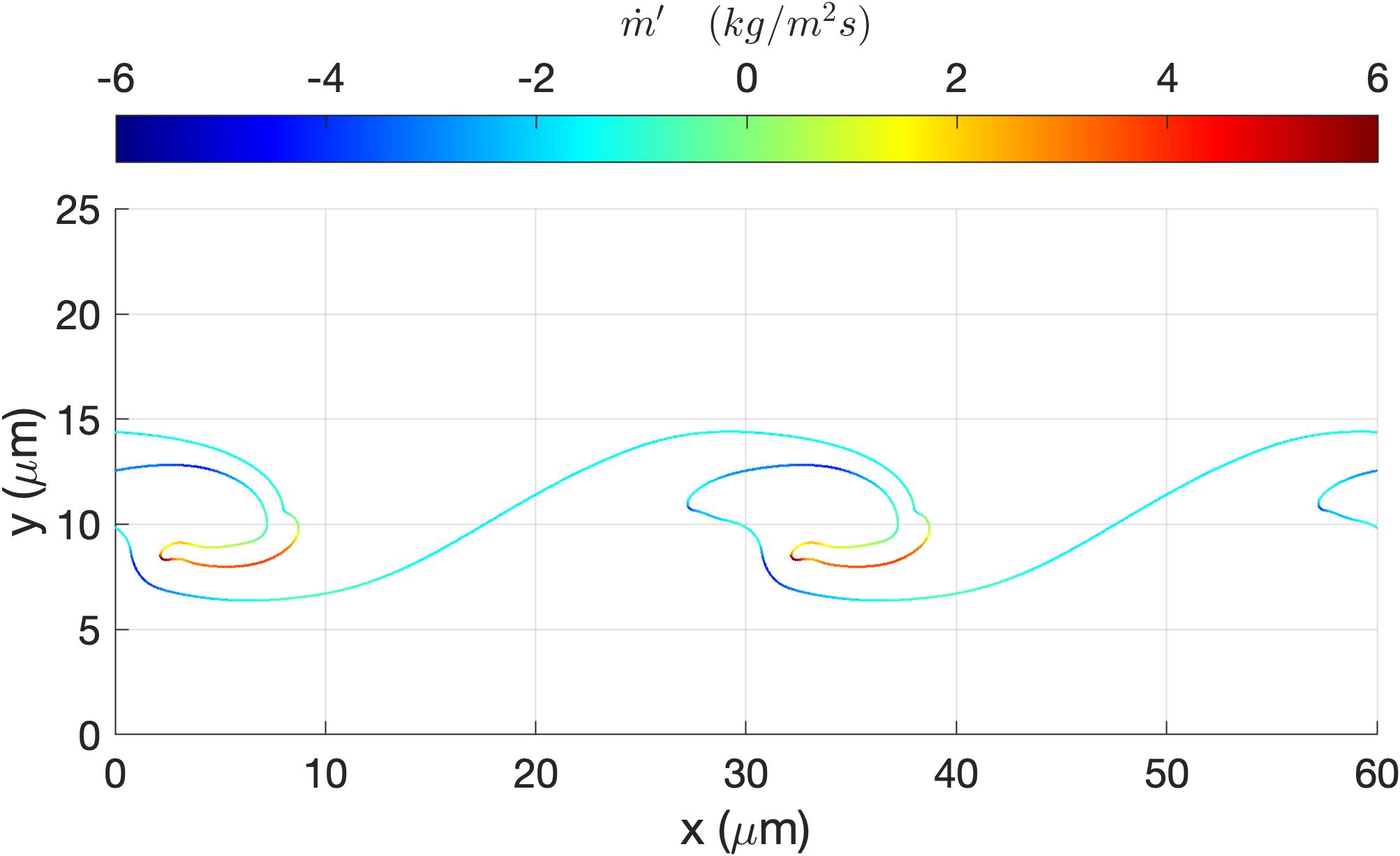}
  \caption{\label{subfig:mflux_4mus_inter}\(\dot{m}'\) at \(t=4\) \(\mu\)s}
\end{subfigure}%
\\
\begin{subfigure}{.43\textwidth}
  \centering
  \includegraphics[width=1.0\linewidth]{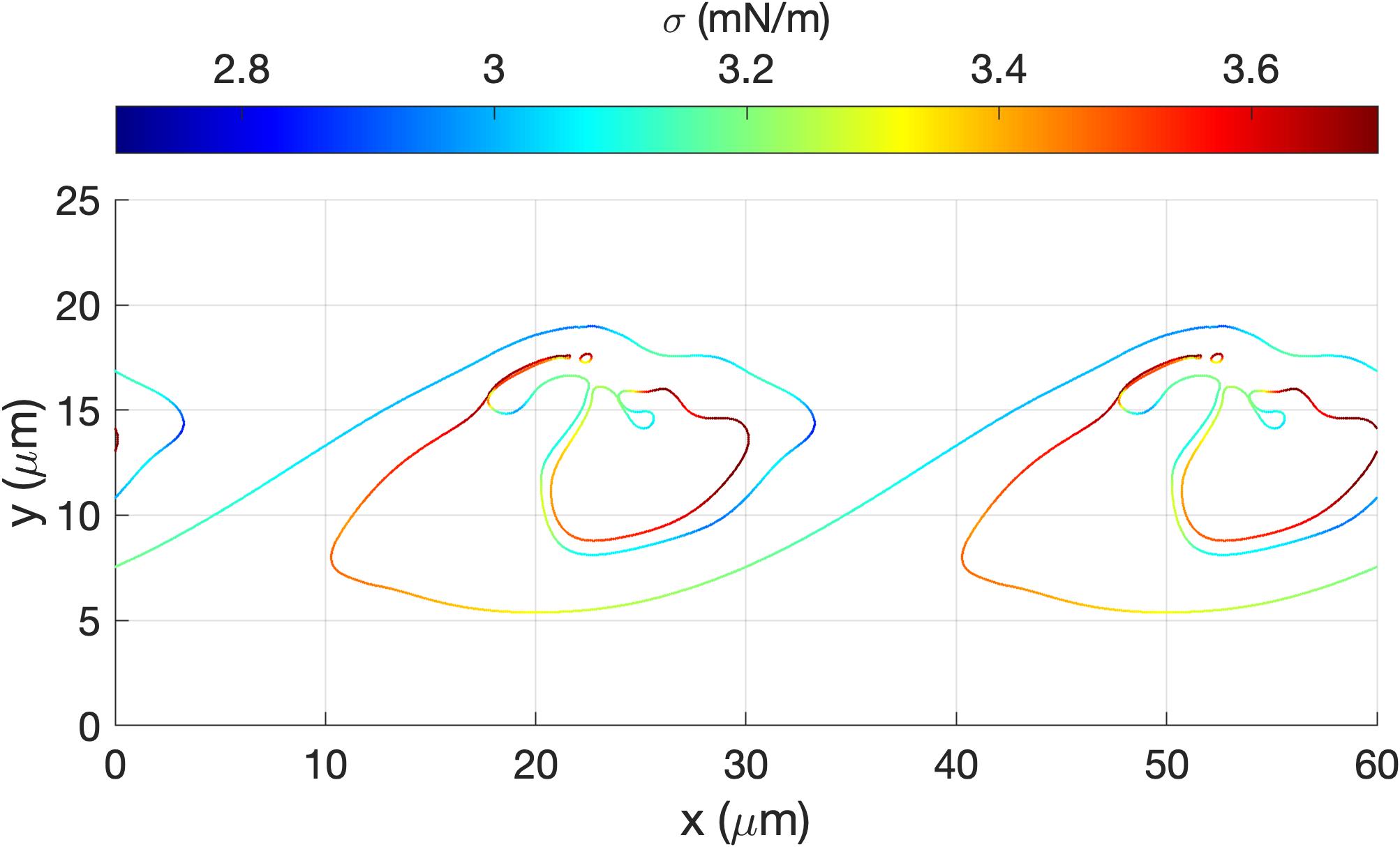}
  \caption{\label{subfig:sigma_6mus_inter}\(\sigma\) at \(t=6\) \(\mu\)s}
\end{subfigure}%
\begin{subfigure}{.43\textwidth}
  \centering
  \includegraphics[width=1.0\linewidth]{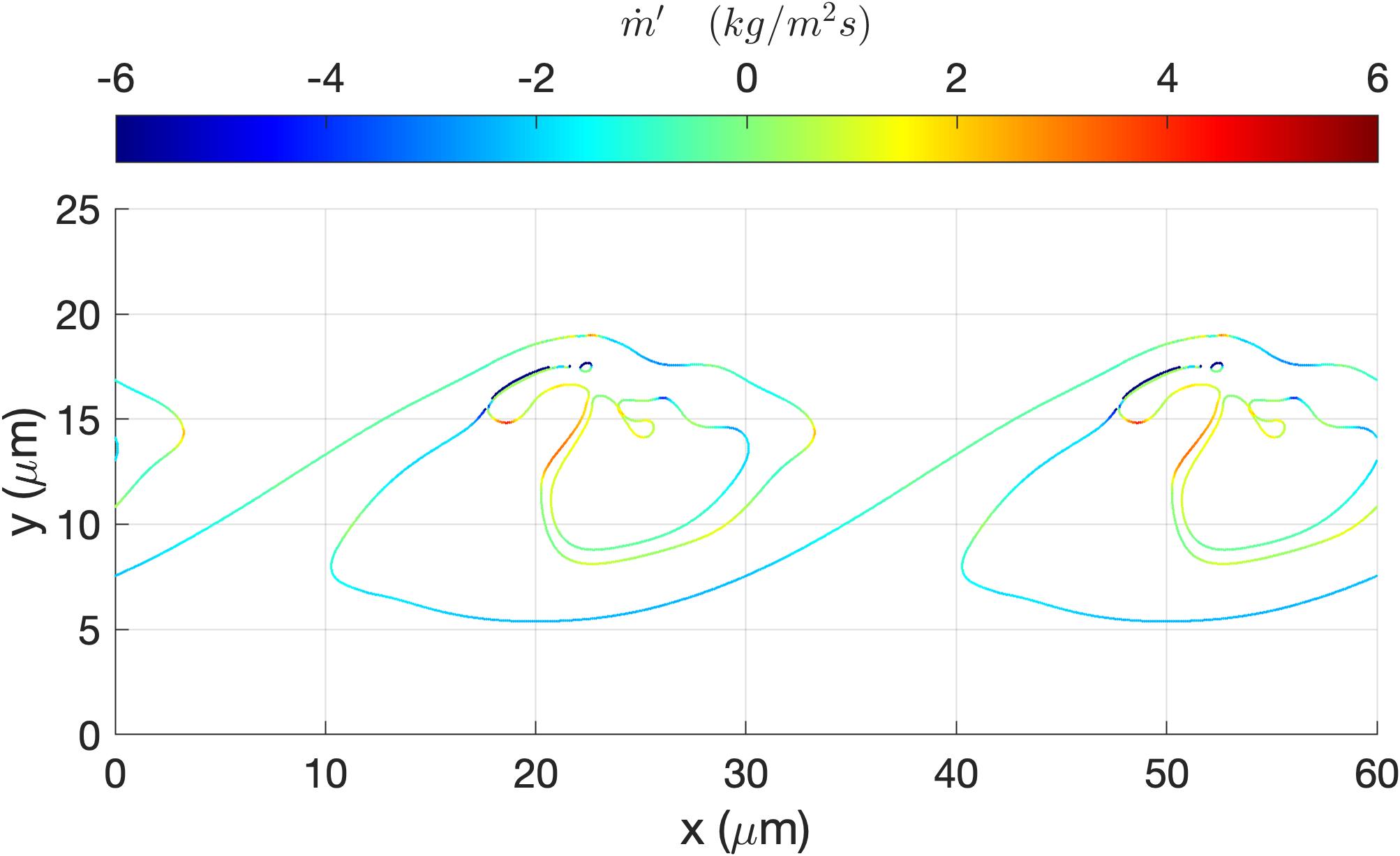}
  \caption{\label{subfig:mflux_6mus_inter}\(\dot{m}'\) at \(t=6\) \(\mu\)s}
\end{subfigure}%
\\
\begin{subfigure}{.43\textwidth}
  \centering
  \includegraphics[width=1.0\linewidth]{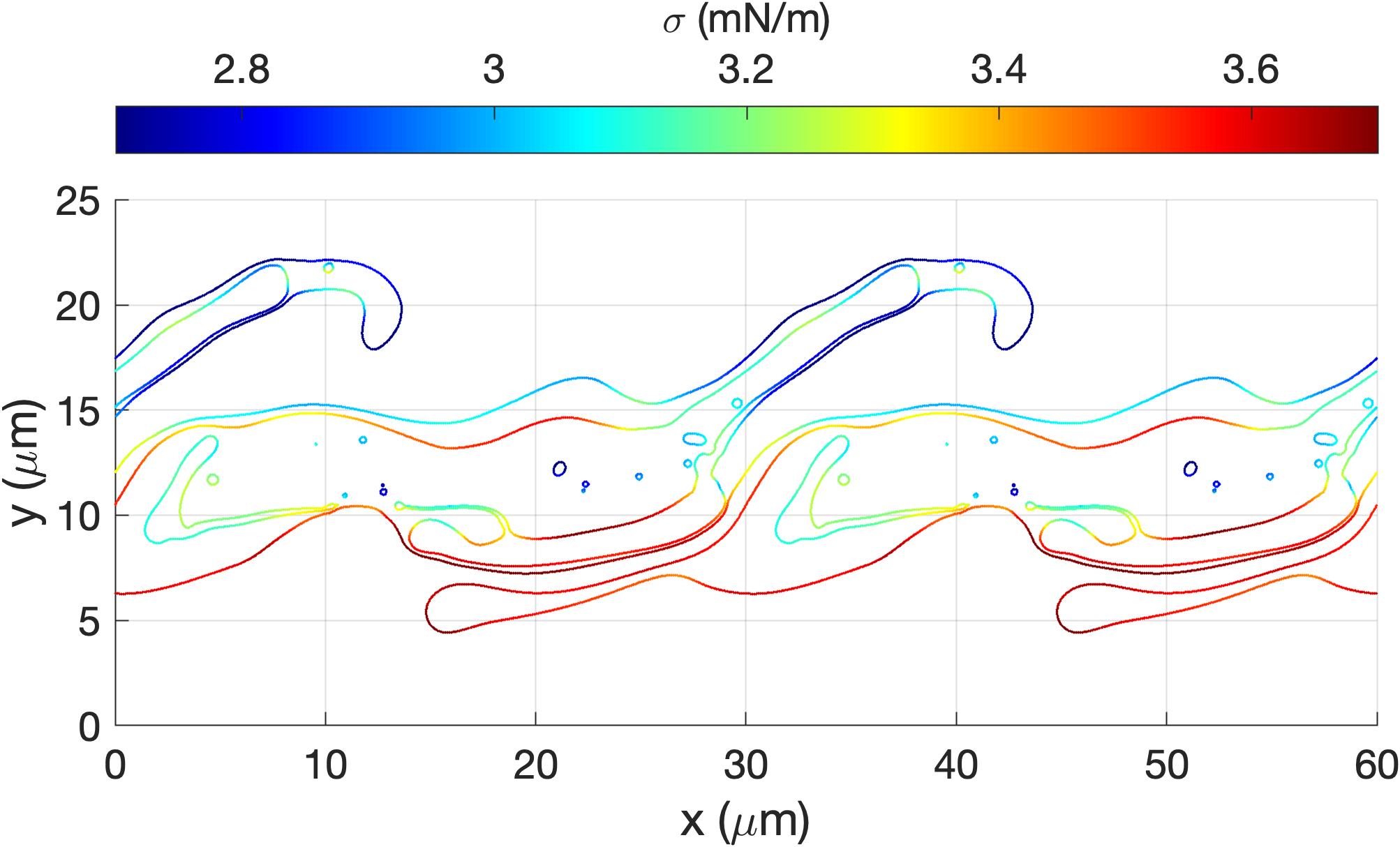}
  \caption{\label{subfig:sigma_8mus_inter}\(\sigma\) at \(t=8\) \(\mu\)s}
\end{subfigure}%
\begin{subfigure}{.43\textwidth}
  \centering
  \includegraphics[width=1.0\linewidth]{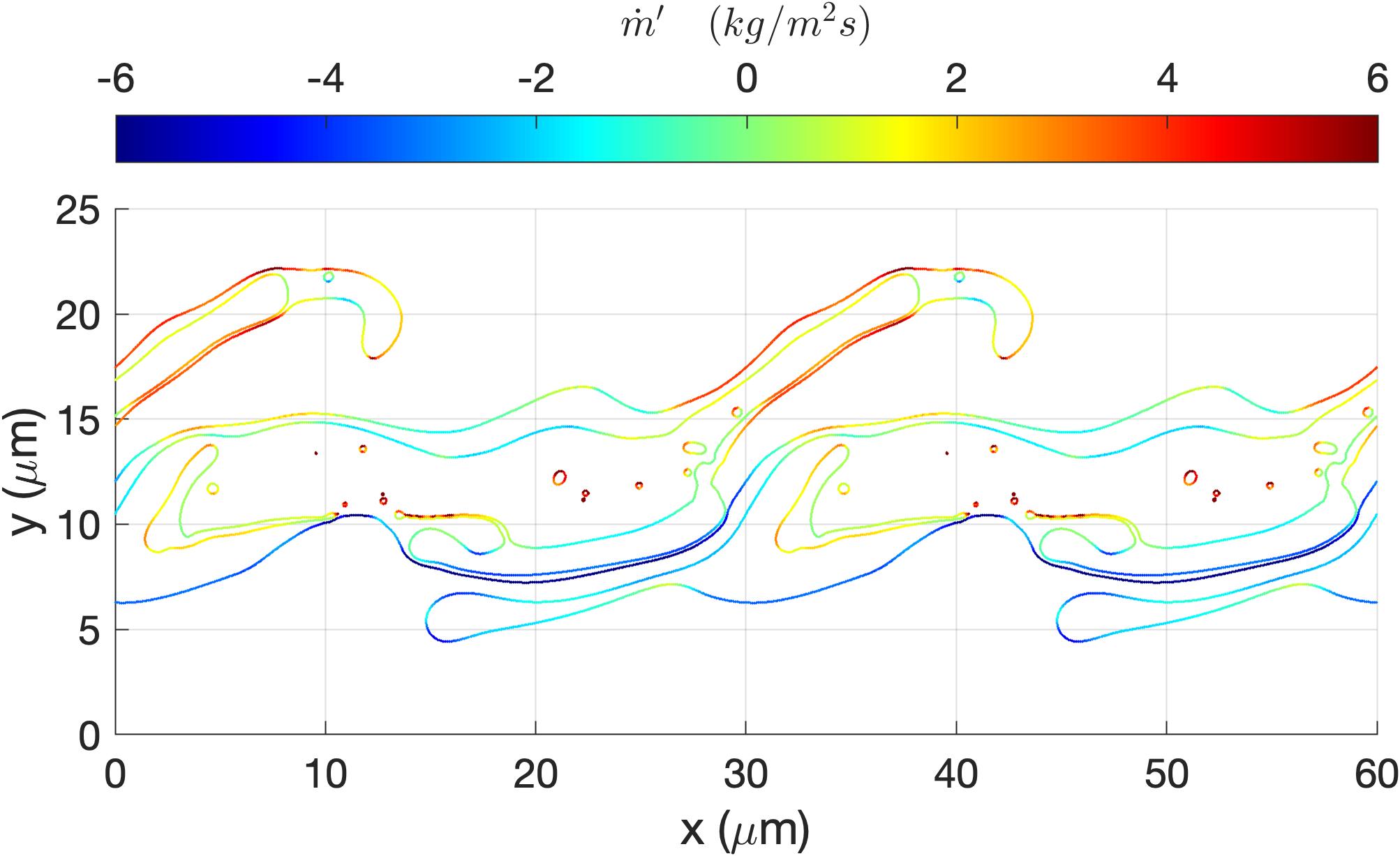}
  \caption{\label{subfig:mflux_8mus_inter}\(\dot{m}'\) at \(t=8\) \(\mu\)s}
\end{subfigure}%
\caption{\label{fig:2djet_sigma_mflux_inter}Surface tension coefficient and net mass flux per unit area along the interface for the two-dimensional planar jet at 150 bar. The interface shape is colored by the value of each respective variable.}
\end{figure*}

Species and energy mixing (i.e., the solution of Eqs.~(\ref{eqn:spcont}) and~(\ref{eqn:energy})) are shown in Figures \ref{fig:2djet_T}, \ref{fig:2djet_YFg} and \ref{fig:2djet_YOl}. Initially (i.e., \(t<1\) \(\mu\)s), mass and thermal diffusion drive the mixing between both phases as seen in simpler studies where the interface does not deform~\cite{poblador2018transient,davis2019development,poblador2021selfsimilar}. Temperature drops substantially in the gas phase while heating in the liquid phase occurs more slowly. In combustion applications, the analysis of the fuel mixing into the oxidizer stream (both in gaseous and liquid states) is important since it determines combustion efficiency. As observed in Figure~\ref{fig:2djet_YFg}, \textit{n}-decane vaporizes and mixes well with oxygen as the atomization cascade develops at these very high pressures. In particular, small discrete domains of high fuel-vapor concentration form in the gas phase at intervals, along the interface. Moreover, smaller liquid fuel parcels appear and extend into the hotter gas freestream, where vaporization rates increase (see Figure~\ref{fig:2djet_sigma_mflux_inter}) and enhance the mixing of the hydrocarbon fuel with the oxidizing species. The higher dissolution of oxygen at high pressures is apparent in Figure~\ref{fig:2djet_YOl}. The growth of the liquid phase mixing region does not change significantly as pressure increases. However, LTE at the interface is satisfied by a higher dissolution of oxygen at supercritical pressures compared to a negligible dissolution of the gas species at subcritical pressures. \par

The interface starts showing substantial deformation very early in time, even when the initial perturbation amplitude is only 1/60th of the perturbation wavelength. From 1 to 5 \(\mu\)s the liquid starts elongating and rolling over itself. This fast deformation is not surprising. The analyzed wavelength is shorter and the magnitude of the shear strain rate is greater than given in prior incompressible flow studies analyzing jets with similar density ratios and Weber numbers~\cite{jarrahbashi2014vorticity,jarrahbashi2016early,zandian2017planar,zandian2018understanding,zandian2019length,zandian2019vorticity}. A previous work of a supercritical axisymmetric liquid jet~\cite{poblador2019axisymmetric} also showed a faster-than-usual interface deformation. The choice of a short perturbation wavelength is made due to the reduced surface tension environment and the strong mixing expected at very high pressures, which induce the growth of such perturbations. Later in time (\(t>5 \mu\)s), the liquid deformation becomes more chaotic. Coalescence of different liquid regions occur, which may capture small gas pockets. Overall, thin and elongated liquid structures develop. These thin ligaments might break up, generating some two-dimensional droplets. \par 

As noted earlier in the text, the three-dimensional picture can be substantially different. A higher presence of droplets is expected since surface tension forces naturally cause neck thinning and breakup. Thus, elongated thin liquid structures might not dominate the picture in the longer term. Moreover, the methodology used in this work relies on a good mesh resolution of the liquid shape. Even though a very fine mesh is used, the evolution of the liquid deformation is affected by numerical errors once very small liquid structures develop or high-curvature regions appear. \par 

\begin{figure*}[h!]
\centering
\begin{subfigure}{.5\textwidth}
  \centering
  \includegraphics[width=0.96\linewidth]{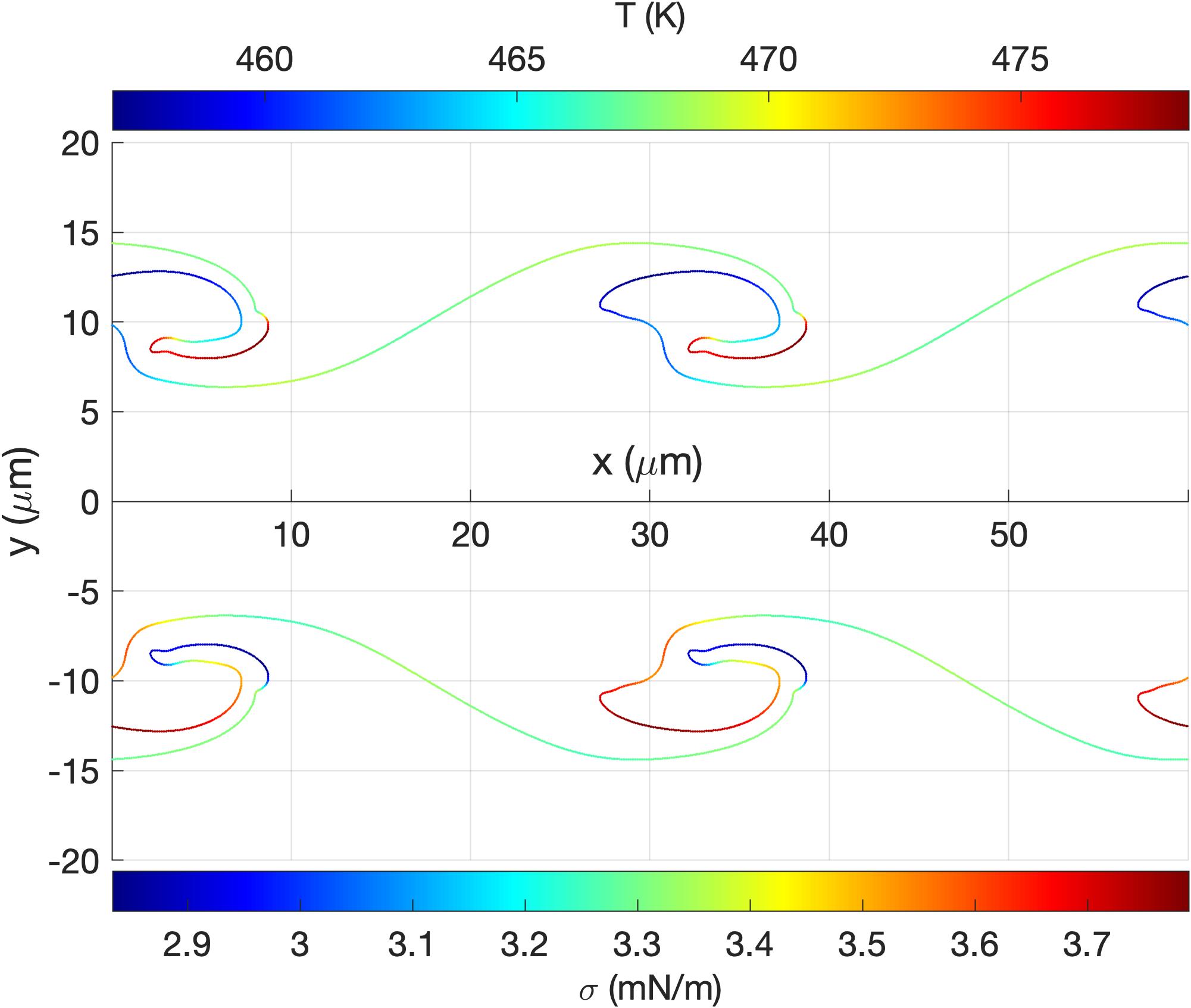}
  \caption{\label{subfig:TpSigma_4mus_inter_new}\(T\) and \(\sigma\)}
\end{subfigure}%
\begin{subfigure}{.5\textwidth}
  \centering
  \includegraphics[width=0.96\linewidth]{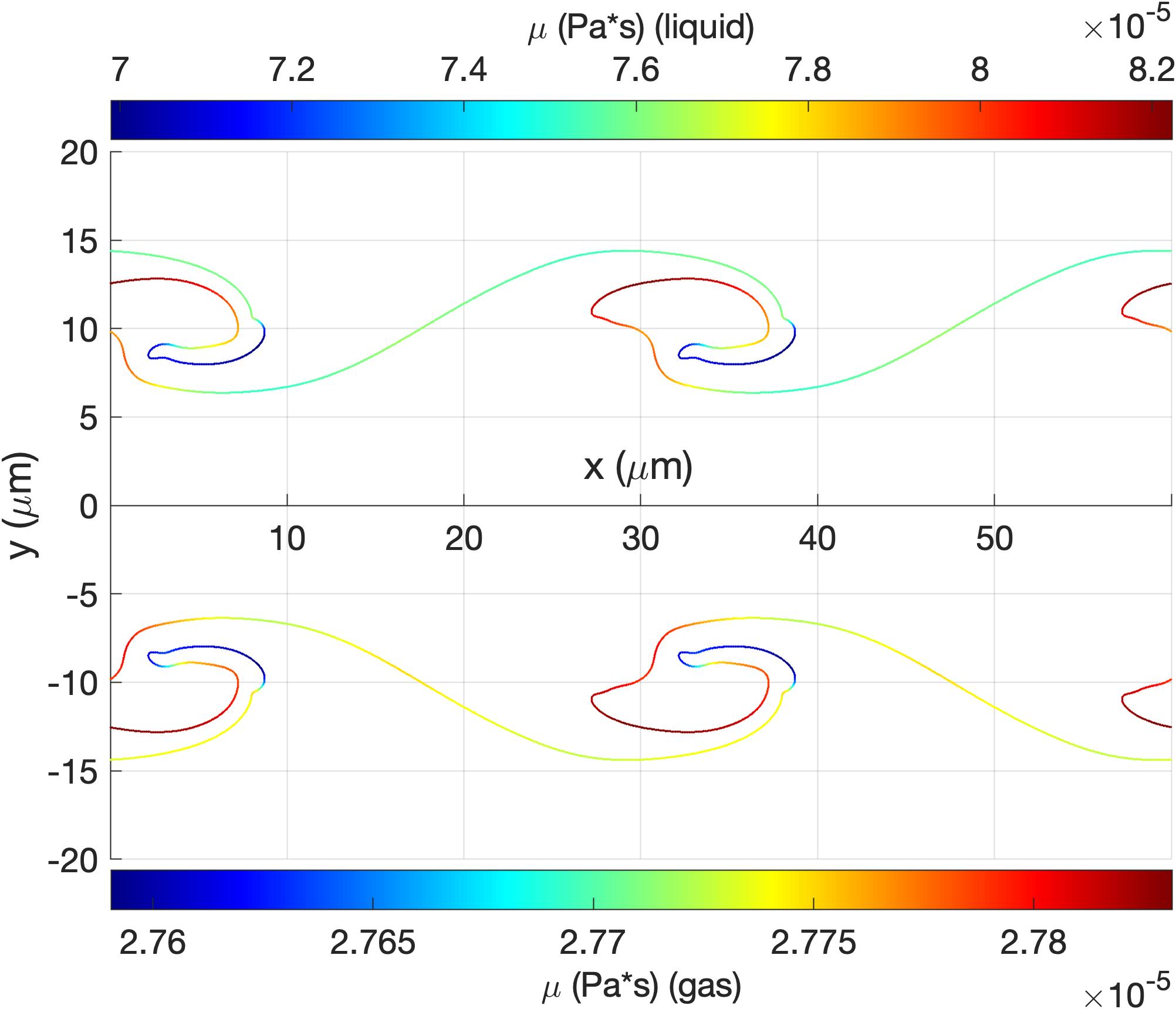}
  \caption{\label{subfig:VIS_4mus_inter_new}\(\mu\)}
\end{subfigure}%
\\
\begin{subfigure}{.5\textwidth}
  \centering
  \includegraphics[width=0.96\linewidth]{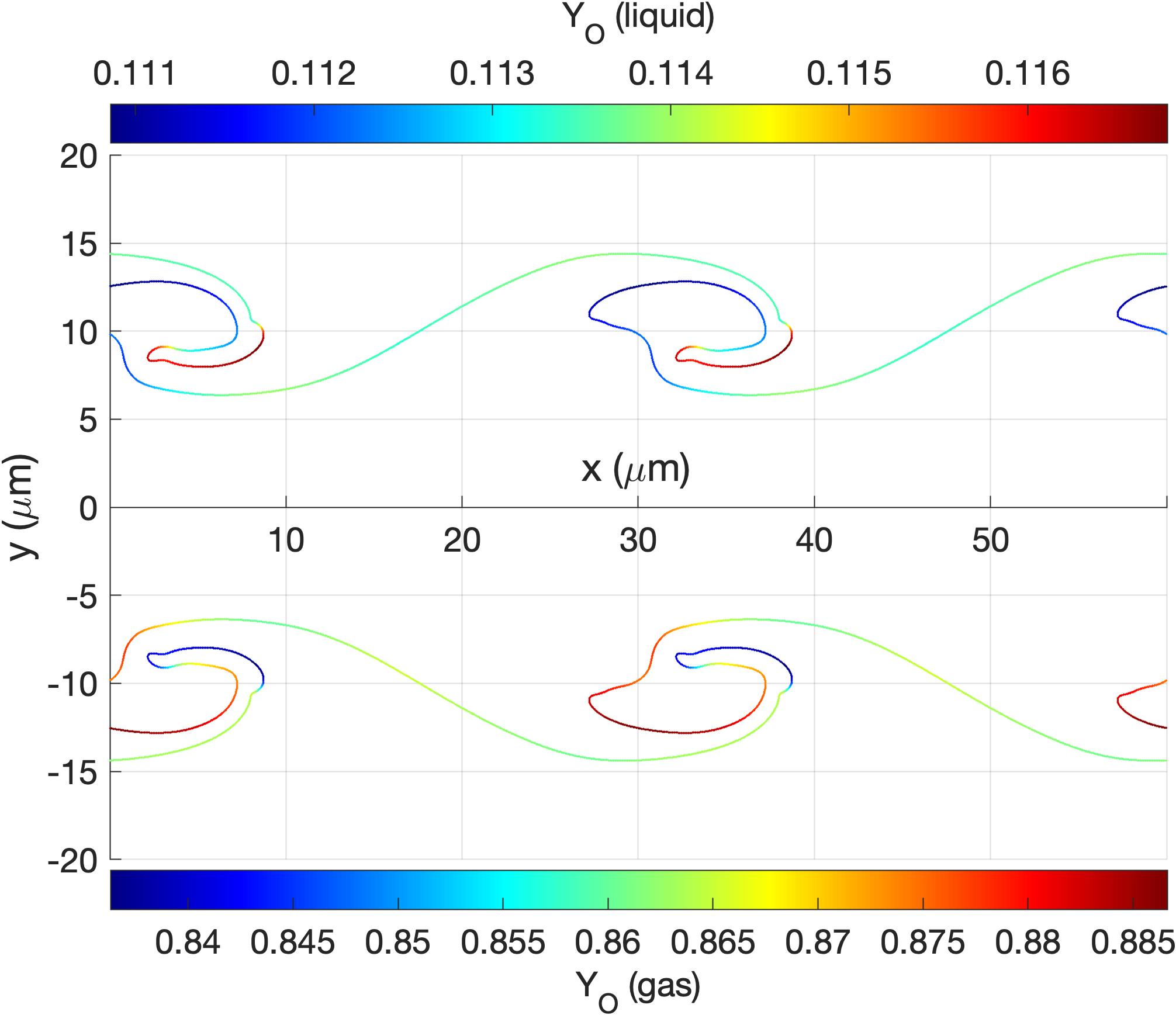}
  \caption{\label{subfig:YO_4mus_inter_new}\(Y_O\)}
\end{subfigure}%
\begin{subfigure}{.5\textwidth}
  \centering
  \includegraphics[width=0.96\linewidth]{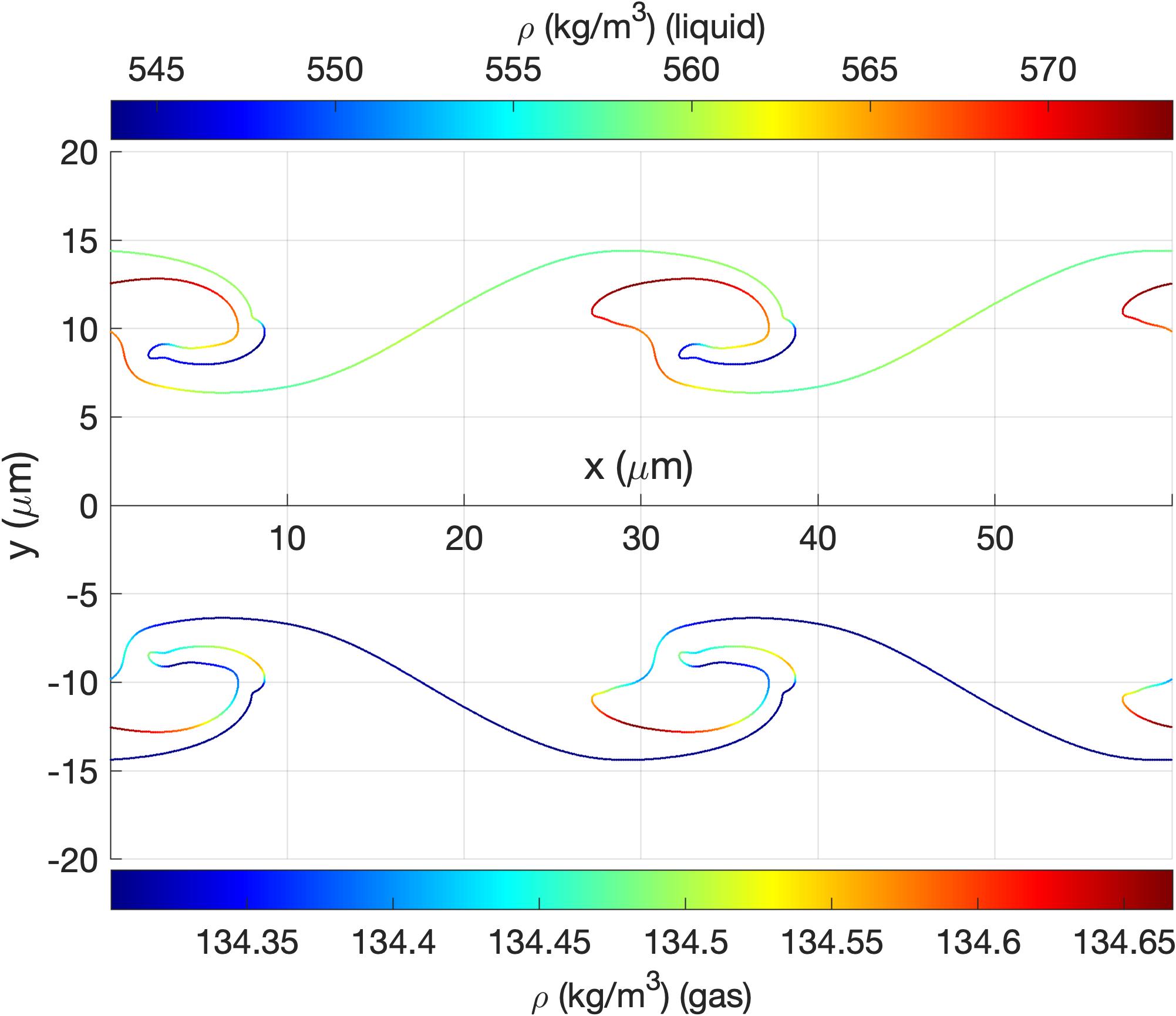}
  \caption{\label{subfig:DEN_4mus_inter_new}\(\rho\)}
\end{subfigure}%
\caption{\label{fig:2djet_various_inter}Variation of various interface properties along the interface for the two-dimensional planar jet at 150 bar and \(t=4\) \(\mu\)s. The symmetry condition is used to mirror the jet and plot different variables on each side. The interface shape is colored by the value of each respective variable.}
\end{figure*}

The mixing process between the liquid and the gas at these very high pressures is responsible for the fast temporal evolution of the liquid-gas interface, as well as the observed  topology. Density and liquid viscosity plots are presented in Figures \ref{fig:2djet_DENg}, \ref{fig:2djet_DENl} and \ref{fig:2djet_VIS}, while interface properties are shown in Figures \ref{fig:2djet_various_inter} and \ref{fig:2djet_sigma_mflux_inter}. \par 

An important feature of high-pressure, two-phase flows is the strong variation of interface properties along the interface. As seen in Figure \ref{fig:2djet_various_inter} for \(t=4\) \(\mu\)s, the interface temperature is higher near compressed gas regions (e.g., crest of the perturbation or regions of strong gas entrainment). Phase-equilibrium at these higher temperatures enhances the dissolution of oxygen into the liquid phase and the vaporization of \textit{n}-decane into the gas phase (see Figure~\ref{subfig:mflux_4mus_inter}). The effect on the interface gas density is minimal, yet the interface liquid density is reduced in these regions. This high temperature, together with an increased concentration of the lighter oxygen species, causes the surface tension coefficient to drop compared to the rest of the interface. On the other hand, regions of compressed liquid (e.g., wave trough) or inside the rolling liquid (captured gas regions) show lower interface temperatures and a higher surface tension coefficient. \par 

Figure~\ref{fig:2djet_sigma_mflux_inter} presents variations of pure interfacial properties (i.e., surface tension coefficient and mass flux per unit area due to phase change) over a wider temporal range. Net mass exchange between both phases varies considerably along the interface. Simpler studies at 150 bar where the interface does not deform show that net condensation occurs~\cite{poblador2018transient,davis2019development,poblador2021selfsimilar}. However, as seen in Figure~\ref{fig:2djet_sigma_mflux_inter}, interface deformation affects \(\dot{m}'\) and net vaporization or net condensation can occur simultaneously at different interface locations. Similarly to the variations in interface temperature and composition, regions of compressed gas (e.g., wave crest, strong gas entrainment) show net vaporization or weaker condensation than regions of compressed liquid. This behavior was also seen and discussed in Poblador-Ibanez and Sirignano~\cite{poblador2019axisymmetric}. Despite the fact that the total mass exchange across the interface over time is relatively negligible in this problem setup (less than 1\% of the liquid mass) and that it does not seem to have a substantial impact on the interface dynamics at high pressures, it remains an important feature of supercritical two-phase flows due to its complex behavior. \par 

The reduced surface-tension environment at supercritical pressures already exists as both phases look more alike near the interface. Faster growth of instabilities with shorter wavelength are expected, but further weakening of surface tension forces near the wave crest enhances the instability growth. Moreover, mixing in the liquid phase has a strong effect on the liquid dynamical properties (i.e., density and viscosity). At \(t=1\) \(\mu\)s, the higher interface temperature near the wave crest enhances the dissolution of oxygen into the liquid. Thus, a region of lower density and gas-like liquid viscosity develops (see Figures \ref{fig:2djet_DENl} and \ref{fig:2djet_VIS}). Actually, these mixing effects can be observed everywhere along the interface, but they are stronger near the wave crest as mentioned. A lower liquid viscosity can be related to less damping of a surface instability due to momentum diffusion.  \par

Additionally, the reductions in density, surface tension, and viscosity makes the liquid more susceptible of being deformed by the inertia of the gas phase, which explains the quick growth of elongated liquid structures. For instance, the deformation of the liquid near the wave crest (i.e., stronger gas inertia and reduced liquid density and viscosity) from \(t=1\) \(\mu\)s to \(t=4\) \(\mu\)s shows the appearance of an elongated two-dimensional lobe that stretches rather fast into the gas phase before vortical motion curves it. \par

\subsection{Interface development and mixing in three dimensions}

\begin{figure*}[h!]
\centering
\begin{subfigure}{.5\textwidth}
  \centering
  \includegraphics[width=1.0\linewidth]{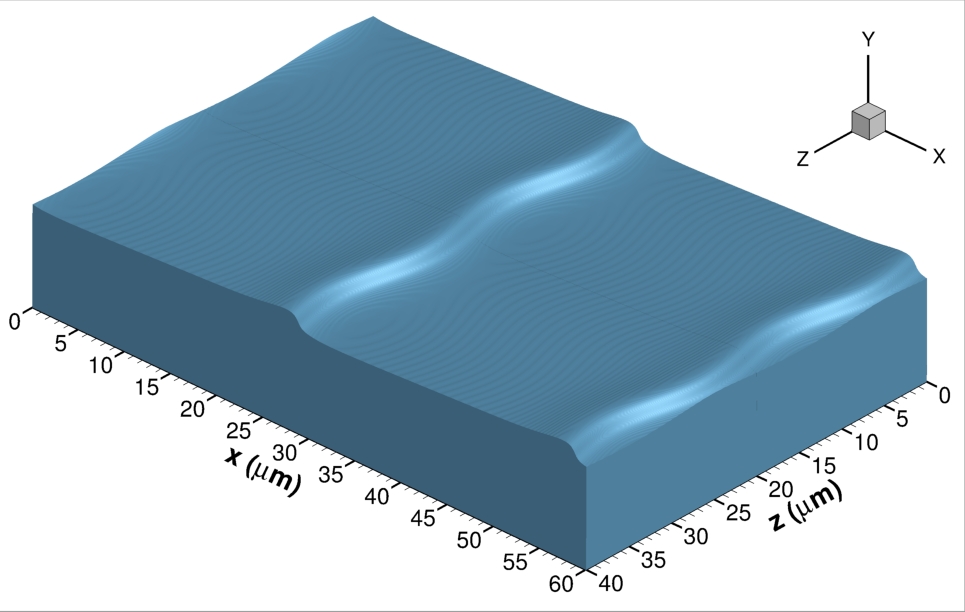}
  \caption{\label{subfig:150_A1_int_1mus}Initial 0.1 \(\mu\)m spanwise amplitude at \(t=1\) \(\mu\)s}
\end{subfigure}%
\begin{subfigure}{.5\textwidth}
  \centering
  \includegraphics[width=1.0\linewidth]{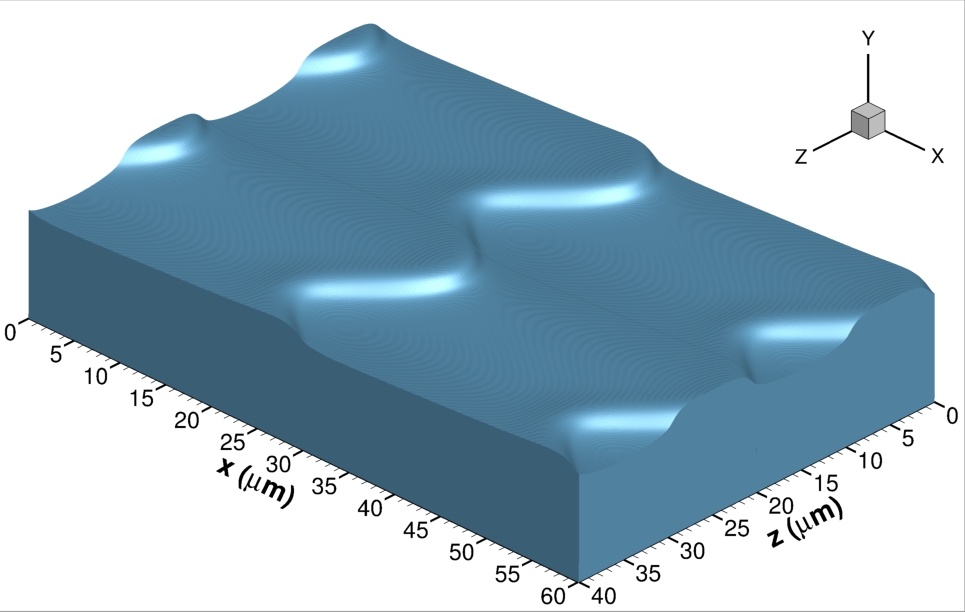}
  \caption{\label{subfig:150_A2_int_1mus}Initial 0.5 \(\mu\)m spanwise amplitude at \(t=1\) \(\mu\)s}
\end{subfigure}%
\\
\begin{subfigure}{.5\textwidth}
  \centering
  \includegraphics[width=1.0\linewidth]{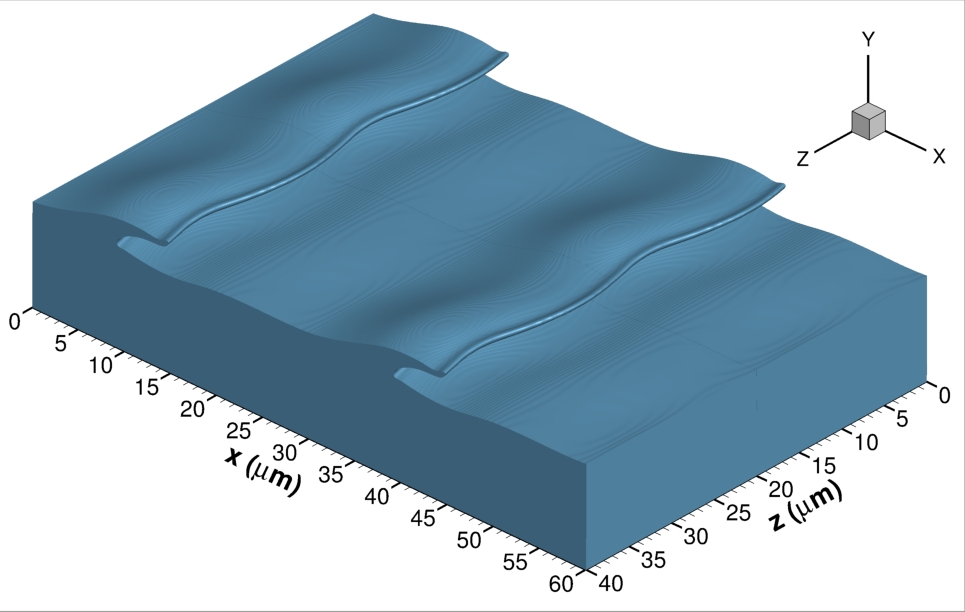}
  \caption{\label{subfig:150_A1_int_2mus}Initial 0.1 \(\mu\)m spanwise amplitude at \(t=2\) \(\mu\)s}
\end{subfigure}%
\begin{subfigure}{.5\textwidth}
  \centering
  \includegraphics[width=1.0\linewidth]{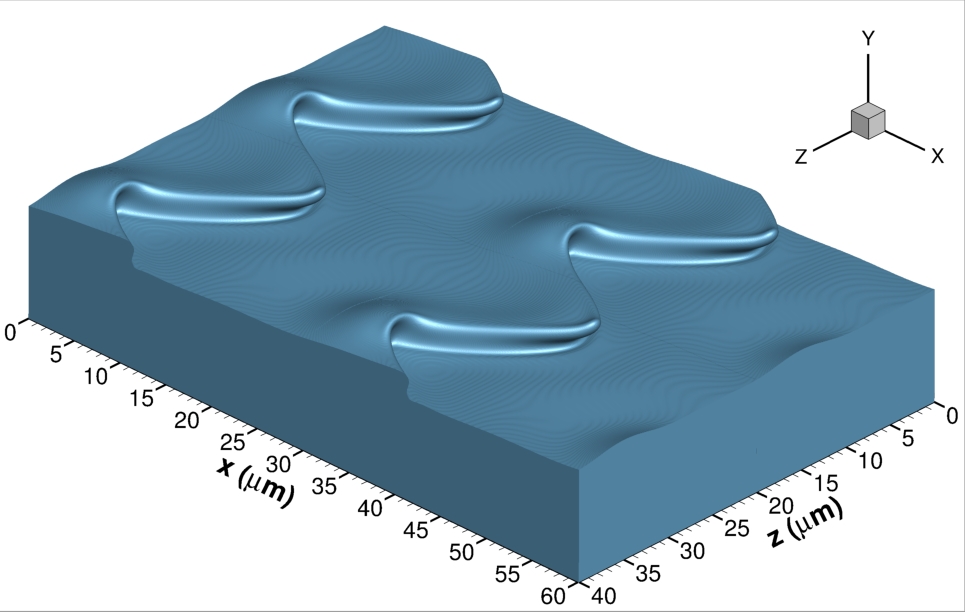}
  \caption{\label{subfig:150_A2_int_2mus}Initial 0.5 \(\mu\)m spanwise amplitude at \(t=2\) \(\mu\)s}
\end{subfigure}%
\\
\begin{subfigure}{.5\textwidth}
  \centering
  \includegraphics[width=1.0\linewidth]{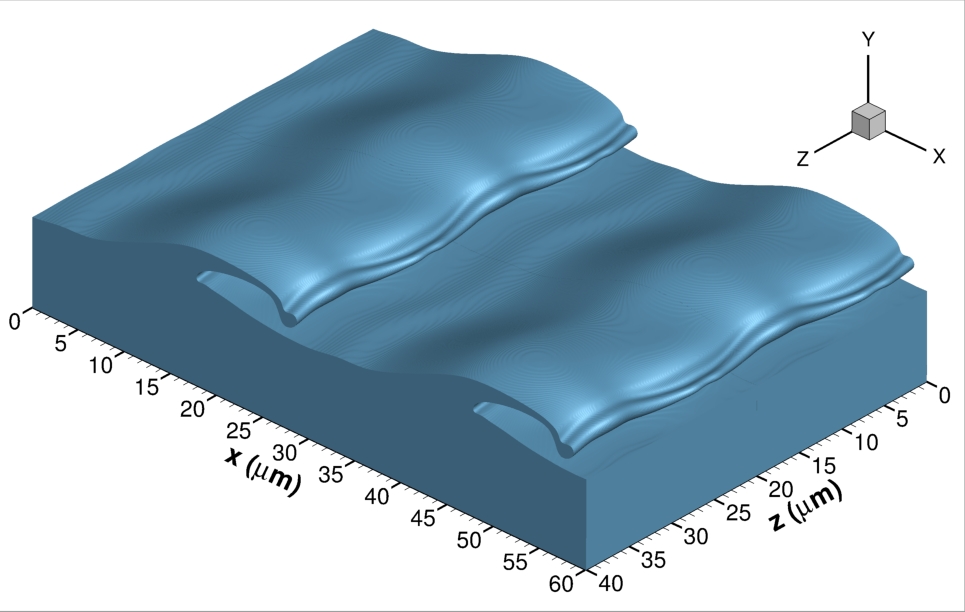}
  \caption{\label{subfig:150_A1_int_3mus}Initial 0.1 \(\mu\)m spanwise amplitude at \(t=3\) \(\mu\)s}
\end{subfigure}%
\begin{subfigure}{.5\textwidth}
  \centering
  \includegraphics[width=1.0\linewidth]{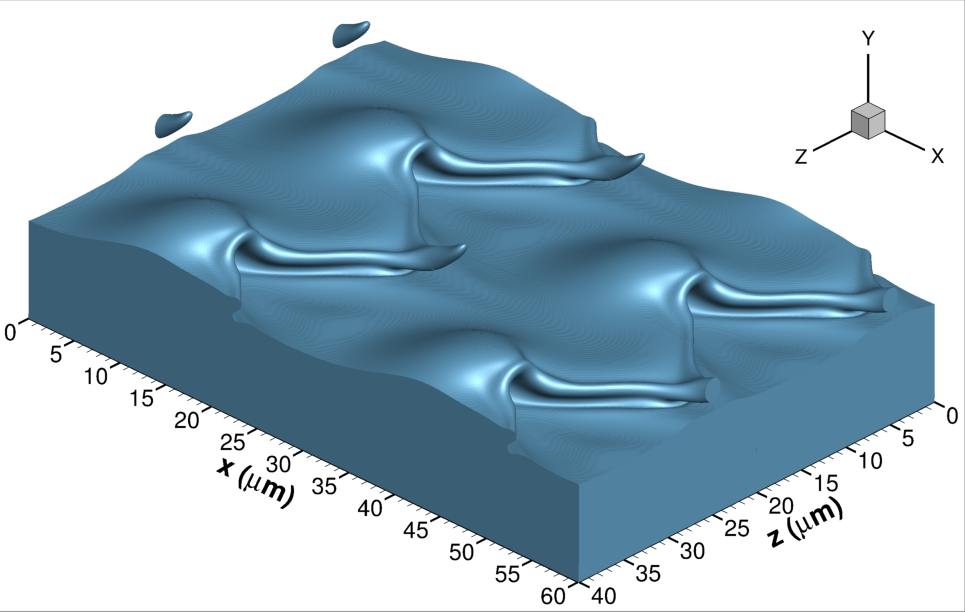}
  \caption{\label{subfig:150_A2_int_3mus}Initial 0.5 \(\mu\)m spanwise amplitude at \(t=3\) \(\mu\)s}
\end{subfigure}%
\caption{\label{fig:3djet_comp}Interface deformation for the three-dimensional planar jet at 150 bar. Two different initial configurations are compared with initial spanwise perturbation amplitude of 0.1 \(\mu\)m and 0.5 \(\mu\)m. The interface location is identified as the iso-surface with \(C=0.5\).}
\end{figure*}

\begin{figure*}[h!]
\centering
\begin{subfigure}{.5\textwidth}
  \centering
  \includegraphics[width=1.0\linewidth]{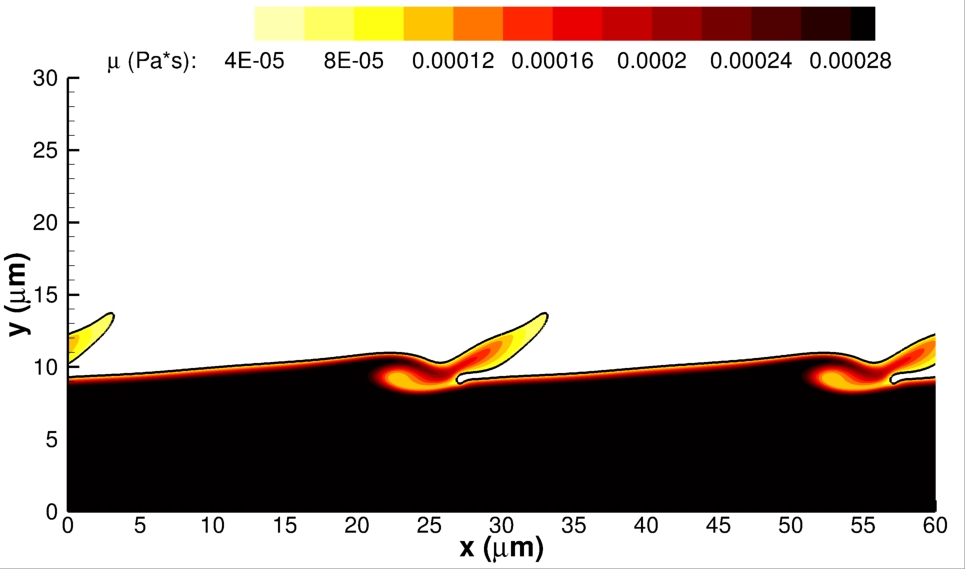}
  \caption{\label{subfig:150_A2_VIS_3mus_Z5mum}\(\mu\) at \(z=5\) \(\mu\)m}
\end{subfigure}%
\begin{subfigure}{.5\textwidth}
  \centering
  \includegraphics[width=1.0\linewidth]{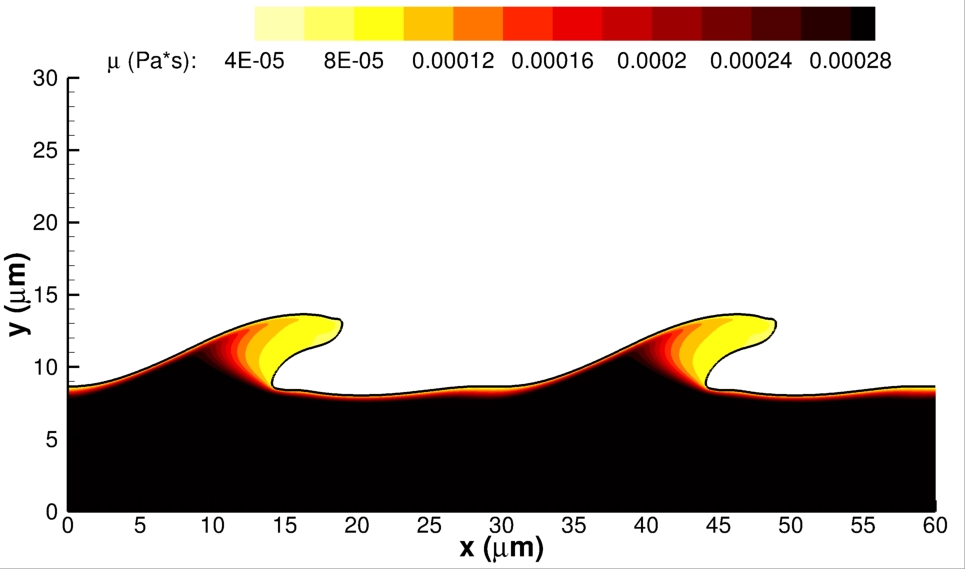}
  \caption{\label{subfig:150_A2_VIS_3mus_Z15mum}\(\mu\) at \(z=15\) \(\mu\)m}
\end{subfigure}%
\caption{\label{fig:3djet_VIS_A2}Viscosity plots in the liquid phase for the three-dimensional planar jet at 150 bar with initial spanwise perturbation amplitude of 0.5 \(\mu\)m. Different spanwise locations are shown at \(t=3\) \(\mu\)s. At this high pressure, the viscosity of the gas mixture remains within the range of 2.8-3.4x10\(^{-5}\) Pa*s. The interface location is highlighted with a solid black curve representing the isocontour with \(C=0.5\).}
\end{figure*}

\begin{figure*}[h!]
\centering
\begin{subfigure}{.5\textwidth}
  \centering
  \includegraphics[width=1.0\linewidth]{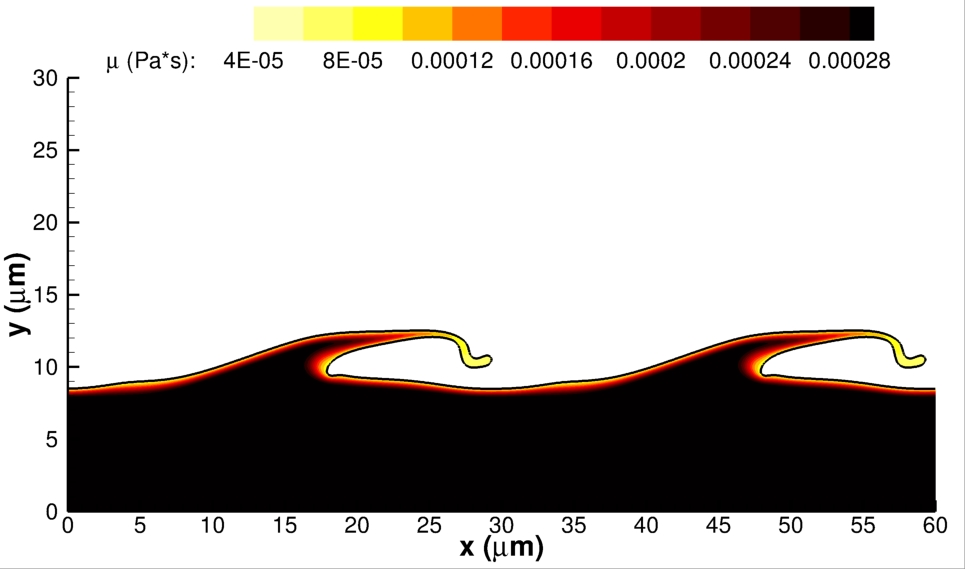}
  \caption{\label{subfig:150_A1_VIS_3mus_Z5mum}\(\mu\) at \(z=5\) \(\mu\)m}
\end{subfigure}%
\begin{subfigure}{.5\textwidth}
  \centering
  \includegraphics[width=1.0\linewidth]{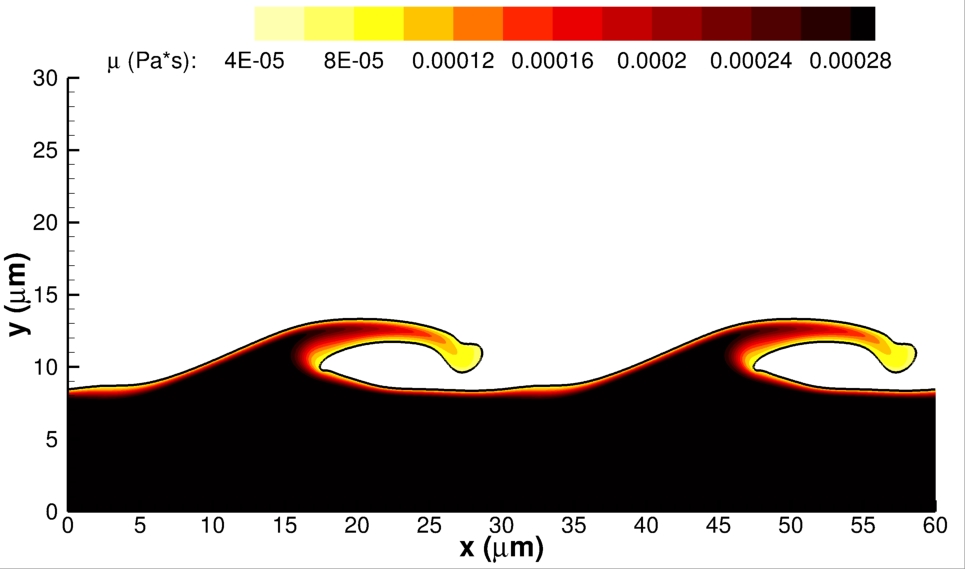}
  \caption{\label{subfig:150_A1_VIS_3mus_Z15mum}\(\mu\) at \(z=15\) \(\mu\)m}
\end{subfigure}%
\caption{\label{fig:3djet_VIS_A1}Viscosity plots in the liquid phase for the three-dimensional planar jet at 150 bar with initial spanwise perturbation amplitude of 0.1 \(\mu\)m. Different spanwise locations are shown at \(t=3\) \(\mu\)s. At this high pressure, the viscosity of the gas mixture remains within the range of 2.8-3.4x10\(^{-5}\) Pa*s. The interface location is highlighted with a solid black curve representing the isocontour with \(C=0.5\).}
\end{figure*}

\begin{figure*}[h!]
\centering
\begin{subfigure}{.5\textwidth}
  \centering
  \includegraphics[trim=0 0 0 140,clip,width=1.0\linewidth]{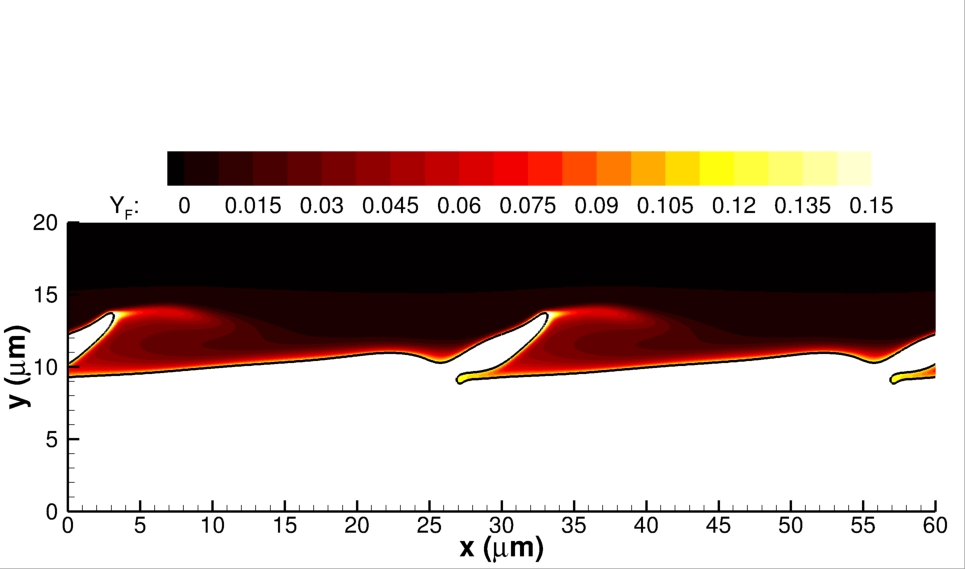}
  \caption{\label{subfig:150_A2_YFg_3mus_Z5mum}\(Y_F\) at \(z=5\) \(\mu\)m}
\end{subfigure}%
\begin{subfigure}{.5\textwidth}
  \centering
  \includegraphics[trim=0 0 0 140,clip,width=1.0\linewidth]{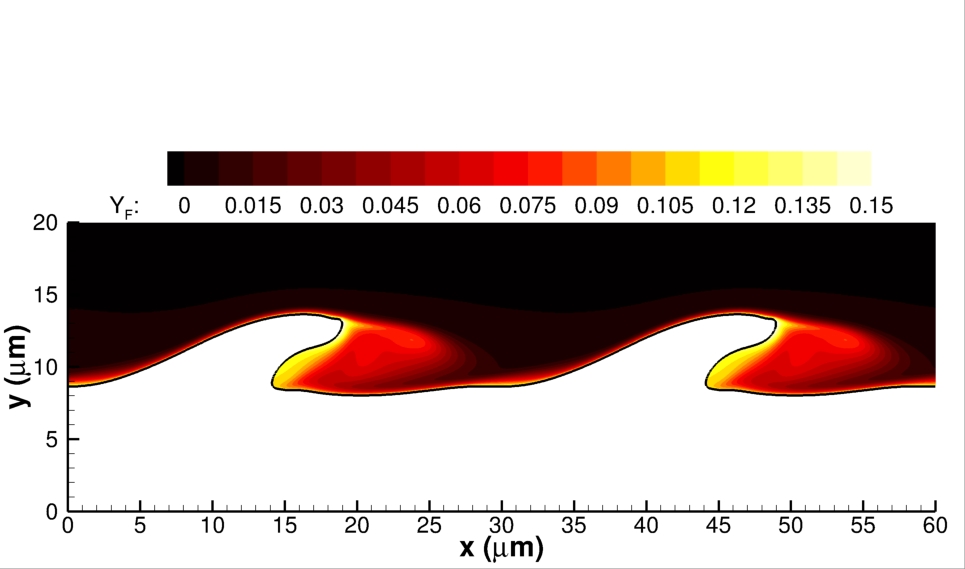}
  \caption{\label{subfig:150_A2_YFg_3mus_Z15mum}\(Y_F\) at \(z=15\) \(\mu\)m}
\end{subfigure}%
\caption{\label{fig:3djet_YFg_A2}\textit{n}-decane mass fraction plots in the gas phase for the three-dimensional planar jet at 150 bar with initial spanwise perturbation amplitude of 0.5 \(\mu\)m. Different spanwise locations are shown at \(t=3\) \(\mu\)s. The interface location is highlighted with a solid black curve representing the isocontour with \(C=0.5\).}
\end{figure*}

\begin{figure*}[h!]
\centering
\begin{subfigure}{.5\textwidth}
  \centering
  \includegraphics[trim=0 0 0 140,clip,width=1.0\linewidth]{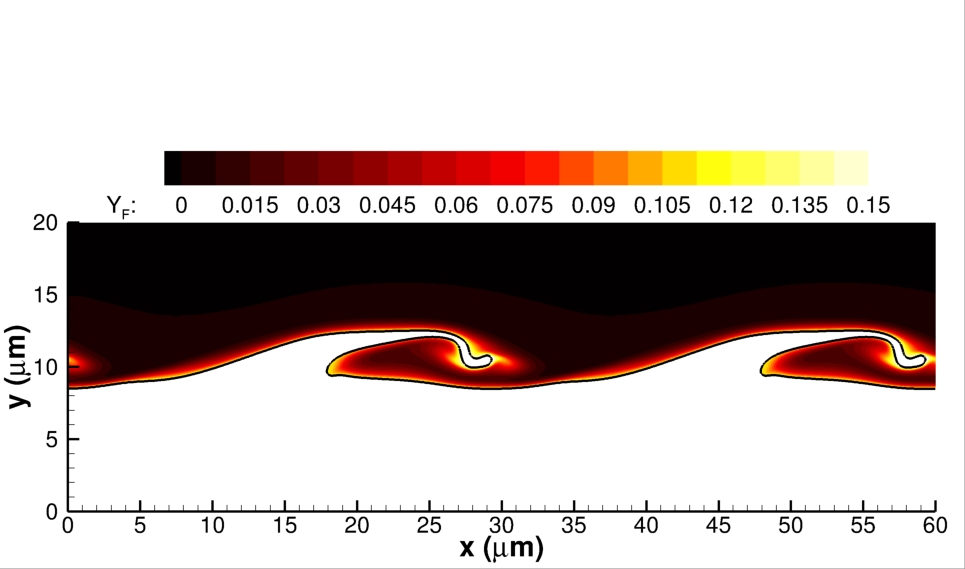}
  \caption{\label{subfig:150_A1_YFg_3mus_Z5mum}\(Y_F\) at \(z=5\) \(\mu\)m}
\end{subfigure}%
\begin{subfigure}{.5\textwidth}
  \centering
  \includegraphics[trim=0 0 0 140,clip,width=1.0\linewidth]{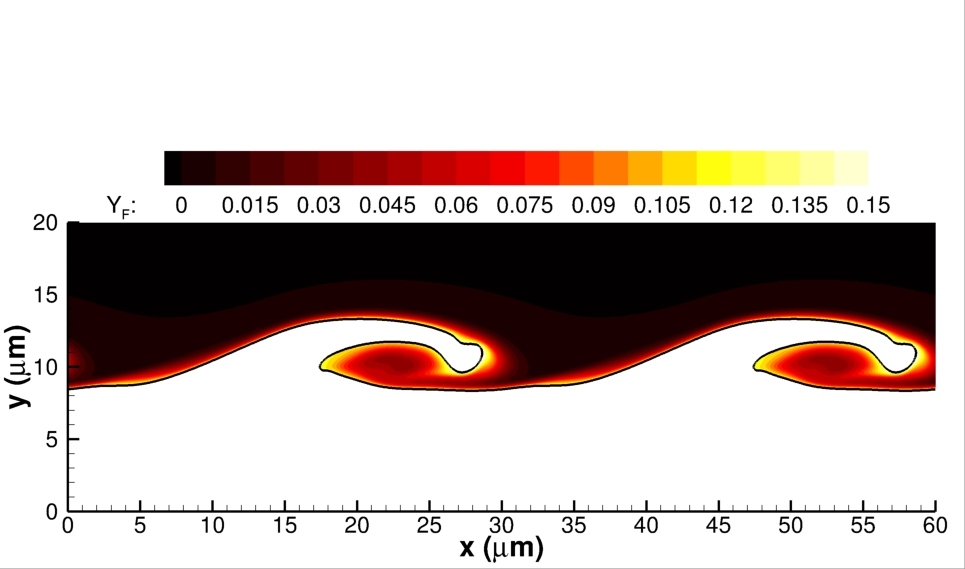}
  \caption{\label{subfig:150_A1_YFg_3mus_Z15mum}\(Y_F\) at \(z=15\) \(\mu\)m}
\end{subfigure}%
\caption{\label{fig:3djet_YFg_A1}\textit{n}-decane mass fraction plots in the gas phase for the three-dimensional planar jet at 150 bar with initial spanwise perturbation amplitude of 0.1 \(\mu\)m. Different spanwise locations are shown at \(t=3\) \(\mu\)s. The interface location is highlighted with a solid black curve representing the isocontour with \(C=0.5\).}
\end{figure*}

\begin{figure*}[h!]
\centering
\begin{subfigure}{.5\textwidth}
  \centering
  \includegraphics[width=1.0\linewidth]{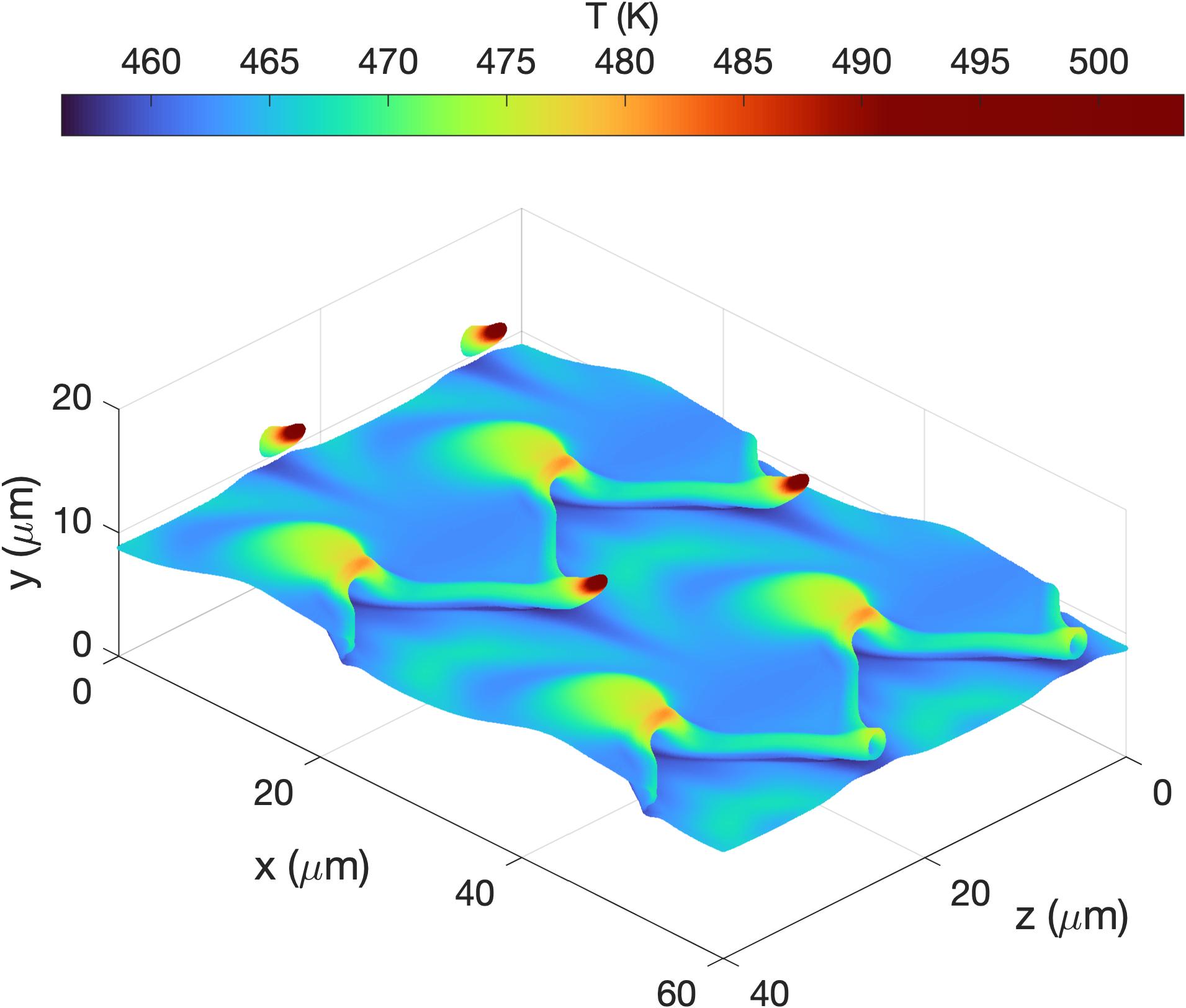}
  \caption{\label{subfig:150_A2_int_T_3mus}\(T\) at \(t=3\) \(\mu\)s}
\end{subfigure}%
\begin{subfigure}{.5\textwidth}
  \centering
  \includegraphics[width=1.0\linewidth]{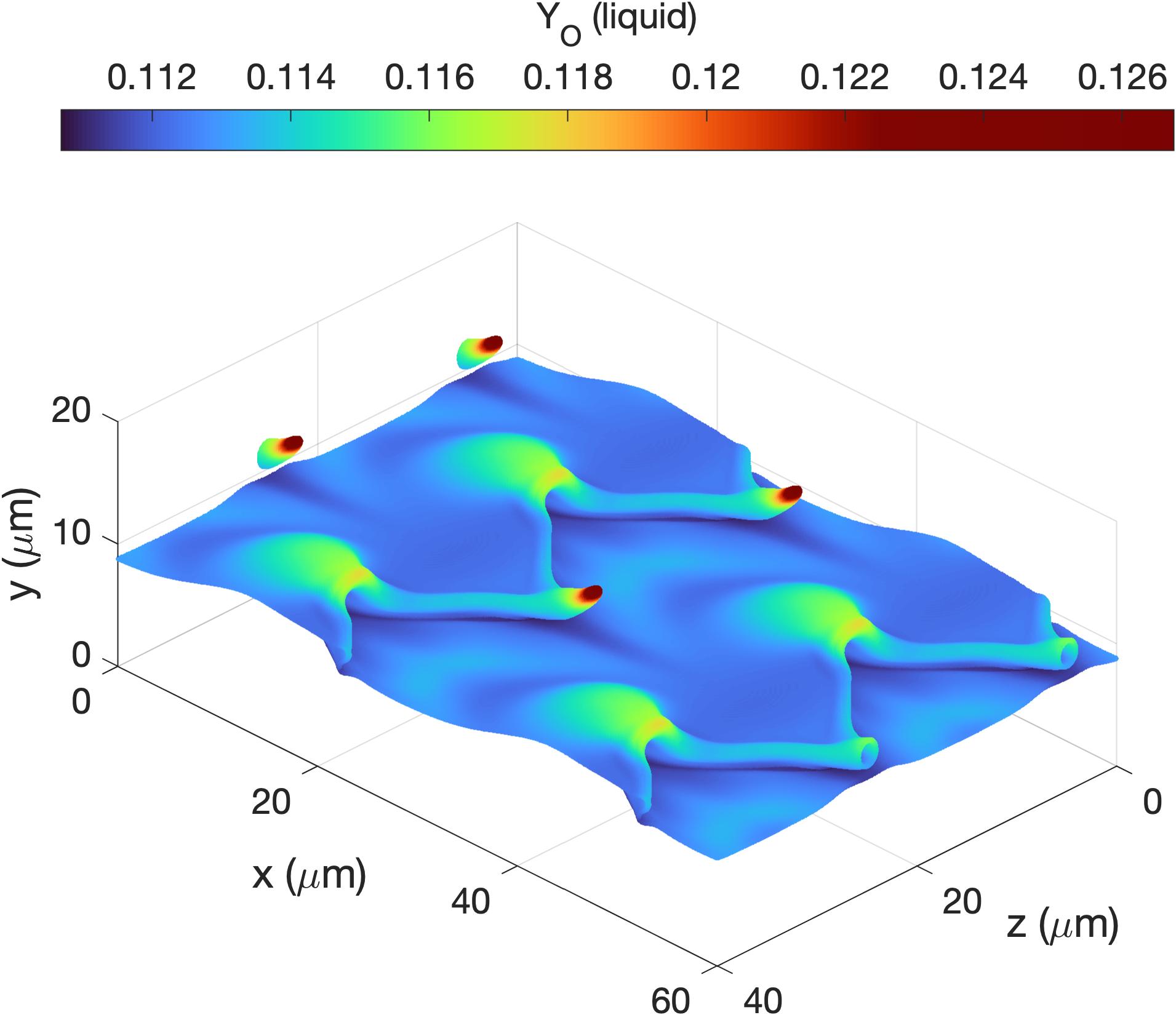}
  \caption{\label{subfig:150_A2_int_YOl_3mus}\(Y_O\) in the liquid at \(t=3\) \(\mu\)s}
\end{subfigure}%
\\
\begin{subfigure}{.5\textwidth}
  \centering
  \includegraphics[width=1.0\linewidth]{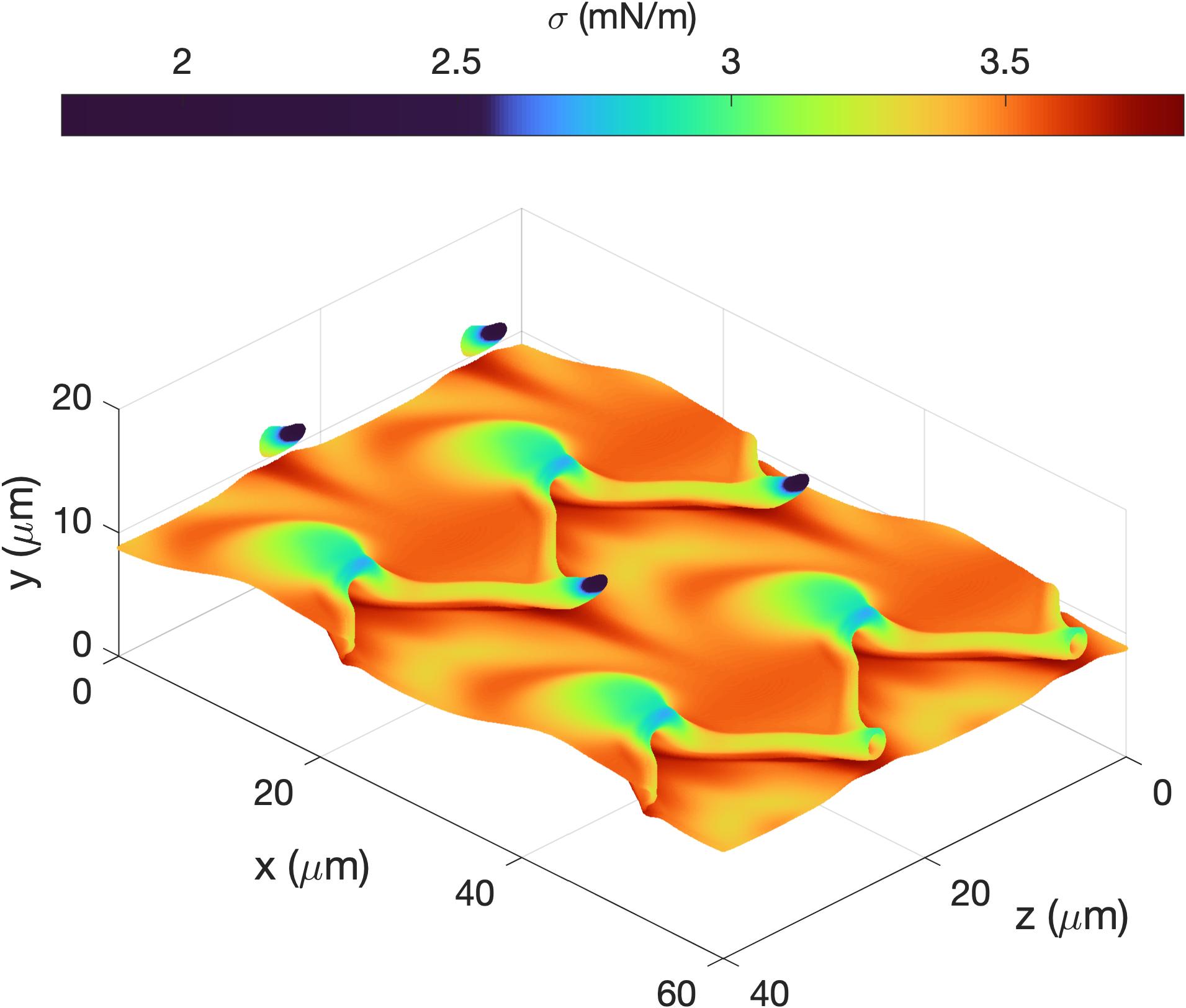}
  \caption{\label{subfig:150_A2_int_sigma_3mus}\(\sigma\) at \(t=3\) \(\mu\)s}
\end{subfigure}%
\begin{subfigure}{.5\textwidth}
  \centering
  \includegraphics[width=1.0\linewidth]{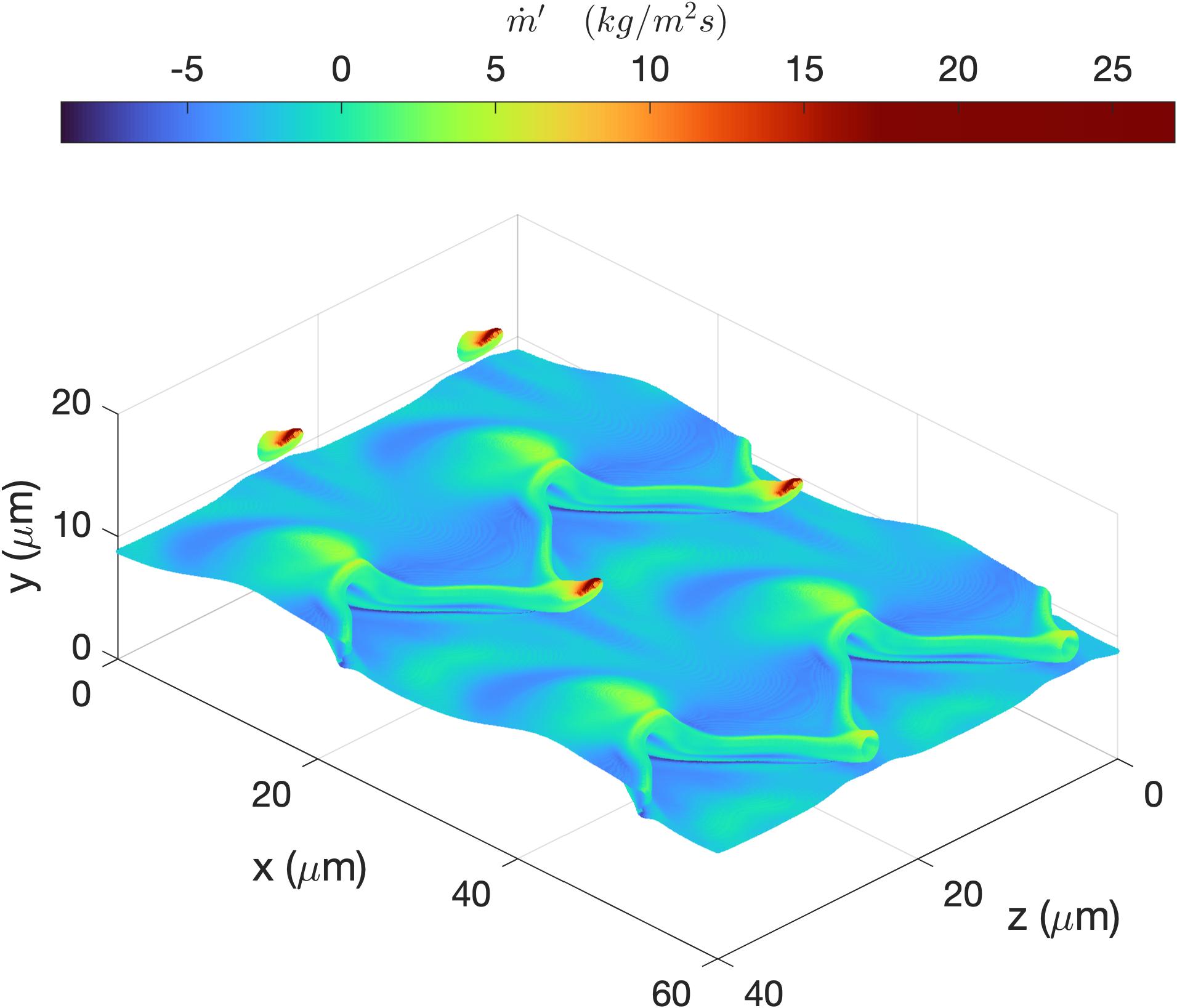}
  \caption{\label{subfig:150_A2_int_mflux_3mus}\(\dot{m}'\) at \(t=3\) \(\mu\)s}
\end{subfigure}%
\caption{\label{fig:3djet_inter}Interface thermodynamic properties for the three-dimensional planar jet at 150 bar and \(t=3\) \(\mu\)s with initial spanwise perturbation amplitude of 0.5 \(\mu\)m. The interface shape is colored by the value of each respective variable. The color gradient in the color map is skewed to represent the variations of interface properties better.}
\end{figure*}

\begin{figure*}[h!]
\centering
\begin{subfigure}{.5\textwidth}
  \centering
  \includegraphics[width=1.0\linewidth]{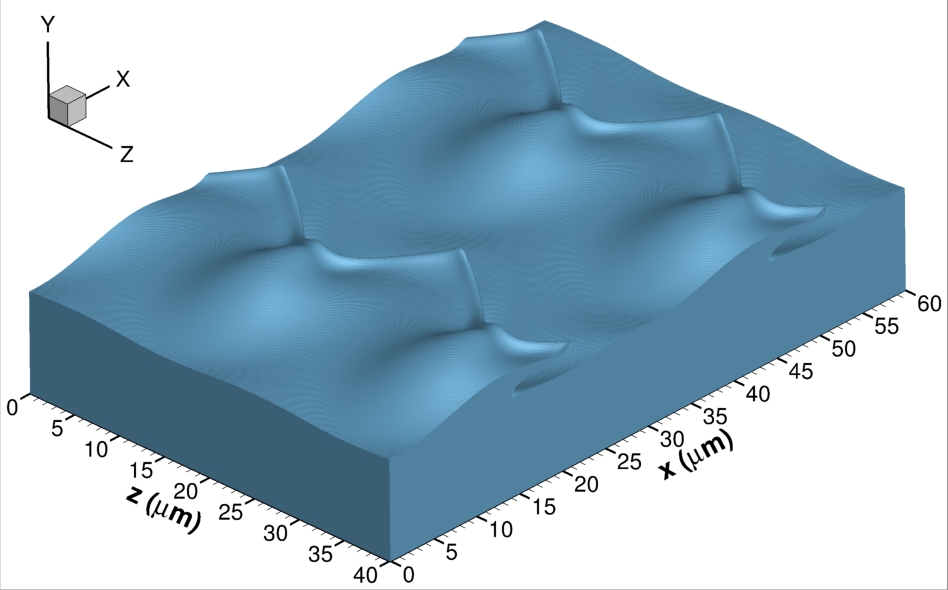}
  \caption{\label{subfig:150_A3_int_2p5mus}\(t=2.5\) \(\mu\)s}
\end{subfigure}%
\begin{subfigure}{.5\textwidth}
  \centering
  \includegraphics[width=1.0\linewidth]{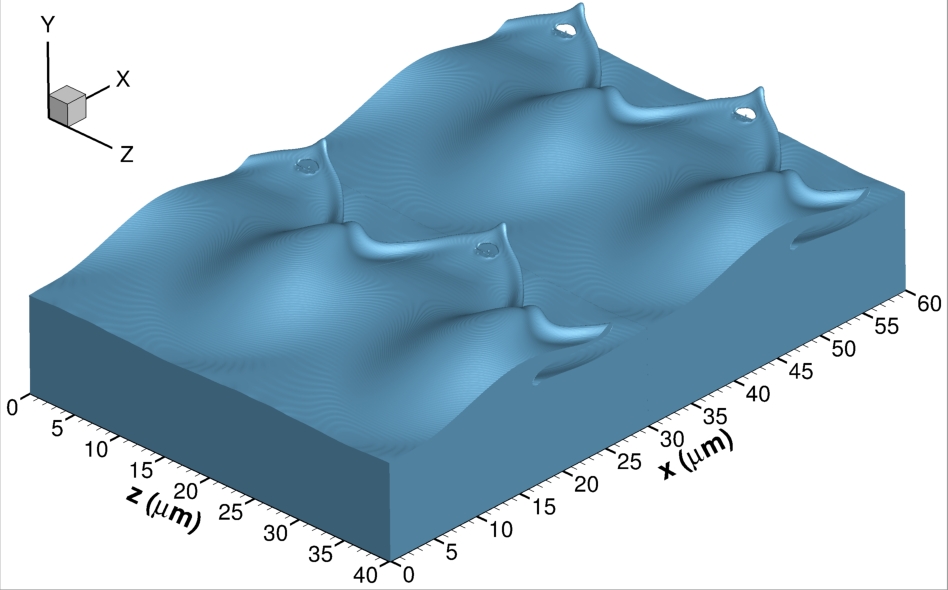}
  \caption{\label{subfig:150_A3_int_2p8mus}\(t=2.8\) \(\mu\)s}
\end{subfigure}%
\\
\begin{subfigure}{.5\textwidth}
  \centering
  \includegraphics[width=1.0\linewidth]{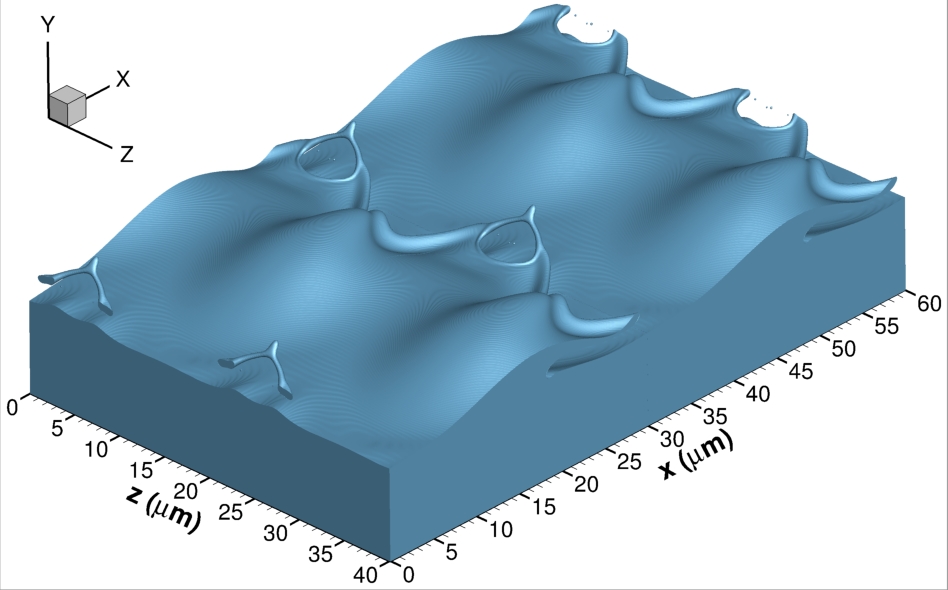}
  \caption{\label{subfig:150_A3_int_3mus}\(t=3\) \(\mu\)s}
\end{subfigure}%
\begin{subfigure}{.5\textwidth}
  \centering
  \includegraphics[width=1.0\linewidth]{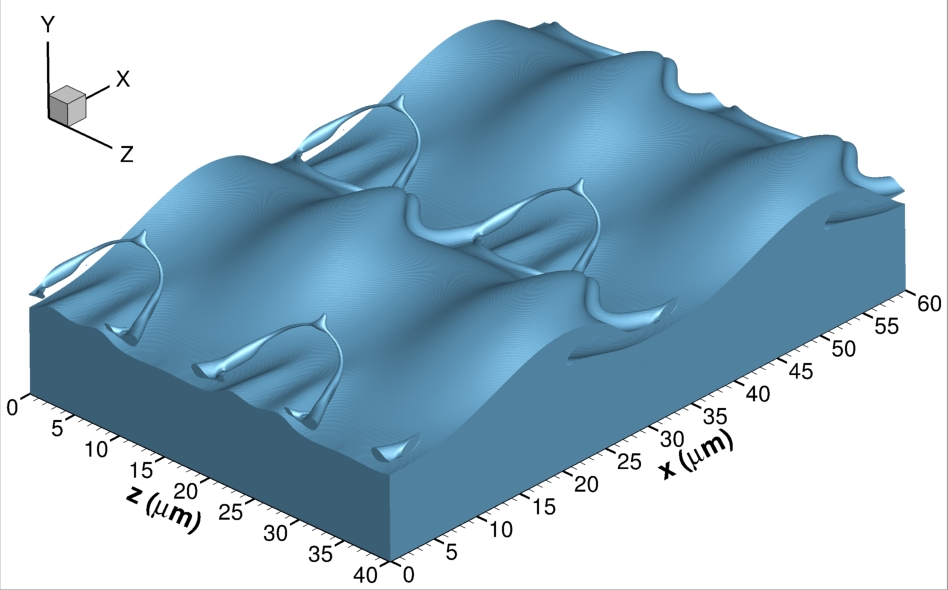}
  \caption{\label{subfig:150_A3_int_3p3mus}\(t=3.3\) \(\mu\)s}
\end{subfigure}%
\caption{\label{fig:3djet_hole}Interface deformation for the three-dimensional planar jet at 150 bar with an initial spanwise perturbation amplitude of 0.3 \(\mu\)m. The interface location is identified as the iso-surface with \(C=0.5\).}
\end{figure*}

The two-dimensional results presented here show the main features and the complexity of liquid injection at supercritical pressures. Three-dimensional results are expected to be more realistic, but at the same time more complex. As mentioned earlier, surface tension forces affect the liquid differently. Although the surface tension coefficient is smaller at these high pressures, a tendency to generate necks or ligament thinning exists, which might yield to a faster formation of droplets. Interface properties vary along the surface in every direction and other regions of enhanced mixing might exist different than those observed in the two-dimensional case. Thus, different liquid deformation patterns may develop. The three-dimensional results are shown from Figure~\ref{fig:3djet_comp} to Figure~\ref{fig:3djet_hole}. Similar to the plots of the two-dimensional results, the domain is extended using the periodic behavior in both the streamwise and spanwise directions for a better visualization of the interface deformation. The analyzed time duration is limited by computational costs. \par

Differences exist in the liquid deformation patterns when changing the initial amplitude of the perturbation in the spanwise direction (see Figure~\ref{fig:3djet_comp}). Three-dimensional effects are enhanced when both the streamwise and spanwise perturbations have an initial amplitude of 0.5 \(\mu\)m. Lobes extend on the liquid surface with a later formation of ligaments stretching into the hotter gas. At the same time as lobes are extending, another perturbation grows in the region connecting two consecutive lobes, capturing gas underneath it in a similar pattern as the two-dimensional case or the three-dimensional case with a smaller initial spanwise amplitude. Figure \ref{fig:3djet_VIS_A2} shows the \(xy\) planes cutting the liquid in two different spanwise locations, highlighting the interface shape and the liquid viscosity. At \(z=5\) \(\mu\)m, the lobe and ligament stretching are featured, whereas the perturbation growing between lobes is shown at \(z=15\) \(\mu\)m. \par 

For smaller spanwise amplitudes (i.e., 0.1 \(\mu\)m), three-dimensional effects take longer to become dominant. Figure~\ref{subfig:150_A1_int_3mus} shows a nearly two-dimensional lobe extending in the spanwise direction and starting to roll on itself, reminiscent of the two-dimensional behavior seen at 3 \(\mu\)s. Again, a vortex located in front of the wave is responsible for pushing upwards the liquid near the tip. However, the overall structure changes for different spanwise locations. Figure~\ref{fig:3djet_VIS_A1} shows the interface shape and viscosity plots for different \(xy\) planes. At \(z=5\) \(\mu\)m, the liquid lobe presents the thinnest cross section with a strongly perturbed lobe's tip forming an L-shape. The progressive deformation and thinning of the lobe in this region suggests the future formation of a hole later in time. On the other hand, a thicker cross section exists at \(z=15\) \(\mu\)m with a rounder lobe's tip. Mixing in the liquid phase is very similar to the two-dimensional case. Nevertheless, its implications in the three-dimensional evolution of the liquid surface still need to be analyzed. \par 

As discussed with the two-dimensional results, proper mixing of the fuel with the oxidizer stream is critical in combustion applications. Figures~\ref{fig:3djet_YFg_A2} and~\ref{fig:3djet_YFg_A1} present the fuel mass fraction in the gas phase for both three-dimensional configurations at \(t=3\) \(\mu\)s and at different spanwise locations. Although results do not extend further in time, small discrete regions of high fuel concentration can already be identified. Later in time it is expected that continuous liquid deformation and breakup into droplets will enhance the fuel mixing with the oxidizer stream. \par 

The variation of fluid properties along the interface is shown in Figure~\ref{fig:3djet_inter}, where results for the case with initial spanwise amplitude of 0.5 \(\mu\)m at 3 \(\mu\)s are shown. The complexity of the solution of the jump conditions and LTE at the interface increases, yet it still follows similar patterns shown in the two-dimensional results. Higher interface temperatures are obtained in regions where the liquid surface has extended towards the hotter gas. This higher interface temperature results in enhanced dissolution of oxygen into the liquid phase, reducing the local liquid density and surface tension coefficient. Moreover, localized net vaporization occurs around these regions at 150 bar. As the ligament stretching from the lobe's tip penetrates the gas phase, the interface temperature increases substantially compared to the rest of the liquid surface. This behavior exaggerates the effects previously discussed (e.g., enhanced dissolution of the gas species), which cause the local liquid density and viscosity to be even more similar to the gas properties. Thus, the ligament may be more easily affected by the gas flowing around it. Regarding the mass exchange at the interface, some regions near the ligament's tip show stronger vaporization occurring compared to the rest of the liquid surface. \par

These three-dimensional results point to the formation of ligaments or finger-like structures extending from the liquid core. The formation of these structures can be related to the presence of hairpin vortices. These vortices develop in high-Reynolds-number shear layers and their importance to two-phase flows was shown by Jarrahbashi et al.~\cite{jarrahbashi2014vorticity,jarrahbashi2016early} and Zandian et al.~\cite{zandian2017planar,zandian2018understanding,zandian2019length,zandian2019vorticity}. More recently, Lagarza-Cort\'{e}s et al.~\cite{lagarza2019large} have shown that hairpin vortices and the formation of finger-like ligaments are also important for transcritical jets. However, no phase interface is considered. The similar large-scale deformation with or without phase interface is expected, as the surface tension force characteristic of two-phase flows is very weak at these high pressures. Nevertheless, surface tension plays an important role in the small-scale liquid breakup and droplet formation. \par  

Zandian et al.~\cite{zandian2017planar} identified different atomization mechanisms for incompressible two-phase flows classifying the liquid injection environment by using the Weber number based on gas properties (i.e., \(We_g=\rho_g U^2 h /\sigma\)) and the Reynolds number based on liquid properties (i.e., \(Re_l=\rho_l U h/\mu_l\)). The jet velocity, \(U\), and the characteristic length, \(h\), (e.g., jet thickness or diameter) are used to define these non-dimensional numbers. The parametrization of real liquid jets is more complex for the reasons presented in this work: thermodynamics (i.e., pressure and temperature) and mixing influence the fluid properties of each phase, as well as the surface tension coefficient along the interface. Therefore, the breakup mechanisms of real liquid jets might not coincide exactly with those presented in Zandian et al.~\cite{zandian2017planar}. \par 

The analyzed configuration has \(We_g \approx 530\) and \(Re_l \approx 1283\), where \(\sigma = 3.4\) mN/m is chosen as a representative value for surface tension during the early stages and \(h=20\) \(\mu\)m (twice the jet half-thickness). For these low values of \(We_g\) and \(Re_l\), atomization for the incompressible case is driven by lobe formation followed by ligament stretching and eventual breakup into droplets. This mechanism is the pattern for the real-fluid case with an initial spanwise amplitude of 0.5 \(\mu\)m. However, the real-fluid case with initial spanwise amplitude of 0.1 \(\mu\)m hints the possible formation of a hole as previously mentioned, which should happen for higher Weber numbers (\(We_g>4000-5000\)) according to Zandian et al.~\cite{zandian2017planar}. \par 

The formation of holes at these transcritical conditions is corroborated in Figure~\ref{fig:3djet_hole}. A third case with the same initial conditions but initial spanwise perturbation amplitude of 0.3 \(\mu\)m shows the extension of a a very thin lobe on the liquid surface. In this case, a finger-like ligament stretching from the tip of the lobe, like in the case with initial spaniwse perturbation amplitude of 0.5 \(\mu\)m, does not form. Instead, the lobe becomes a thin sheet with reduced density and viscosity which is easily affected by the surrounding gas. As it extends, vortical motion in front of the lobe pushes gas underneath it. As a result, the lobe bends and submerges into the gas phase. Later, the thin sheet is perforated due to the gas flowing above and under it. As the large-scale instability continues to grow, the hole extends and forms a very thin bridge or filament. \par

The hole formation can be influenced by the mesh resolution (e.g., onset of the perforation). This mesh-induced breakup is a problem that exists in classical two-phase problems such as bag breakup in secondary atomization. However, the flow dynamics of the configuration presented in Figure~\ref{fig:3djet_hole} support the physically-induced formation of the hole as explained above. \par 

It is plausible that the main liquid jet dynamics are dominated by the properties of the liquid core and the ambient gas. Nevertheless, fluid properties near the interface differ considerably from the liquid core or the ambient gas. If they were considered as reference values, \(We_g\) and \(Re_l\) would be about 40\% and 320\% larger, respectively. This increase in \(We_g\) does not take into account the reduction of the surface tension coefficient in some areas. If included, the local gas Weber number can be up to 60\% higher than based on the ambient gas. \par 

The variation of fluid properties around the interface alone might not explain the formation of holes in conditions different than those reported by Zandian et al.~\cite{zandian2017planar}. One possible explanation could be the choice of characteristic length in the Weber and Reynolds numbers. Instead of using the jet thickness, the perturbation wavelength might be a better representation. If used, the Weber number of the configuration presented here is much higher and more in line with the cases presenting hole formation in Zandian et al.~\cite{zandian2017planar}. Moreover, changing the spanwise perturbation amplitude has a strong impact on the deformation cascade process. Clearly, a broader range of configurations, including higher Weber number cases, need to be analyzed to understand and classify atomization domains for real liquid jets. \par 

\subsection{Pressure effects and real-fluid considerations}

\begin{figure*}[h!]
\centering
\begin{subfigure}{.25\textwidth}
  \centering
  \includegraphics[width=1.0\linewidth]{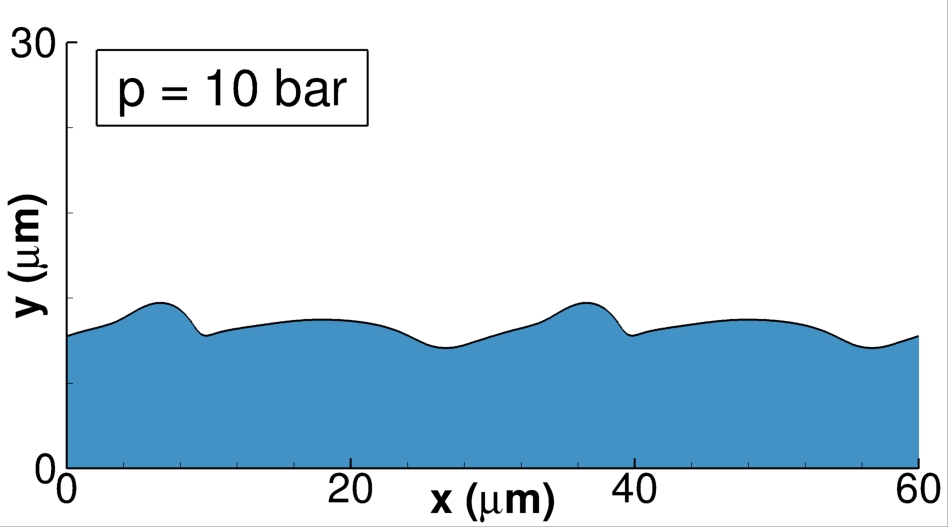}
  \caption{\label{subfig:10_C_2mus}10 bar at \(t=2\) \(\mu\)s}
\end{subfigure}%
\begin{subfigure}{.25\textwidth}
  \centering
  \includegraphics[width=1.0\linewidth]{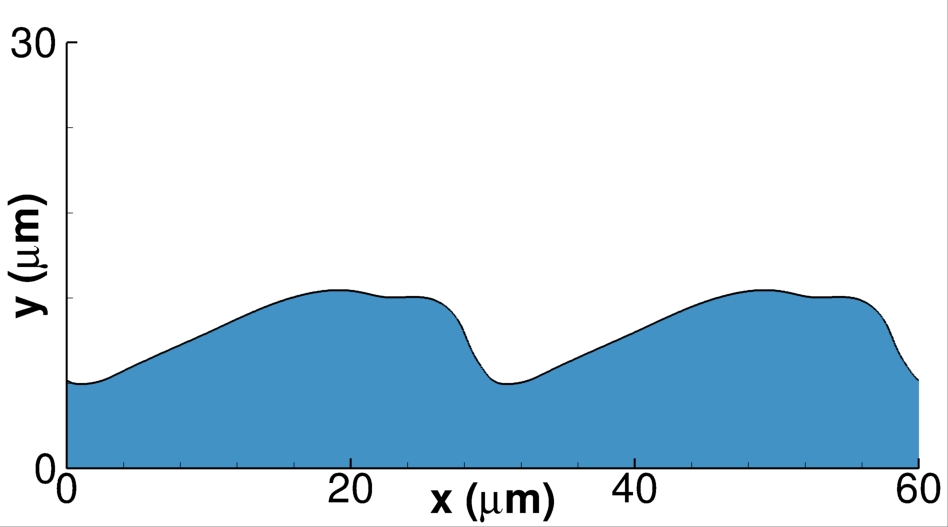}
  \caption{\label{subfig:10_C_4mus}10 bar at \(t=4\) \(\mu\)s}
\end{subfigure}%
\begin{subfigure}{.25\textwidth}
  \centering
  \includegraphics[width=1.0\linewidth]{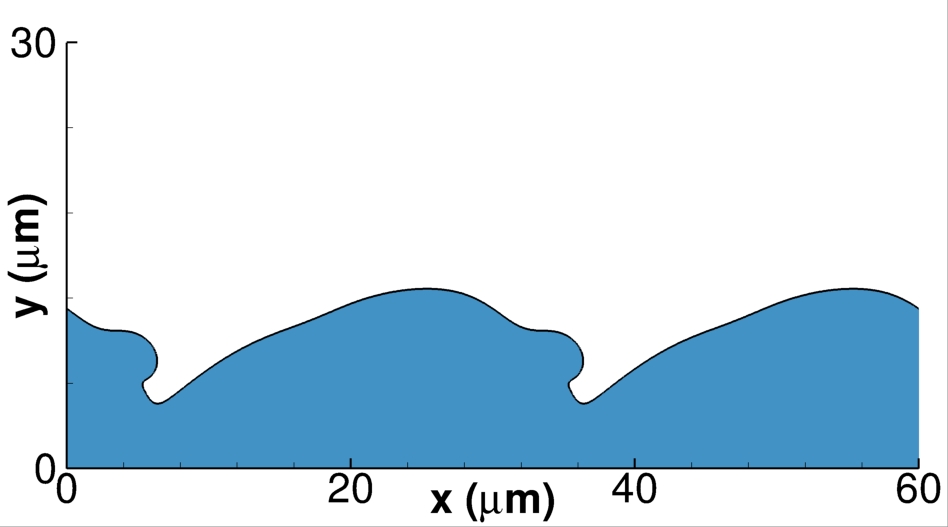}
  \caption{\label{subfig:10_C_6mus}10 bar at \(t=6\) \(\mu\)s}
\end{subfigure}%
\begin{subfigure}{.25\textwidth}
  \centering
  \includegraphics[width=1.0\linewidth]{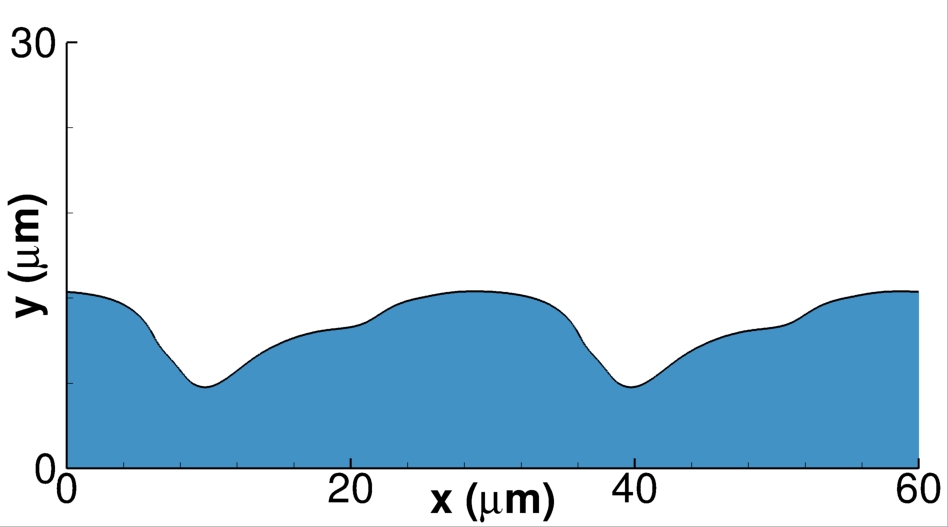}
  \caption{\label{subfig:10_C_8mus}10 bar at \(t=8\) \(\mu\)s}
\end{subfigure}%
\\
\begin{subfigure}{.25\textwidth}
  \centering
  \includegraphics[width=1.0\linewidth]{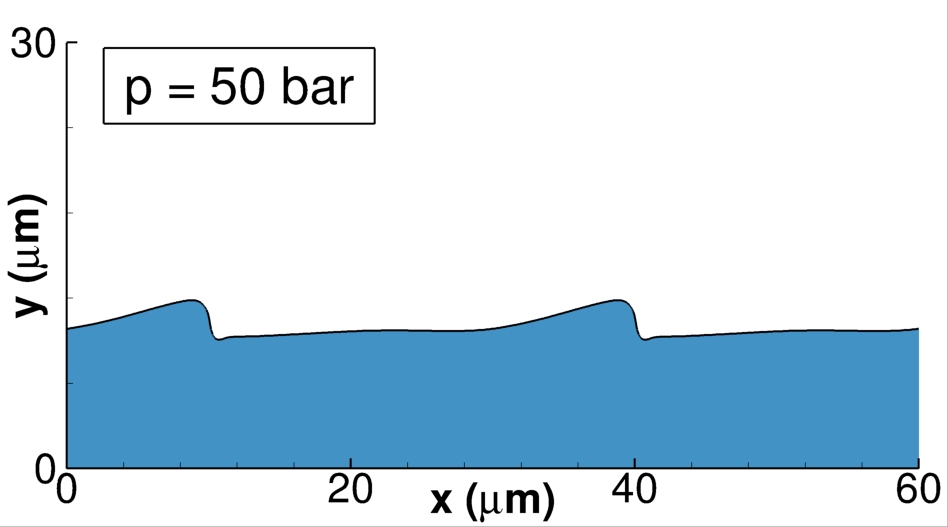}
  \caption{\label{subfig:50_C_2mus}50 bar at \(t=2\) \(\mu\)s}
\end{subfigure}%
\begin{subfigure}{.25\textwidth}
  \centering
  \includegraphics[width=1.0\linewidth]{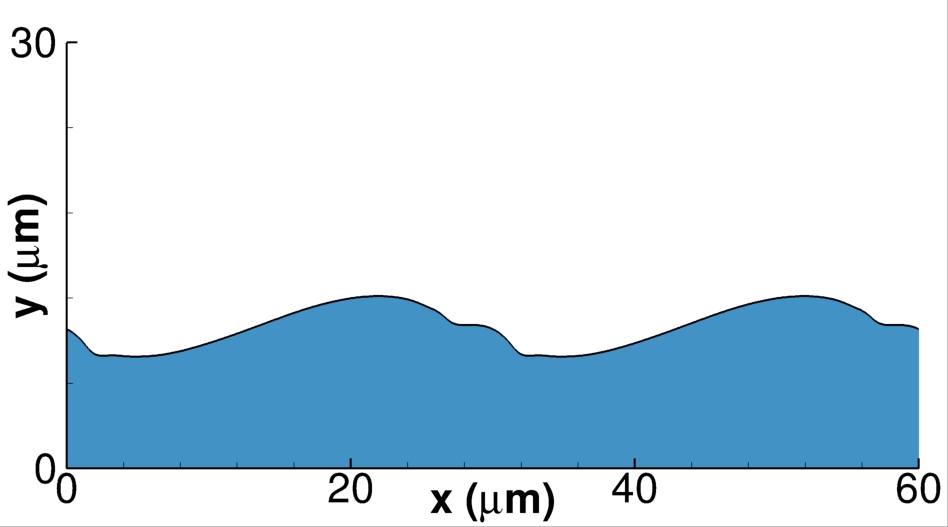}
  \caption{\label{subfig:50_C_4mus}50 bar at \(t=4\) \(\mu\)s}
\end{subfigure}%
\begin{subfigure}{.25\textwidth}
  \centering
  \includegraphics[width=1.0\linewidth]{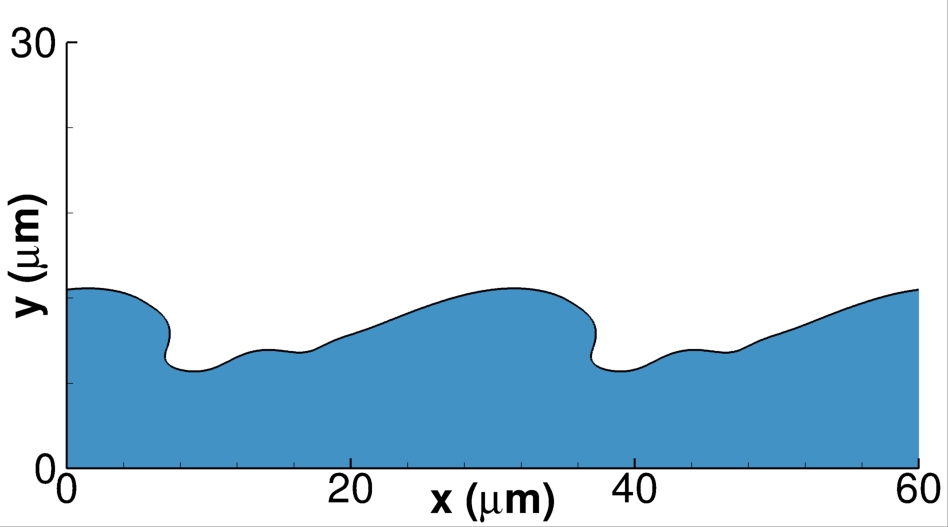}
  \caption{\label{subfig:50_C_6mus}50 bar at \(t=6\) \(\mu\)s}
\end{subfigure}%
\begin{subfigure}{.25\textwidth}
  \centering
  \includegraphics[width=1.0\linewidth]{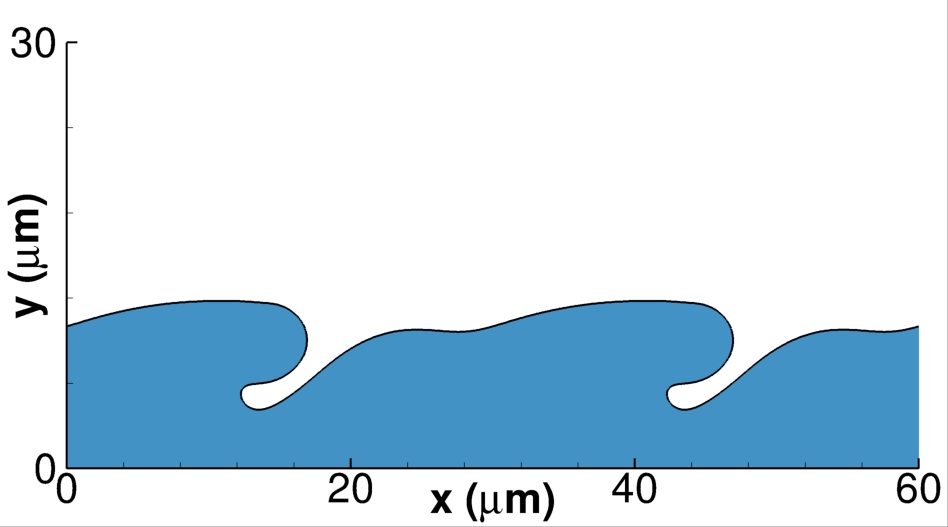}
  \caption{\label{subfig:50_C_8mus}50 bar at \(t=8\) \(\mu\)s}
\end{subfigure}%
\\
\begin{subfigure}{.25\textwidth}
  \centering
  \includegraphics[width=1.0\linewidth]{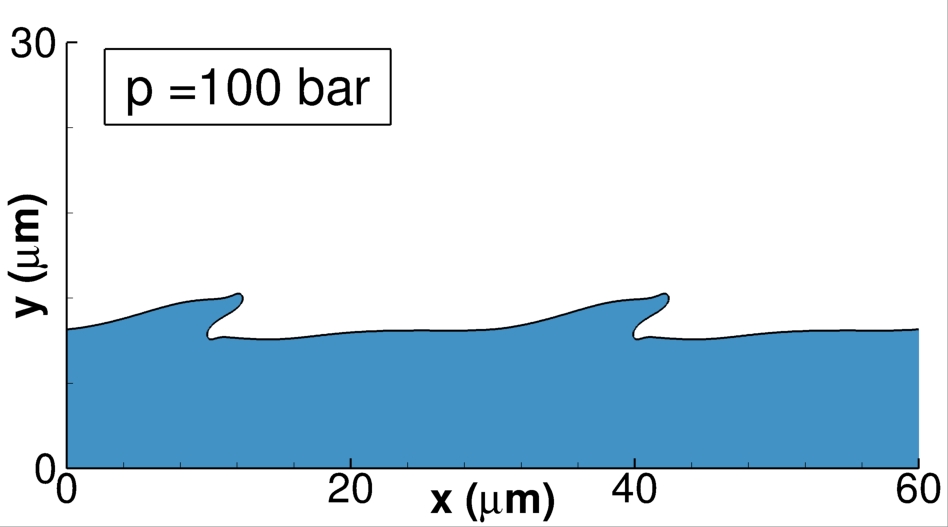}
  \caption{\label{subfig:100_C_2mus}100 bar at \(t=2\) \(\mu\)s}
\end{subfigure}%
\begin{subfigure}{.25\textwidth}
  \centering
  \includegraphics[width=1.0\linewidth]{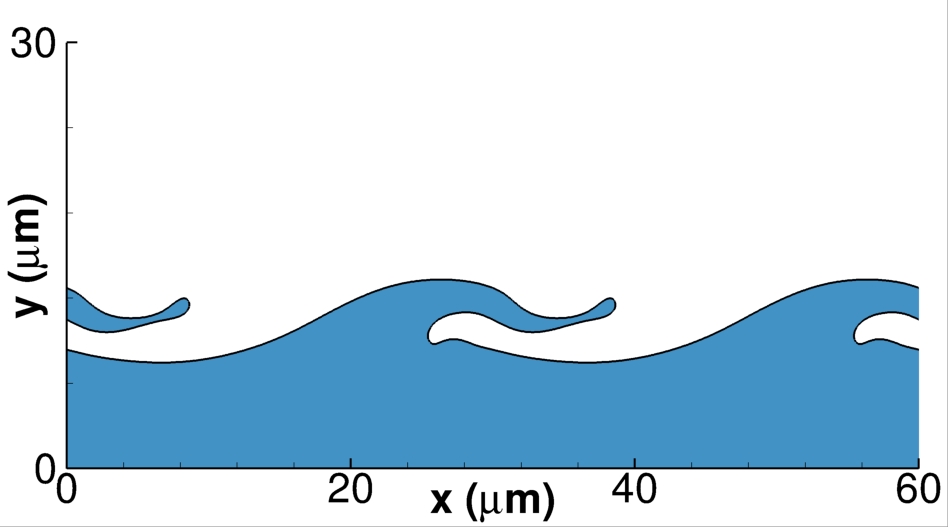}
  \caption{\label{subfig:100_C_4mus}100 bar at \(t=4\) \(\mu\)s}
\end{subfigure}%
\begin{subfigure}{.25\textwidth}
  \centering
  \includegraphics[width=1.0\linewidth]{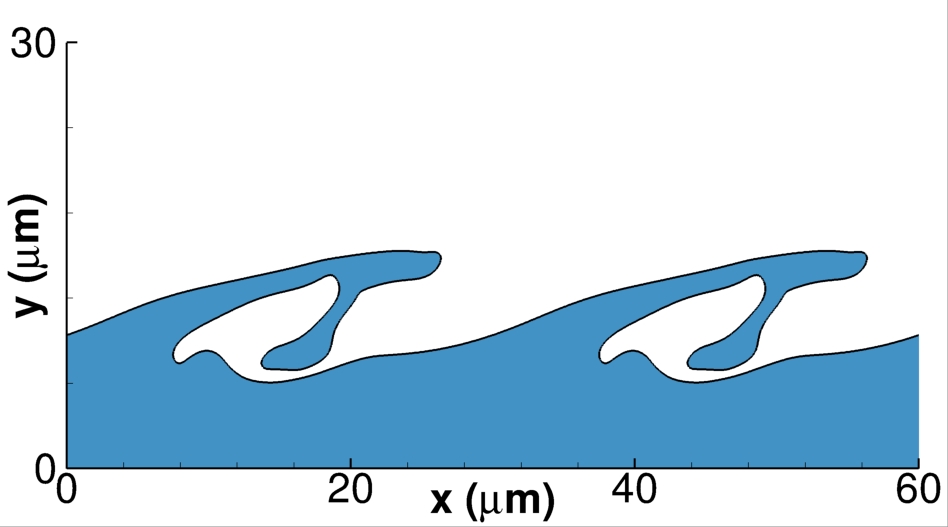}
  \caption{\label{subfig:100_C_6mus}100 bar at \(t=6\) \(\mu\)s}
\end{subfigure}%
\begin{subfigure}{.25\textwidth}
  \centering
  \includegraphics[width=1.0\linewidth]{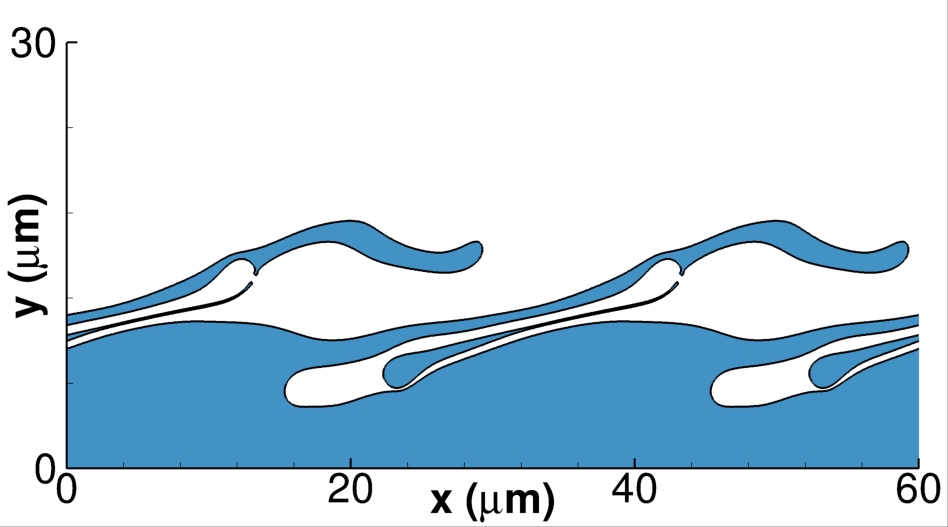}
  \caption{\label{subfig:100_C_8mus}100 bar at \(t=8\) \(\mu\)s}
\end{subfigure}%
\\
\begin{subfigure}{.25\textwidth}
  \centering
  \includegraphics[width=1.0\linewidth]{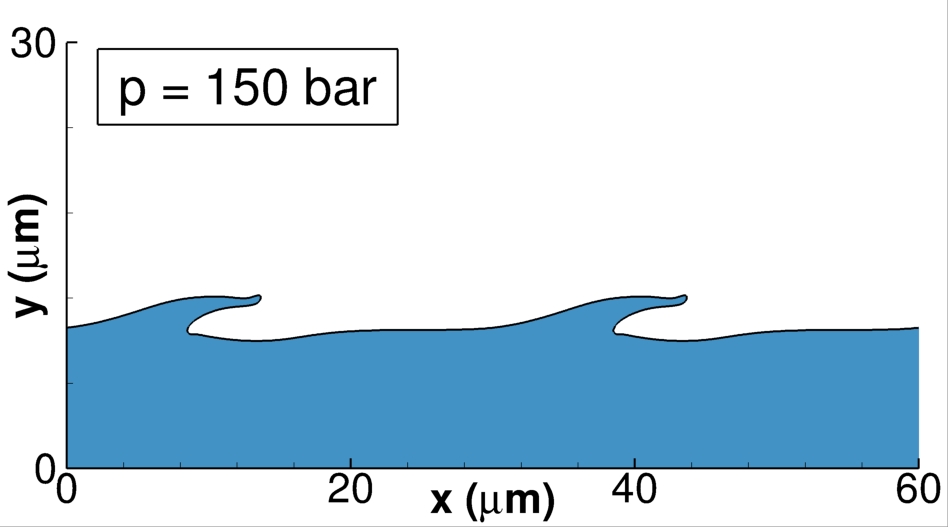}
  \caption{\label{subfig:150_C_2mus}150 bar at \(t=2\) \(\mu\)s}
\end{subfigure}%
\begin{subfigure}{.25\textwidth}
  \centering
  \includegraphics[width=1.0\linewidth]{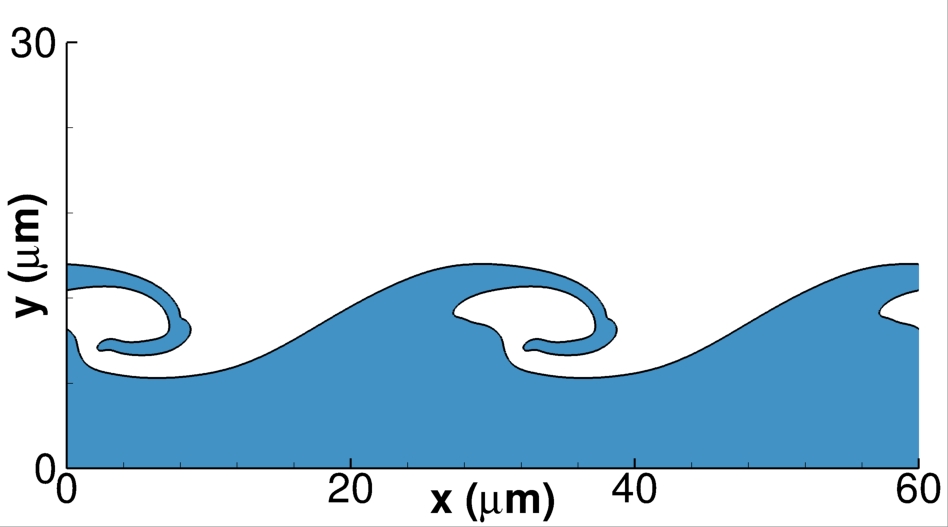}
  \caption{\label{subfig:150_C_4mus}150 bar at \(t=4\) \(\mu\)s}
\end{subfigure}%
\begin{subfigure}{.25\textwidth}
  \centering
  \includegraphics[width=1.0\linewidth]{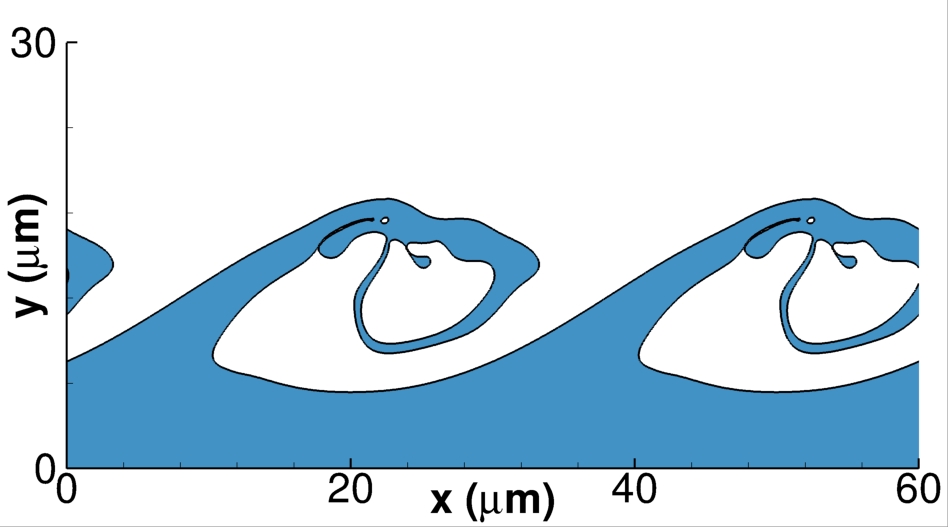}
  \caption{\label{subfig:150_C_6mus}150 bar at \(t=6\) \(\mu\)s}
\end{subfigure}%
\begin{subfigure}{.25\textwidth}
  \centering
  \includegraphics[width=1.0\linewidth]{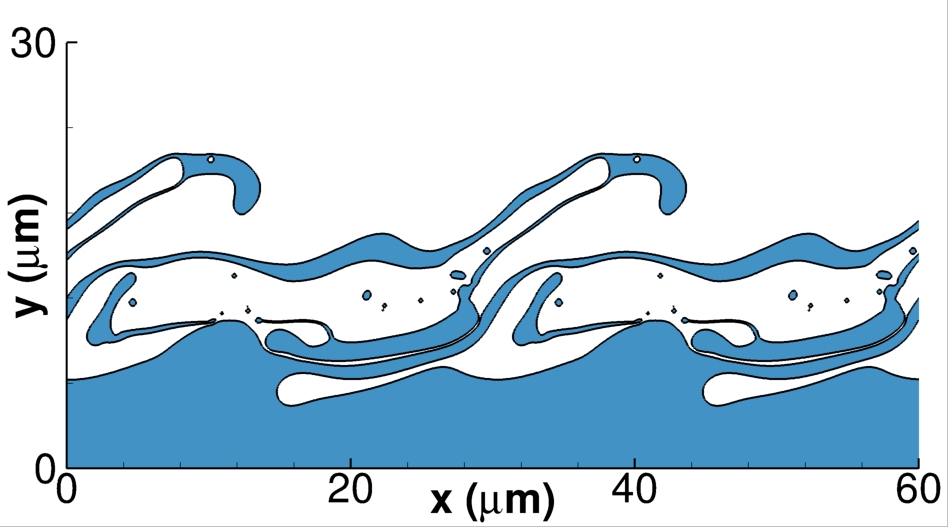}
  \caption{\label{subfig:150_C_8mus}150 bar at \(t=8\) \(\mu\)s}
\end{subfigure}%
\\
\begin{subfigure}{.25\textwidth}
  \centering
  \includegraphics[width=1.0\linewidth]{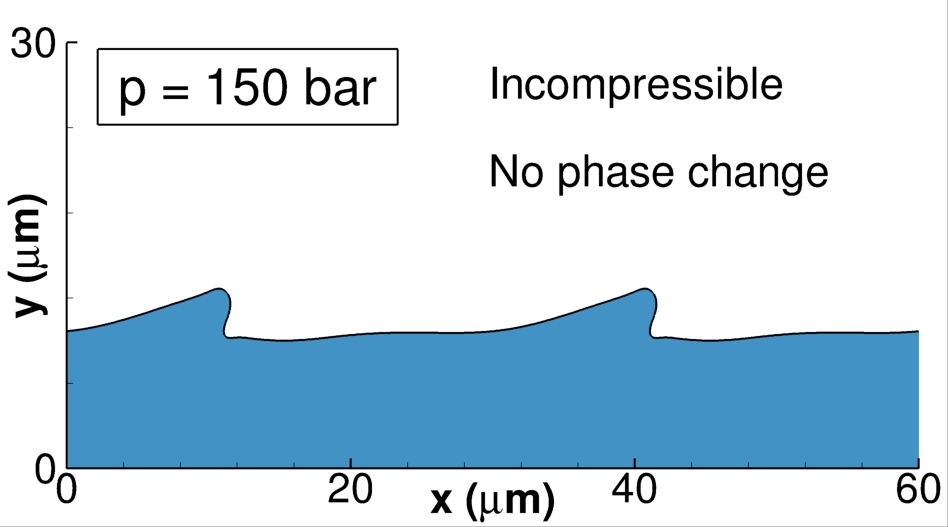}
  \caption{\label{subfig:150_C_2mus_incomp}150 bar at \(t=2\) \(\mu\)s (incompressible)}
\end{subfigure}%
\begin{subfigure}{.25\textwidth}
  \centering
  \includegraphics[width=1.0\linewidth]{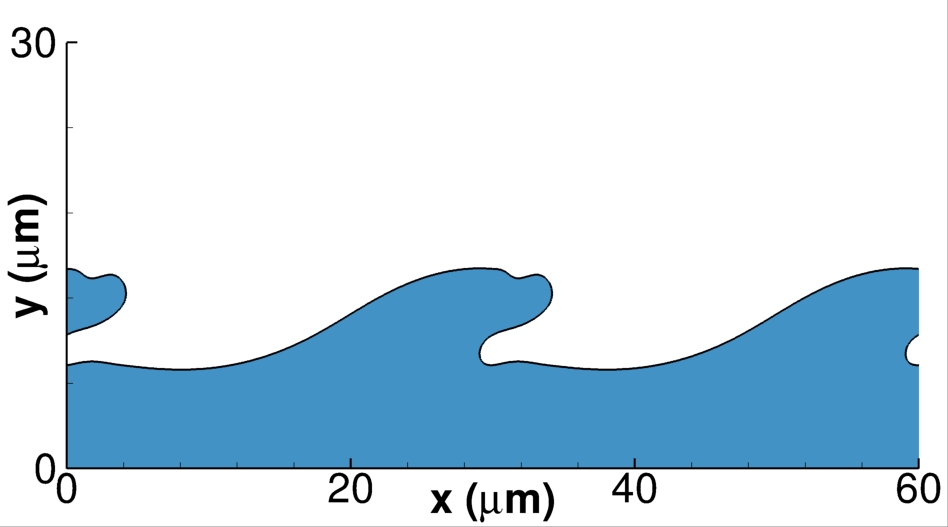}
  \caption{\label{subfig:150_C_4mus_incomp}150 bar at \(t=4\) \(\mu\)s (incompressible)}
\end{subfigure}%
\begin{subfigure}{.25\textwidth}
  \centering
  \includegraphics[width=1.0\linewidth]{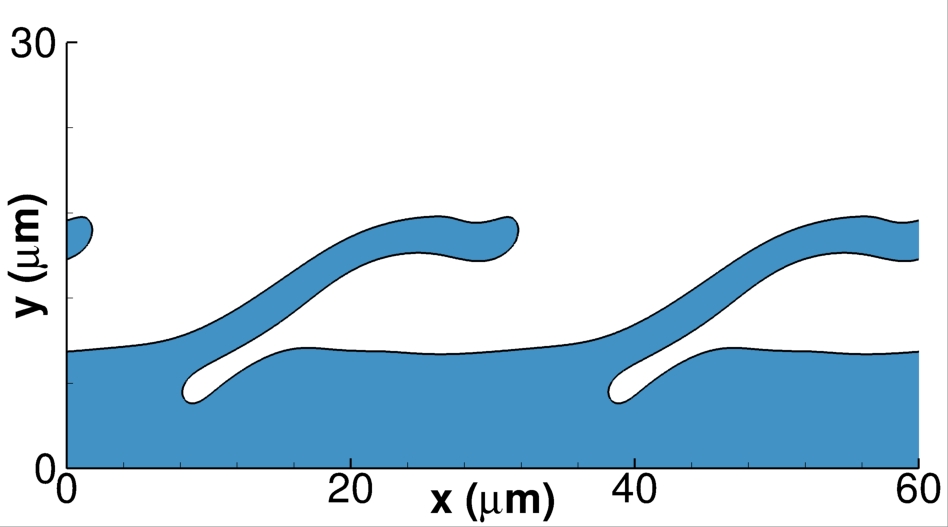}
  \caption{\label{subfig:150_C_6mus_incomp}150 bar at \(t=6\) \(\mu\)s (incompressible)}
\end{subfigure}%
\begin{subfigure}{.25\textwidth}
  \centering
  \includegraphics[width=1.0\linewidth]{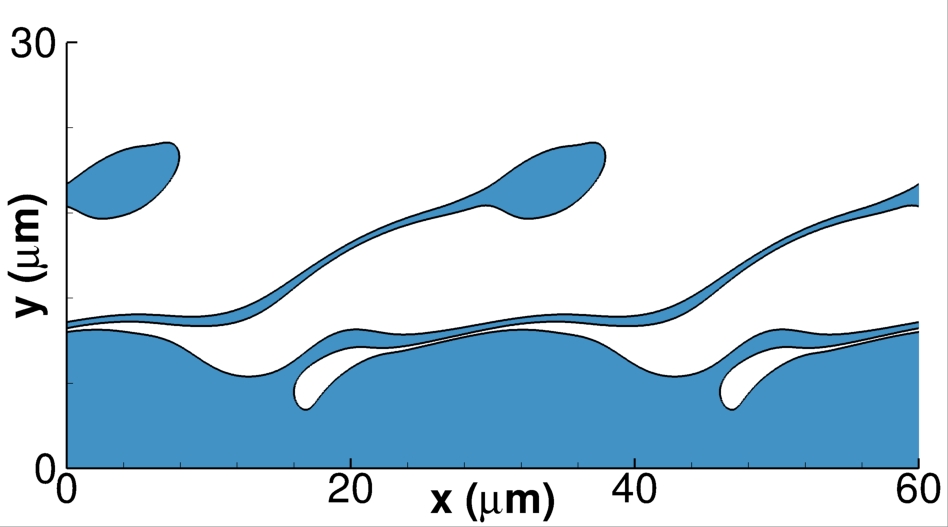}
  \caption{\label{subfig:150_C_8mus_incomp}150 bar at \(t=8\) \(\mu\)s (incompressilbe)}
\end{subfigure}%
\caption{\label{fig:2djet_pressures}Pressure effects on the two-dimensional planar jet deformation using the real-fluid model with phase change. The figures show the liquid phase with the interface location highlighted with a solid black curve representing the isocontour with \(C=0.5\). Subfigures (a) to (d): 10 bar; subfigures (e) to (h): 50 bar; subfigures (i) to (l): 100 bar; subfigures (m) to (p): 150 bar; and subfigures (q) to (t): 150 bar without phase change and incompressible.}
\end{figure*}

Further evidence of the faster growth of the perturbations and the deformation of the liquid surface at high pressures or engine-relevant conditions is provided in Figure~\ref{fig:2djet_pressures}. The effects of increasing pressure on the liquid deformation of the two-dimensional jet are shown using the same binary mixture of oxygen and \textit{n}-decane with the same initial conditions presented in this work. The liquid phase is shown using the volume fraction field. For more details on the range of fluid properties involved at each pressure, the reader is referred to either Davis et al.~\cite{davis2019development} or Poblador-Ibanez et al.~\cite{poblador2021selfsimilar} where the same pressures are analyzed using the same thermodynamic model. \par

Different pressures are shown where the complete numerical model is used (i.e., real-fluid thermodynamic model with phase change). At low pressures (i.e., 10 and 50 bar) the interface deforms under the initial shear strain and the action of higher surface tension forces and momentum transfer by mass exchange (especially at the subcritical pressure of 10 bar). Gas density is rather low (i.e., about 7 kg/m\(^3\) at 10 bar and 34.5 kg/m\(^3\) at 50 bar) while the liquid density is of the order of 600 kg/m\(^3\). Moreover, the variations of fluid properties across mixing regions are negligible at 10 bar and weak at 50 bar. Therefore, fluid properties remain almost constant. At these lower pressures, short-wavelength instabilities fail to grow in the time scales analyzed in this work. On the other hand, higher pressures (i.e., 100 and 150 bar) present reduced surface tension forces, higher gas density and a stronger variation of fluid properties across mixing layers as mentioned earlier. This situation results in substantial deformation of the interface. \par

The observed pressure effects could be argued as an increased importance of inertial terms on the two-phase dynamics as gas density increases and the surface tension coefficient drops. However, Weber and Reynolds numbers are still low in order to justify such a chaotic deformation at 150 bar (\(We_g \approx 530\) and \(Re_l \approx 1283\)). Subfigures~\ref{subfig:150_C_2mus_incomp} to~\ref{subfig:150_C_8mus_incomp} highlight the importance of the interface thermodynamic model (i.e., LTE) and the variation of fluid properties in the mixing regions. These subfigures correspond to the same 150-bar configuration, but without phase change nor mixing. Each phase is treated as incompressible and the gas phase remains pure oxygen and the liquid phase remains pure \textit{n}-decane. No LTE is imposed at the interface and only the momentum equation becomes relevant. Without mixing, the surface tension coefficient is higher and the formation of elongated ligaments is delayed and reduced. The higher gas inertia at 150 bar deforms the liquid surface, but not to the same extent as the complete model. That is, the liquid phase remains denser and more viscous than with the presence of oxygen in the liquid mixture. Therefore, liquid injection at high pressures is dominated not only by the increase in gas density and the reduction of surface tension at higher pressures, but also by the enhanced mixing in both phases, which affects drastically the fluid properties and the interface thermodynamic state. \par 

\subsection{Vorticity dynamics}

A clearer picture of deformation patterns can be obtained by using the dynamical vortex identification method \(\lambda_\rho\)~\cite{yao2018toward}. In a compressible flow, vortices are identified by finding local pressure minimums in a plane of a modified pressure Hessian tensor. That is, a vortex is identified as a connected region with two positive eigenvalues of the pressure Hessian. Here, the dynamical terms defining the negative modified pressure Hessian tensor are analyzed; thus, connected regions with two negative eigenvalues define a vortex. This requirement translates to identifying regions with a negative second eigenvalue \(\lambda_\rho\) (or \(\lambda_2\) in incompressible flows~\cite{jeong1995identification}). Terms related to unsteady straining of the fluid and the dilatation part in the viscous strain are neglected since they may create pressure minimums without the presence of a vortex. Moreover, viscous effects are also neglected as they can eliminate pressure minimums in locations with vortical motion. \par 

\begin{figure*}[h!]
\centering
\begin{subfigure}{.45\textwidth}
  \centering
  \includegraphics[width=1.0\linewidth]{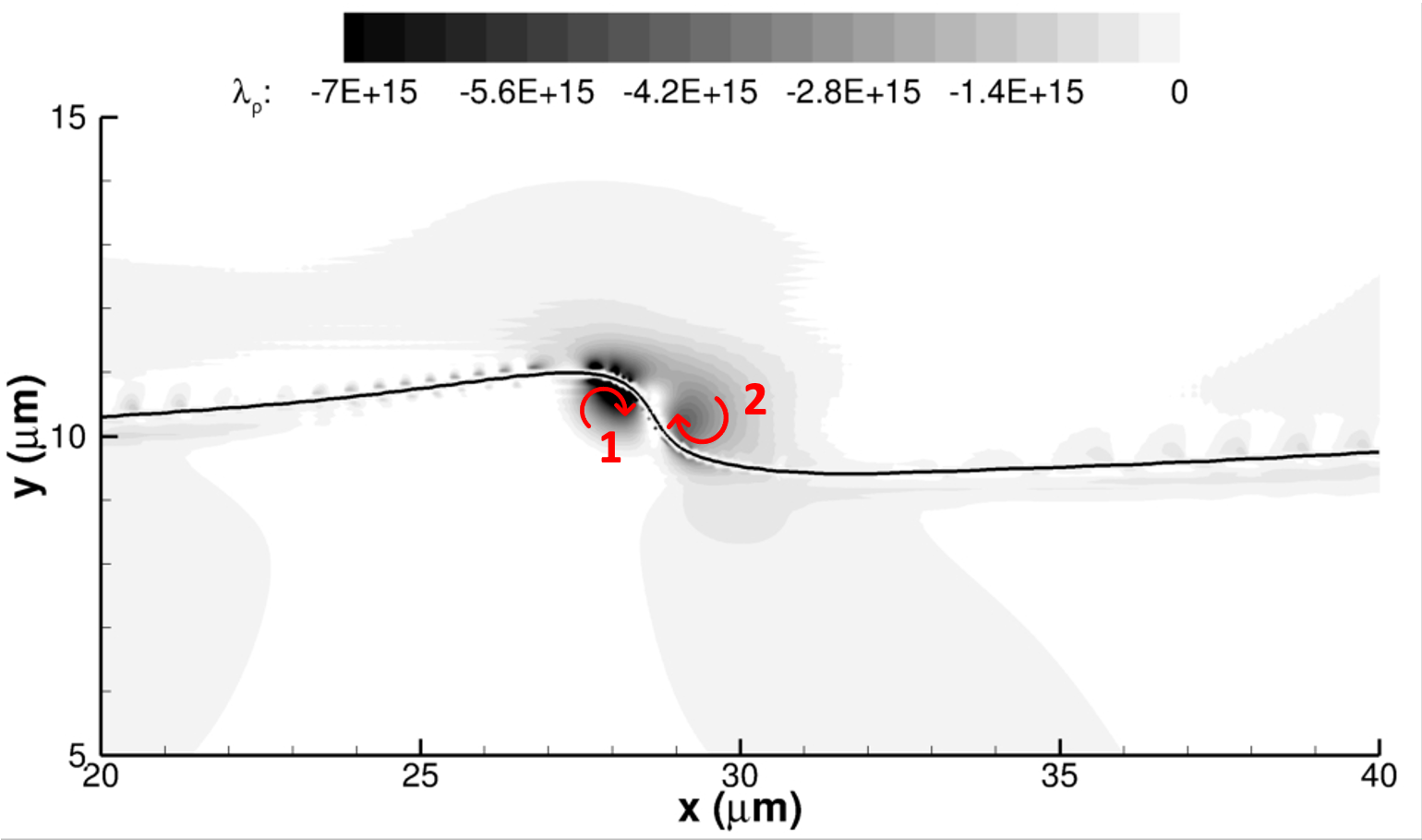}
  \caption{\label{subfig:vortex_1mus}\(\lambda_\rho\) at \(t=1\) \(\mu\)s}
\end{subfigure}%
\begin{subfigure}{.45\textwidth}
  \centering
  \includegraphics[width=1.0\linewidth]{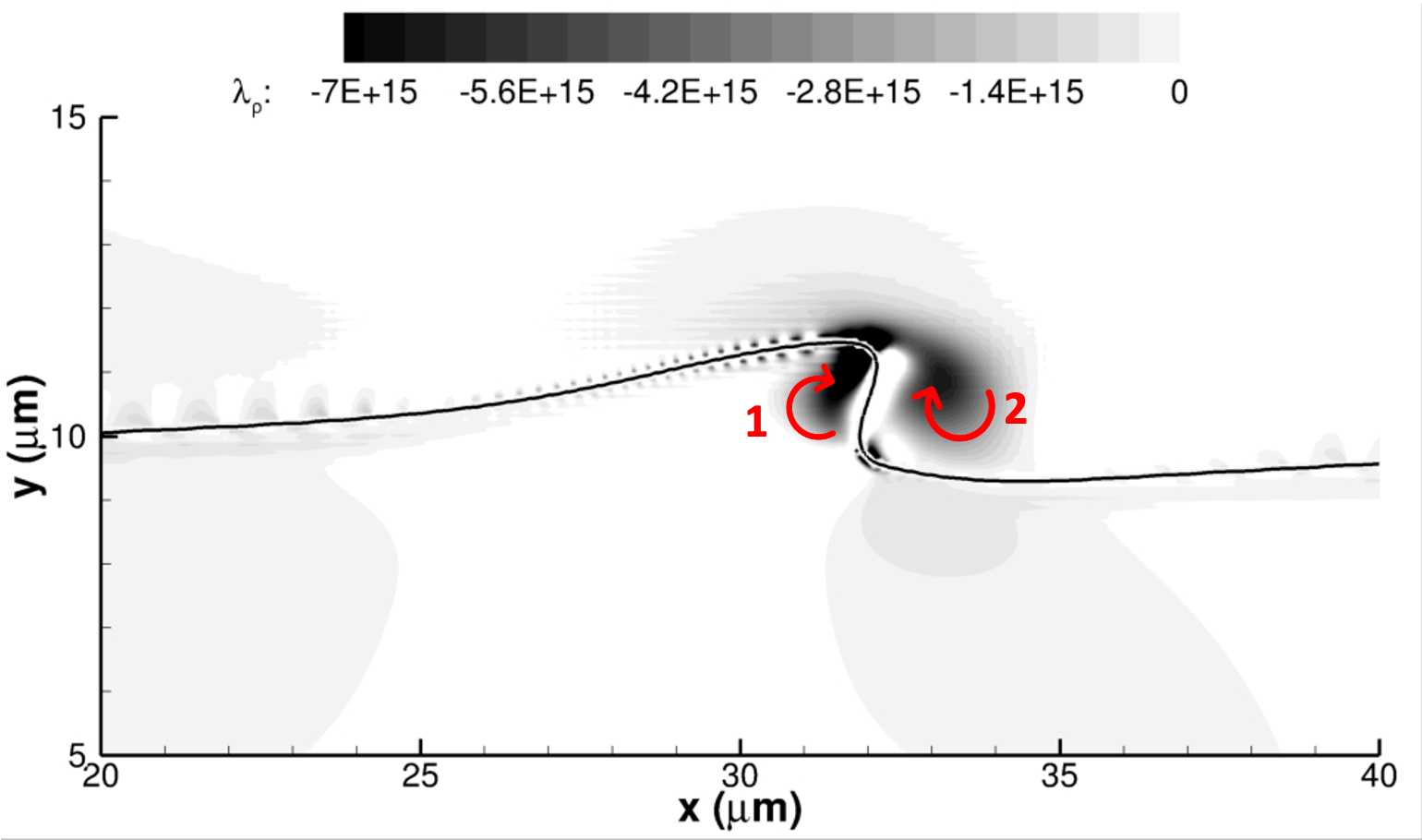}
  \caption{\label{subfig:vortex_1p3mus}\(\lambda_\rho\) at \(t=1.3\) \(\mu\)s}
\end{subfigure}%
\\
\begin{subfigure}{.45\textwidth}
  \centering
  \includegraphics[width=1.0\linewidth]{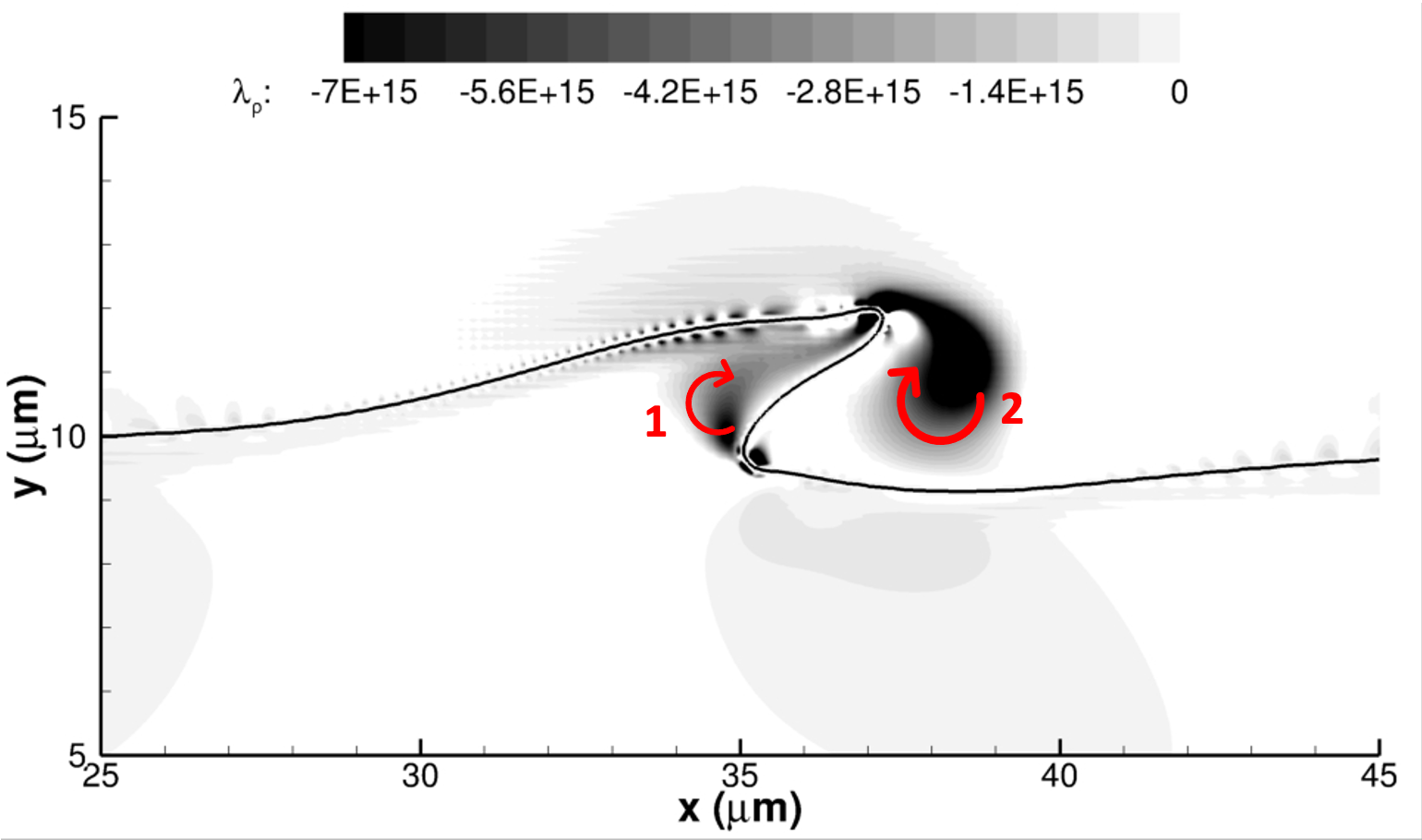}
  \caption{\label{subfig:vortex_1p6mus}\(\lambda_\rho\) at \(t=1.6\) \(\mu\)s}
\end{subfigure}%
\begin{subfigure}{.45\textwidth}
  \centering
  \includegraphics[width=1.0\linewidth]{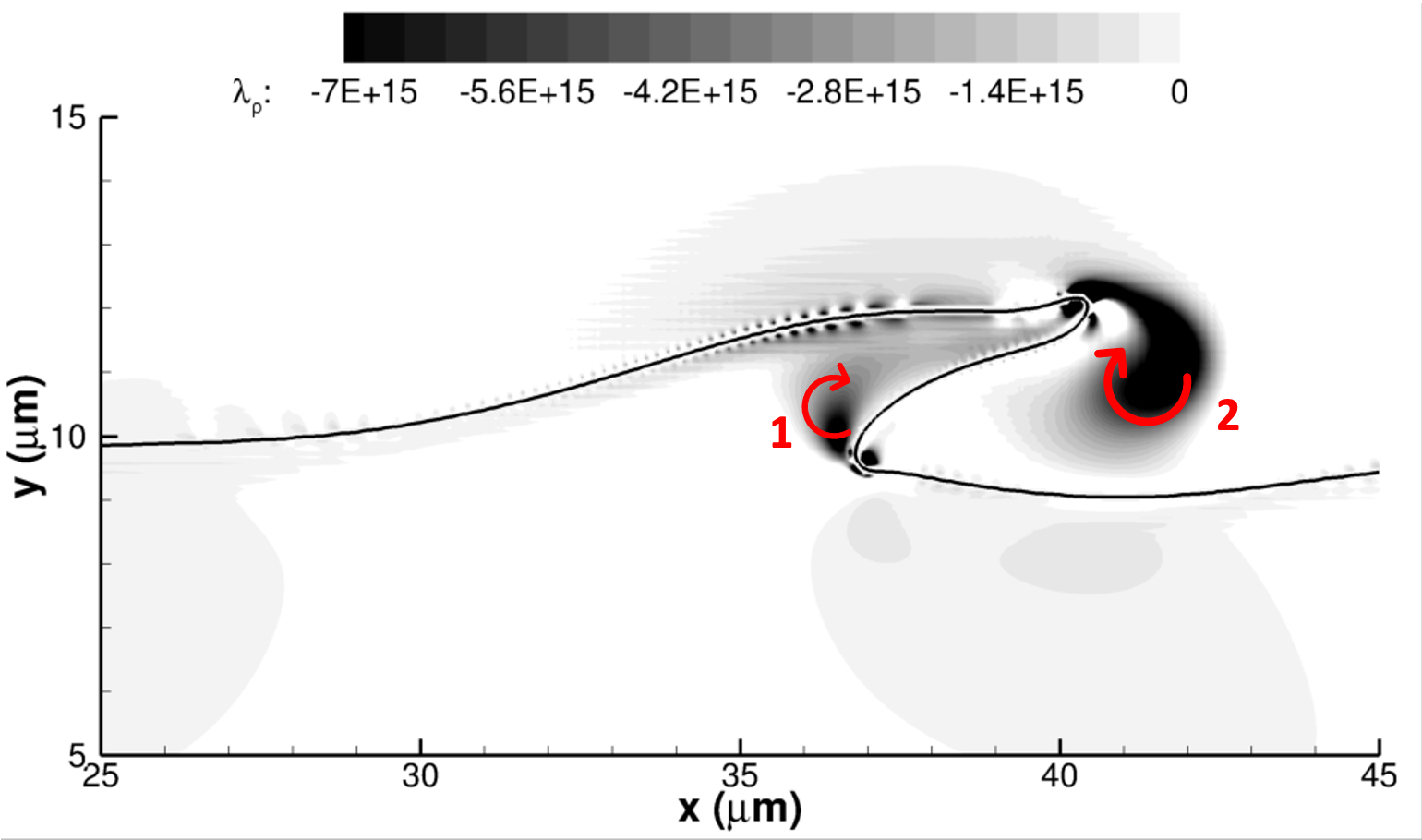}
  \caption{\label{subfig:vortex_1p8mus}\(\lambda_\rho\) at \(t=1.8\) \(\mu\)s}
\end{subfigure}%
\\
\begin{subfigure}{.45\textwidth}
  \centering
  \includegraphics[width=1.0\linewidth]{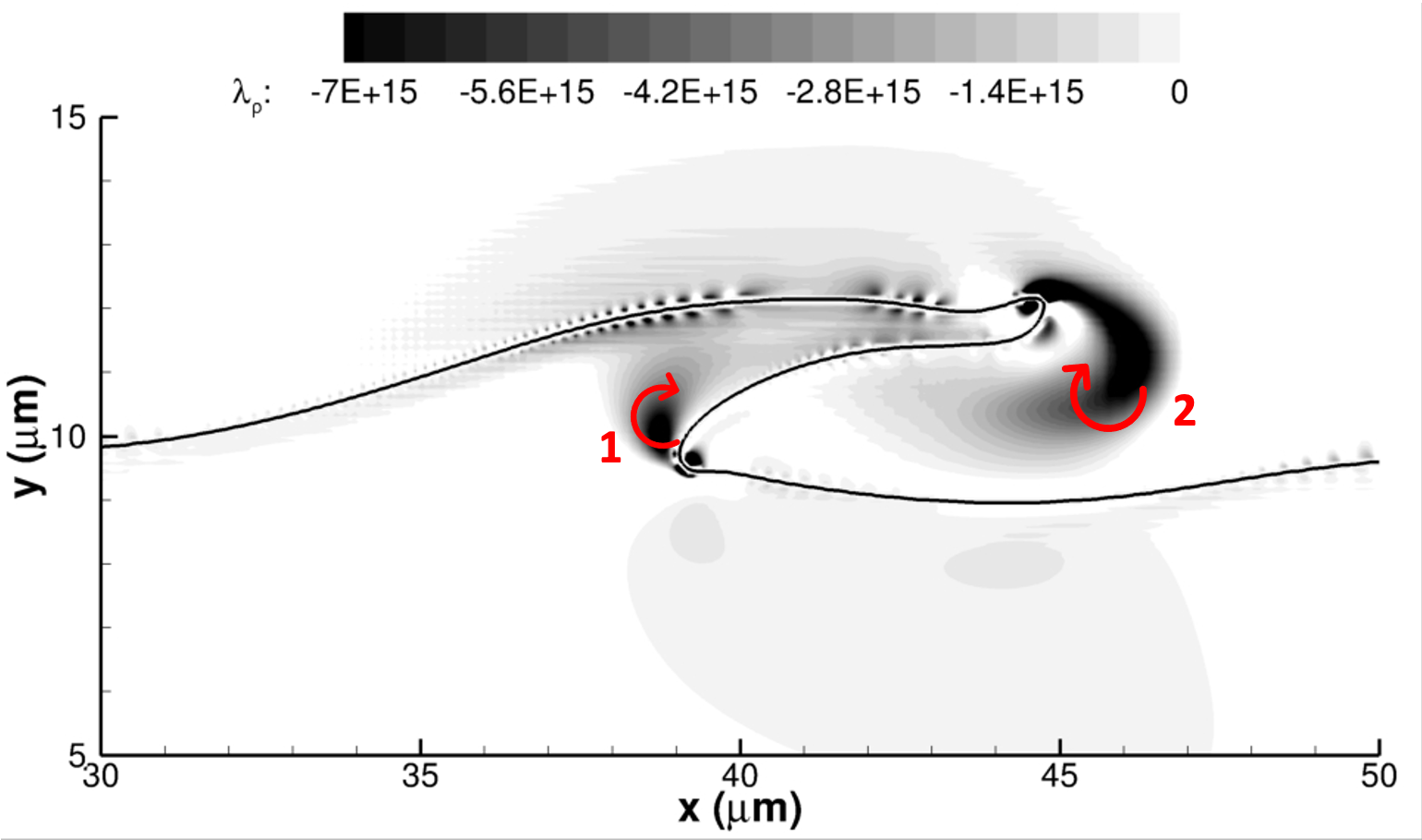}
  \caption{\label{subfig:vortex_2mus}\(\lambda_\rho\) at \(t=2\) \(\mu\)s}
\end{subfigure}%
\begin{subfigure}{.45\textwidth}
  \centering
  \includegraphics[width=1.0\linewidth]{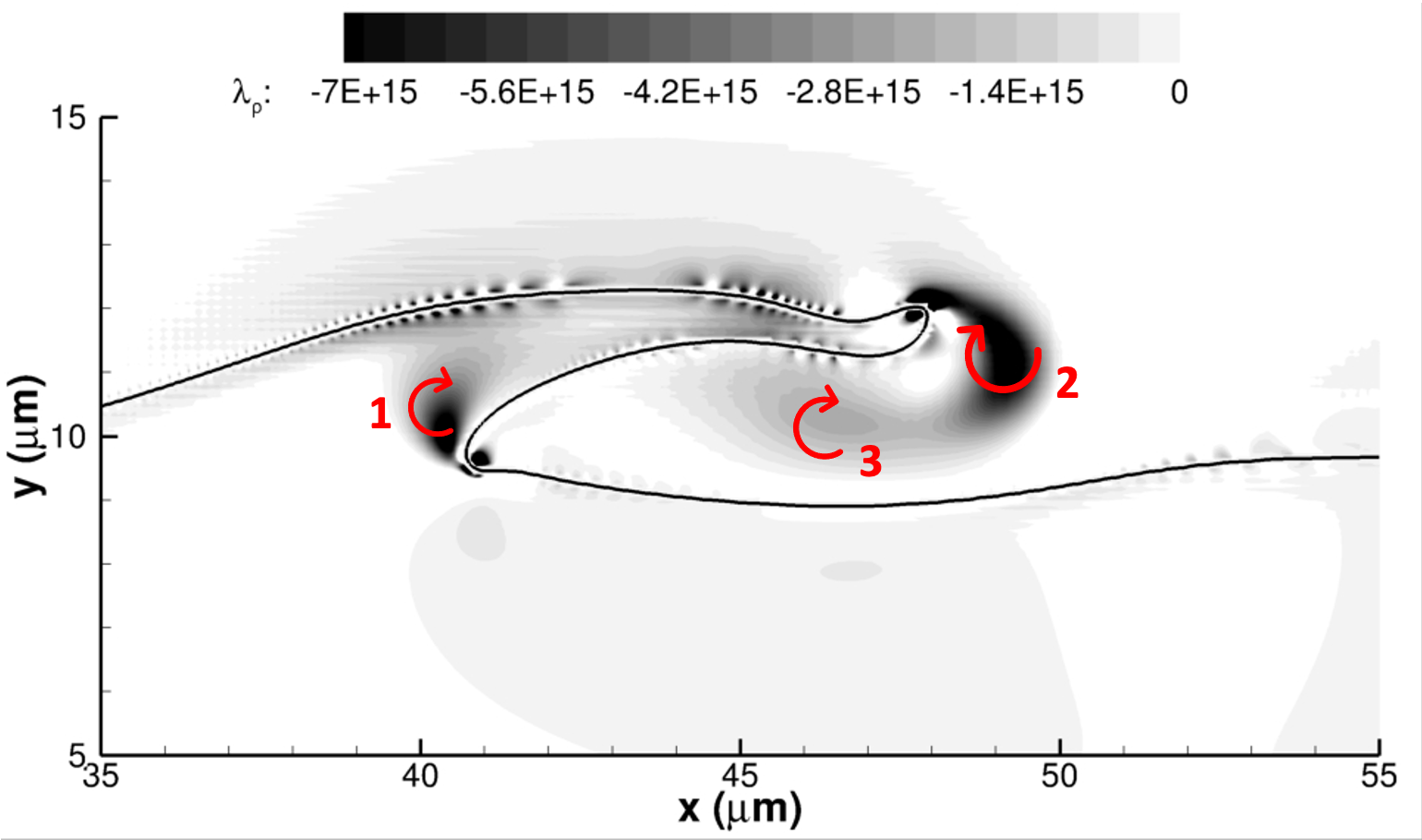}
  \caption{\label{subfig:vortex_2p2mus}\(\lambda_\rho\) at \(t=2.2\) \(\mu\)s}
\end{subfigure}%
\\
\begin{subfigure}{.45\textwidth}
  \centering
  \includegraphics[width=1.0\linewidth]{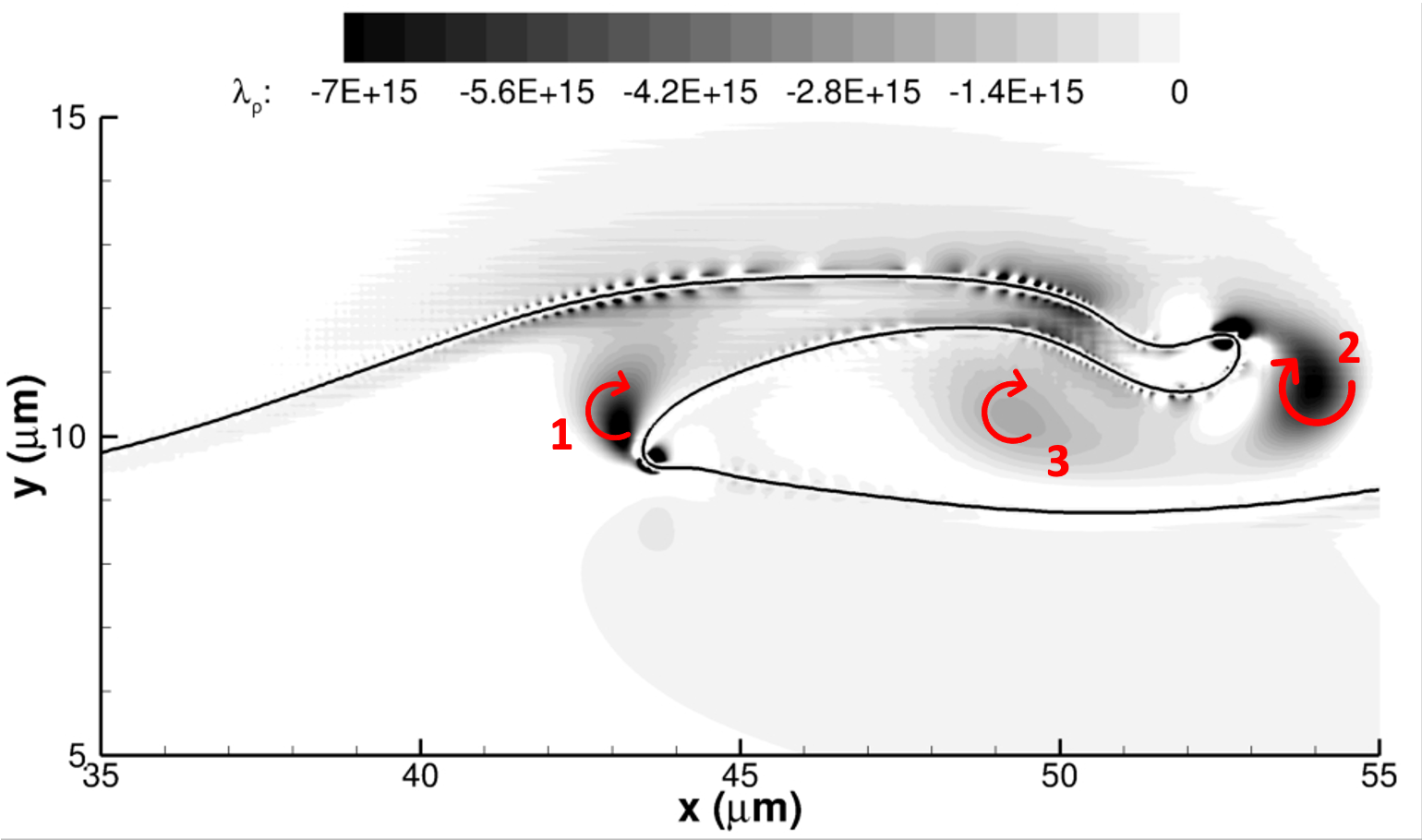}
  \caption{\label{subfig:vortex_2p5mus}\(\lambda_\rho\) at \(t=2.5\) \(\mu\)s}
\end{subfigure}%
\begin{subfigure}{.45\textwidth}
  \centering
  \includegraphics[width=1.0\linewidth]{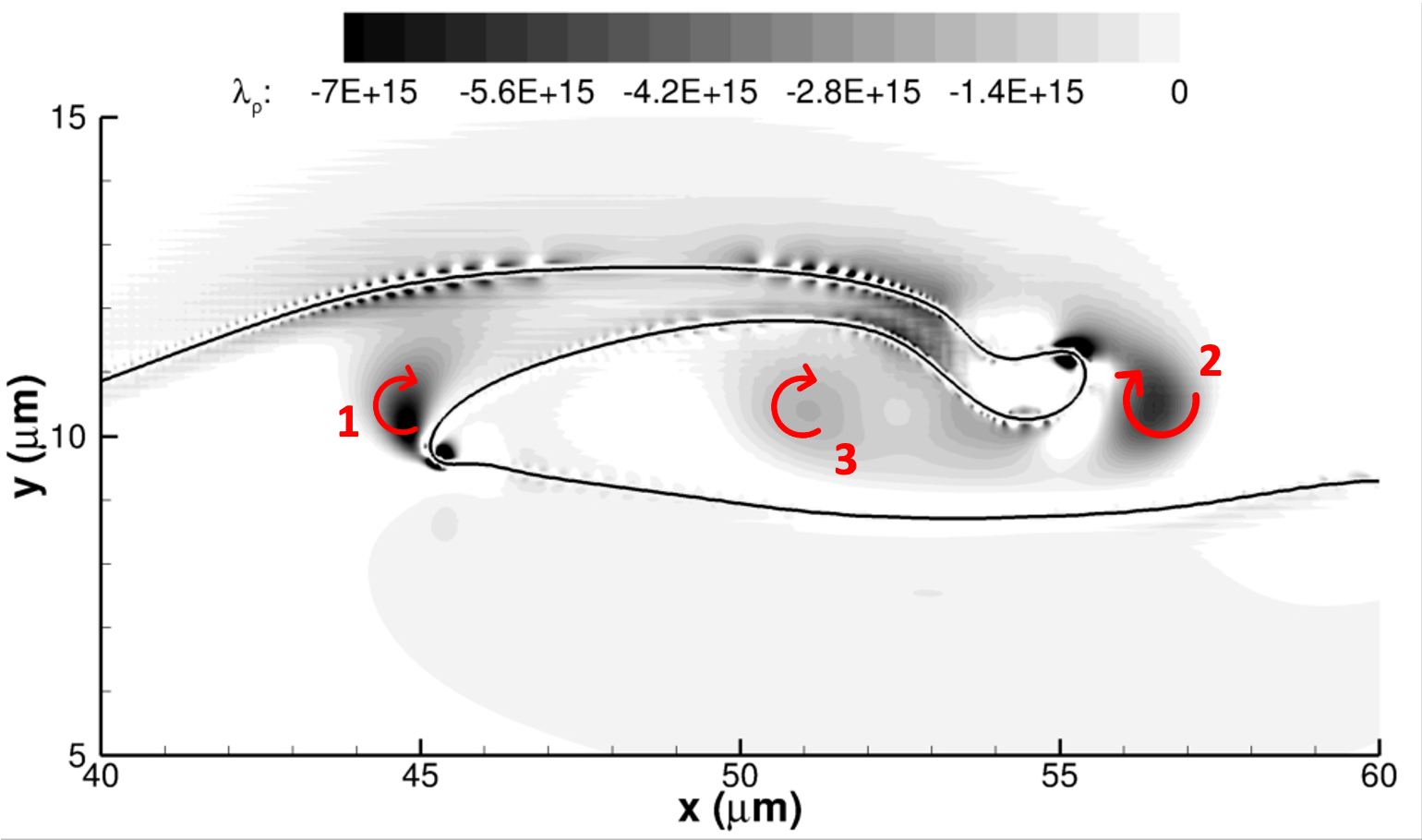}
  \caption{\label{subfig:vortex_2p7mus}\(\lambda_\rho\) at \(t=2.7\) \(\mu\)s}
\end{subfigure}%
\caption{\label{fig:2djet_lambda}\(\lambda_\rho\) contours for the two-dimensional planar jet at 150 bar. The red arrows point the rotation direction of the vortex. The interface location is highlighted with a solid black curve representing the isocontour with \(C=0.5\).}
\end{figure*}

The contours of \(\lambda_\rho\) in the two-dimensional jet are shown in Figure \ref{fig:2djet_lambda} where the locations of the main vortices from \(t=1\) \(\mu\)s to \(t=2.7\) \(\mu\)s are analyzed. The \(\lambda_\rho\) method does not provide information on the rotation direction of the vortex. Thus, the velocity field is used to determine the direction and annotate the plots. Some noise exists in the plots of \(\lambda_\rho\), especially near the interface due to the sharp averaging of fluid properties. In the future, spatial filtering might be applied to provide a cleaner picture of the \(\lambda_\rho\) contours. \par 

\begin{figure*}[h!]
\centering
\begin{subfigure}{.65\textwidth}
  \centering
  \includegraphics[width=1.0\linewidth]{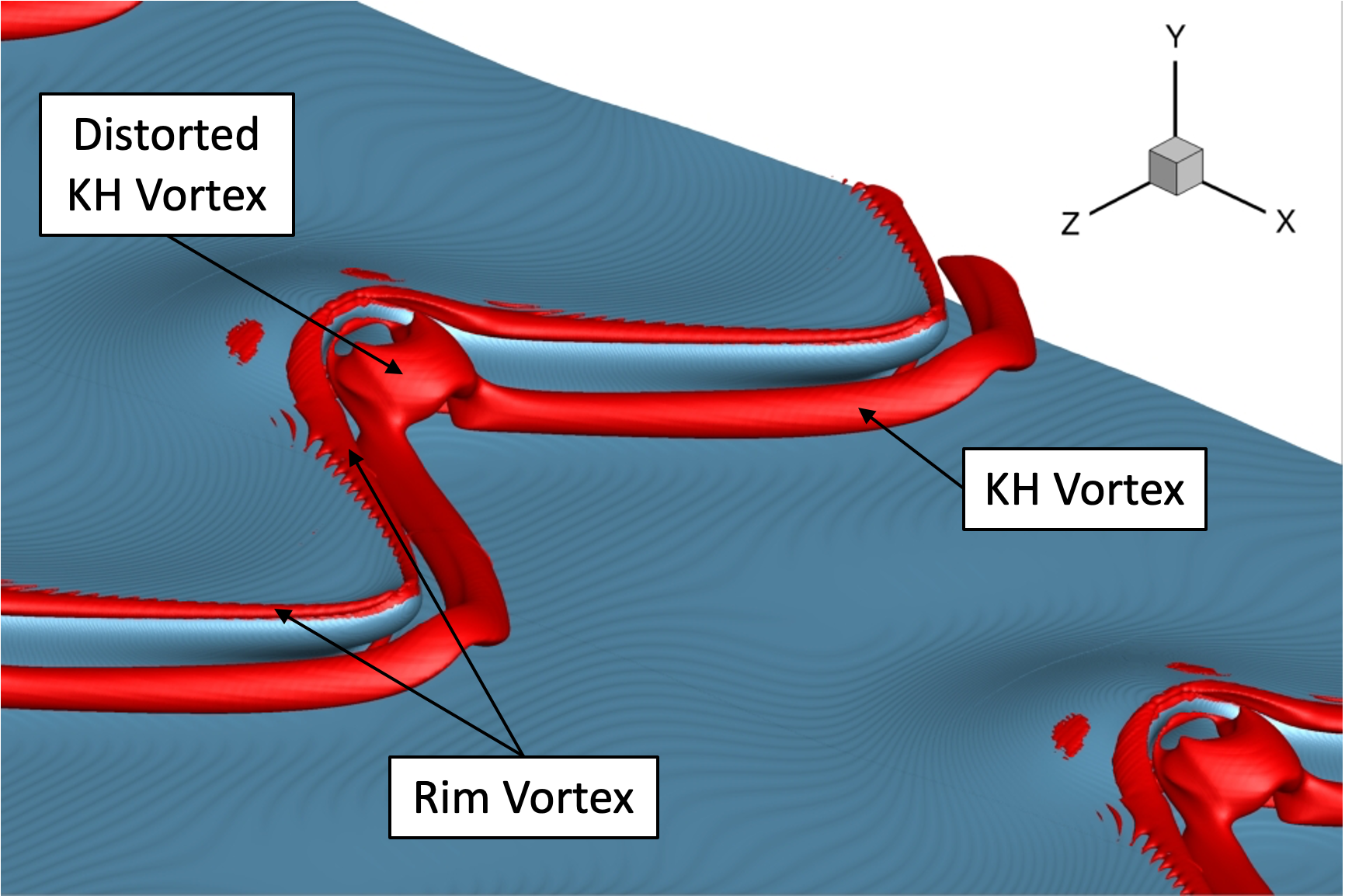}
  \caption{\label{subfig:150_A2_L2_2mus}\(t=2\) \(\mu\)s}
\end{subfigure}%
\\
\begin{subfigure}{.65\textwidth}
  \centering
  \includegraphics[width=1.0\linewidth]{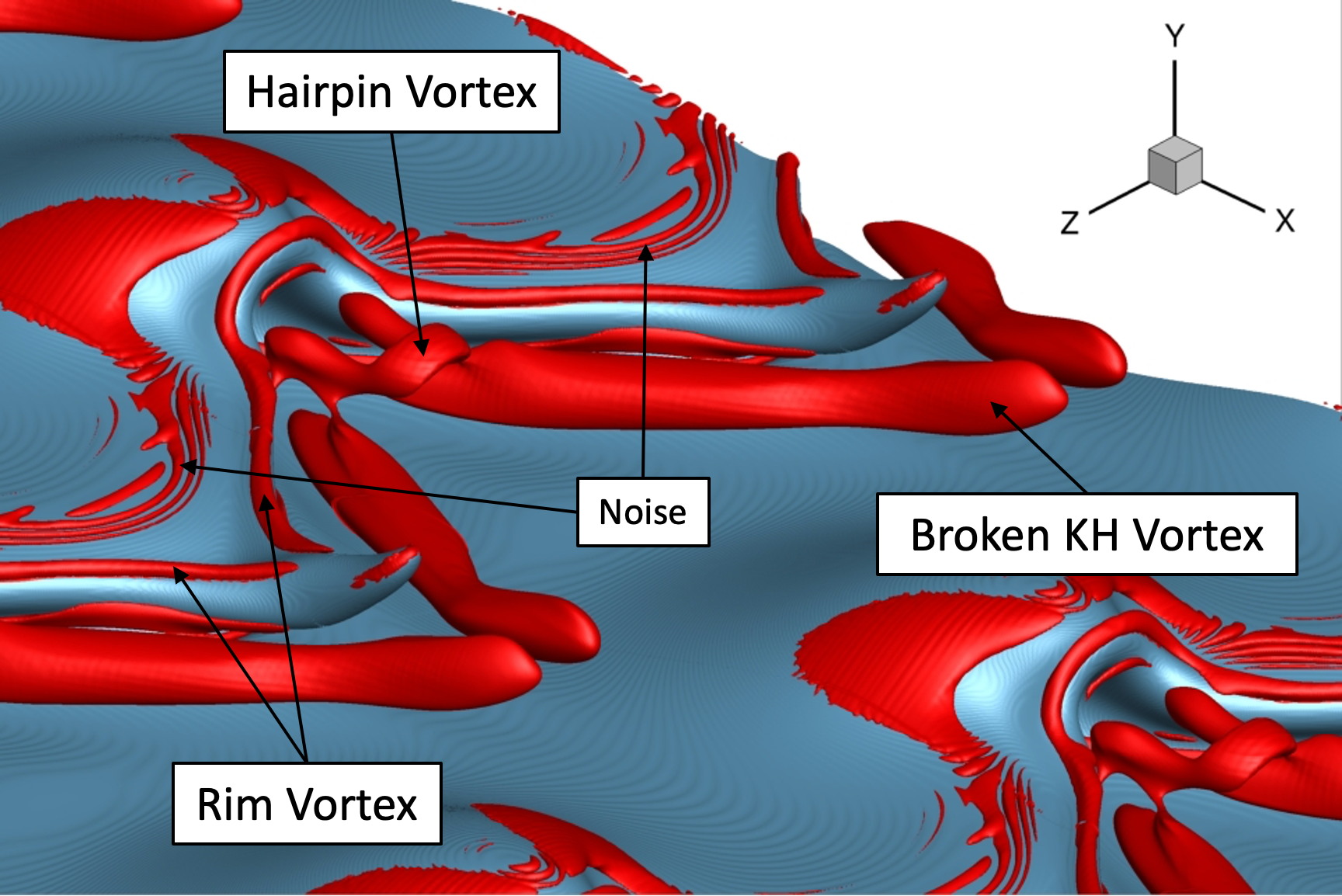}
  \caption{\label{subfig:150_A2_L2_3mus}\(t=3\) \(\mu\)s}
\end{subfigure}%
\caption{\label{fig:3djet_vortex}\(\lambda_\rho=-1\times 10^{15}\) kg/(m\(^3\)s\(^2\)) iso-surfaces (in red) for the three-dimensional planar jet at 150 bar with initial spanwise perturbation amplitude of 0.5 \(\mu\)m. Two different instants of time are shown (i.e., 2 \(\mu\)s and 3 \(\mu\)s). The interface location is identified as the iso-surface with \(C=0.5\).}
\end{figure*}

Initially, two clockwise vortices appear, one in each phase. Vortex 1 belongs to the liquid phase and remains attached to the inner liquid curving point during the analyzed times. This vortex pushes liquid from the core into the two-dimensional lobe and tends to reduce the angle between the lobe and the liquid core surface. Vortex 2 is created as a result of the faster gas flowing over the liquid. As the gas inertia stretches the lobe, vortex 2 remains located in front of the lobe's tip (downstream of the wave). This vortex can be understood as a Kelvin-Helmholtz (KH) vortex and its induced motion is responsible for rolling over the liquid lobe and for the gas entrainment under the wave seen in other figures (e.g., Figure~\ref{fig:2djet_T}). Moreover, vortex 2 is able to push upwards the tip of the lobe as it stretches. This liquid region is less dense and substantially less viscous than the liquid core, thus it can be affected more easily by the gas phase dynamics. Vortex 2 stretches under the lobe and between 2 \(\mu\)s and 2.2 \(\mu\)s, vortex 3 detaches from vortex 2 and moves directly underneath the lobe. Then, its clockwise rotation accentuates the rolling motion of the lobe. The small vortex appearing at the very tip of the lobe will be identified as the ``rim vortex" in the three-dimensional computations shown in Figure~\ref{fig:3djet_vortex}. It can be understood as an extension of vortex 2 with the same clockwise rotation. \par 

As done for the two-dimensional jet, the \(\lambda_\rho\) method is also used to identify vortices in the three-dimensional jet that might explain the deformation patterns of the liquid surface. Some preliminary results are shown in Figure~\ref{fig:3djet_vortex}, where two instants of time (i.e., 2 \(\mu\)s and 3 \(\mu\)s) for the three-dimensional planar jet with initial spanwise amplitude of 0.5 \(\mu\)m are presented. The \(\lambda_\rho\) noise in three dimensions is more troublesome than in two dimensions since the visualization of iso-surfaces of \(\lambda_\rho\) defining vortices can be affected. \par

Looking at vortical structures in the gas phase, the KH vortex inducing the stretching and rolling of the lobes deforms substantially with time. The rotation of this vortex is defined negative following the positive \(z\)-direction. At \(t=2\) \(\mu\)s, the KH vortex can be identified downstream of the lobe (see Figure~\ref{subfig:150_A2_L2_2mus}). The KH vortex is distorted in the region connecting two consecutive lobes, where a strong recirculation occurs as gas penetrates underneath the liquid surface. Moreover, another vortex is identified following the edge or rim of the lobe formation. This type of vortex is also observed at the upper side of the lobe's tip in the two-dimensional results presented in Figure~\ref{fig:2djet_lambda}. \par

Later at \(t=3\) \(\mu\)s, the initial KH vortex has deformed completely. The distortion near the region connecting two consecutive lobes transitions to a hairpin-type vortex, while the vortex breaks downstream of the lobe's tip. The upward motion induced by the two vortices surrounding the ligament pushes the liquid into the hotter gas phase. As previously discussed, this liquid region presents a lower density and viscosity and can be more easily affected by the gas motion. Some visualization noise can be observed. \par

\section{Conclusions}

The early deformation of real liquid jets at supercritical pressures has been presented in this paper. A recently developed numerical foundation detailed in Poblador-Ibanez and Sirignano~\cite{poblador2021vof} has been used to perform a temporal study of a two-dimensional planar liquid jet composed of \textit{n}-decane at 450 K and submerged into hotter oxygen at 550 K. The ambient pressure of 150 bar is supercritical for the injected liquid. Based on Poblador-Ibanez and Sirignano~\cite{poblador2019axisymmetric}, an initial perturbation with wavelength 30 \(\mu\)m and amplitude 0.5 \(\mu\)m is imposed, with the gas moving at 30 m/s. \par

Two-dimensional results already show some of the main features of liquid injection at supercritical pressures. The enhanced dissolution of the lighter gas species into the liquid at these high pressures causes a reduced surface tension environment with liquid density and viscosity dropping near the interface. Fluid properties vary along the interface, affecting the surface tension coefficient, which is the lowest near the initial perturbation crest. These features generate fast growing instabilities characterized by a short wavelength. The perturbations are easily affected by the gas inertia in liquid regions with higher oxygen concentration. Over time, elongated and thin liquid structures are generated, with eventual breakup into two-dimensional droplets. With a hotter gas and cooler liquid, condensation can occur, especially along portions of the wavy interface where the gas is expanded and the liquid is compressed. The fuel vapor also mixes considerably with the oxidizer stream. Small discrete domains of high fuel-vapor concentration form in the gas phase at intervals, along the interface, in both the two- and three-dimensional computations. This is an important observation for combustion applications at high pressures. \par 

Since supercritical liquid injection is a two-phase problem, vortex dynamics explain the liquid deformation process as detailed in previous incompressible works~\cite{jarrahbashi2014vorticity,jarrahbashi2016early,zandian2017planar,zandian2018understanding,zandian2019length,zandian2019vorticity}. At high pressures, this analysis must include the evolution of the liquid fluid properties as they play a role in the deformability of the liquid. \par 

The three-dimensional picture becomes more complex, as surface tension force acts differently from the minimizing of surface area for two-dimensional structures. A higher tendency for neck formation and breakup is expected, albeit surface tension forces are weaker at high pressures. Moreover, the complexity of the interface thermodynamics increases as fluid properties vary along the surface and mixing generates regions with substantially different mixture properties. Limited results are presented in this paper comparing three initial configurations with different spanwise perturbation amplitudes. Three-dimensional effects are shown and discussed, which are stronger for larger initial amplitudes in the spanwise direction. \par 

Future work includes analysis of various three-dimensional configurations with varying ambient pressure, jet size, and initial interface perturbation amplitude. The goal is to determine the main characteristics of the early deformation cascade process, study the role of vortex dynamics and identify parameter values to classify atomization regimes at supercritical pressures. \par


%
%

%

\section*{Acknowledgments}
The authors are grateful for the support of the NSF grant with Award Number 1803833 and Dr. Ron Joslin as Scientific Officer. The authors are also grateful for the helpful discussions with Prof. Antonino Ferrante and his student Pablo Trefftz-Posada regarding the development of the VOF methodology used in this work. \par 

This work utilized the infrastructure for high-performance and high-throughput computing, research data storage and analysis, and scientific software tool integration built, operated, and updated by the Research Cyberinfrastructure Center (RCIC) at the University of California, Irvine (UCI). The RCIC provides cluster-based systems, application software, and scalable storage to directly support the UCI research community. https://rcic.uci.edu \par


\bibliography{ilass_bib_full}

 \end{document}